%% file: insider_v2.tex
\documentclass[11pt,letterpaper]{article}
\usepackage[utf8]{inputenc}
\usepackage[T1]{fontenc}
\usepackage{lmodern}
\usepackage{amsmath,amssymb}
\usepackage{graphicx}
\usepackage{booktabs}
\usepackage{array}
\usepackage{geometry}
\usepackage{caption}
\usepackage{placeins}
\usepackage[round,authoryear]{natbib}
\usepackage{microtype}
\usepackage[hidelinks]{hyperref}
\usepackage{bookmark}

\hypersetup{pdftitle={Insider Purchases Far Below the 52-Week High: Decomposing the Disclosure Reaction in Microcap Equities},
            pdfauthor={Hangyi Zhao},
            pdfkeywords={insider trading; Form 4; microcap stocks; 52-week high; announcement reaction; event study; abnormal returns; gradient boosting; data quality}}

\input{outputs/metrics.tex}

\input{outputs/extension_metrics.tex}
\input{outputs/d18_metrics.tex}

\newcommand{\tabfoot}{\footnotesize\linespread{1}\selectfont\setlength{\tabcolsep}{3pt}\renewcommand{\arraystretch}{1.0}}
\newcommand{\tabtiny}{\scriptsize\linespread{1}\selectfont\setlength{\tabcolsep}{2.5pt}\renewcommand{\arraystretch}{1.0}}
\newcommand{\CAR}{\ensuremath{\mathrm{CAR}_{[1,30]}}}
\newcommand{\arxivid}{arXiv:2602.06198}
\makeatletter\newcommand{\tabinput}{\@@input}\makeatother
\newcommand{\shrinkbox}[1]{\resizebox{\ifdim\width>\textwidth\textwidth\else\width\fi}{!}{#1}}

\title{Insider Purchases Far Below the 52-Week High:\\Decomposing the Disclosure Reaction in Microcap Equities\thanks{This version supersedes the February 2026 version, posted under the title ``Insider Purchase Signals in Microcap Equities: Gradient Boosting Detection of Abnormal Returns''; Appendix~\ref{app:changes} summarizes the changes.}}
\author{Hangyi Zhao\\\texttt{hyz0815@stanford.edu}}
\date{This version: September 2026 \\ First version: February 2026 (\arxivid{}v1)}

\begin{document}
\maketitle

\begin{abstract}
\noindent Purchases reported under transaction code P on SEC Form~4 by insiders of U.S.\ equities with an estimated filing-date capitalization of \$30 million to \$500 million (\ssN{} lines, \ssNTickers{} issuers, \ssYearsRange) are followed by a first-day abnormal return that rises steeply with the stock's distance below its 52-week high: \rxQoneArToneMean{} in the quintile farthest below the high against \rxQfiveArToneMean{} nearest it (two-way clustered $t$ = \rxDiffArToneTTwo); random non-event days of the same issuers show \plQoneCeboArToneMean. Five tests with decision rules fixed in advance characterize the gradient. Most of it is scale: the beaten-down stocks are \dxSigmaRatio{} times as volatile, and with a full set of controls the raw gap fails its pre-specified bar (\dxCtrlQoneCoefPp{} points, $t$ = \dxCtrlQoneTTwo). Per unit of the stock's own volatility the reaction is \dxZRatio{} times as large far below the high ($t$ = \dxZDiffTTwo), \drMatchedZRatio{} to \drMedianZRatio{} on other estimators, though a variance-weighted slope shows none. The gradient is larger than for insider sales by the same issuers and for positive earnings surprises as a class; against the strongest surprises the difference is imprecise. Dropping purchases with a concurrent 8-K leaves the raw gradient intact ($t$ = \dnUnflaggedGapTTwo), but the per-risk gradient no longer clears the controls ($t$ = \dnCtrlUnflaggedZQoneTTwo). The reaction runs for two to three sessions; the 29-day drift is imprecise (two-way $t$ = \rxDiffDriftThirtyTTwo) and a calendar-time portfolio that skips the first day earns no significant alpha. The analysis quantifies sensitivity to price adjustment and benchmark specification: mixing price bases misassigns run-up buckets, and carrying the estimation-window intercept supplies \mechQoneQfiveCarryShareOfVOneGap{} of the 30-day gradient under that benchmark. A gradient-boosting classifier (test AUC \clXgbAUCthree) is indistinguishable from logistic regression. A separate large-cap extension schedules \ExtClusterFlowAnn{} a year of buyer-cluster flow but fails every matched-comparison gate.
\end{abstract}


\section{Introduction}\label{sec:intro}

This paper examines how the market reaction to reported insider purchases varies with distance below the 52-week high, how much of that variation reflects volatility and benchmark construction, and whether returns persist beyond the initial disclosure response. Corporate insiders generally report their trades on SEC Form~4 within two business days \citep{sec2002,sec_form4}. Purchases are followed by positive abnormal returns, especially in small firms \citep{jaffe1974,seyhun1986,lakonishok2001,jeng2003}, but a return measured over several weeks can combine an immediate response to public information, subsequent drift and the expected-return benchmark's treatment of the pre-event price path. Separating these components is necessary both to interpret the disclosure and to assess whether an outside investor can profit from it.

The sample contains \ssN{} transaction-code-P purchase lines from \ssNTickers{} U.S.\ issuers over \ssYearsRange, with estimated filing-date capitalization between \$30 million and \$500 million. Transactions come from Form~4 filings compiled by OpenInsider, prices from Yahoo Finance, and most share counts from point-in-time SEC disclosures. Code P covers purchases that are not necessarily discretionary open-market trades; the source audit distinguishes reported venues and ownership forms. The sample is also conditional on the availability of price histories in 2026. These restrictions bound the interpretation to the retained microcap sample.

The central result is a steep gradient in the disclosure reaction. On the first trading day after the filing day, mean abnormal return is \rxQoneArToneMean{} in the quintile farthest below the 52-week high and \rxQfiveArToneMean{} nearest it (two-way issuer-and-month clustered $t$ = \rxDiffArToneTTwo). Random non-event days of the same issuers show a next-day abnormal return of \plQoneCeboArToneMean{} in the farthest quintile. Acceptance times help locate the response: filings accepted before the close are priced on the filing day, while those accepted after it are priced on the next session, mostly through an overnight gap. The reaction also survives a \$5 price floor and a \$1 million dollar-volume floor (Section~\ref{sec:reaction}).

Most of the raw gradient reflects volatility. Stocks farthest below the high are \dxSigmaRatio{} times as volatile as those nearest it, and the raw gap falls to \dxCtrlQoneCoefPp{} points with the full control set (two-way $t$ = \dxCtrlQoneTTwo), below the pre-specified $|t| \geq 3$ threshold. Per unit of pre-event volatility, the reaction is \dxZRatio{} times as large in the farthest quintile ($t$ = \dxZDiffTTwo). Alternative estimators give ratios from \drMatchedZRatio{} to \drMedianZRatio, although a variance-weighted slope shows no gradient. The volatility-scaled gap clears the controls in the full sample but not among purchases free of concurrent 8-K news. It is larger than the corresponding gradients for insider sales and positive earnings surprises as a class; against the strongest surprises, the difference is imprecise. The five principal tests have decision rules fixed before they were run, after the day-one pattern was known; the alternative estimators, standardized cross-event comparisons and concurrent-news screen are post hoc (Section~\ref{sec:identification}). These results support a conditional disclosure association without isolating a causal information channel.

The reaction extends over the first two to three sessions. Returns over event days 2--30 have a much less precise quintile difference (two-way $t$ = \rxDiffDriftThirtyTTwo), and a calendar-time portfolio that skips the first day earns no significant alpha. For after-close filings, the overnight component is unavailable to an investor who first observes the filing after that close. A multi-week event window therefore combines a precisely measured early response with a drift whose size this panel cannot establish.

Two measurement choices affect this decomposition. Comparing an adjusted daily close with an as-reported Form~4 price mixes share bases: later splits and dividends reassign observations across disclosure-run-up buckets. Under either price basis, the mean run-up premium is statistically indistinguishable from zero on the present sample. Separately, carrying the estimation-window intercept into expected returns contributes \mechQoneQfiveCarryShareOfVOneGap{} of the 30-day quintile gradient under that benchmark. Non-event days reproduce \plGapVOneOverEventVOne{} times the events' gradient with the intercept included, but show a gap indistinguishable from zero under the loadings-only benchmark. The primary specification uses the latter, with the intercept-inclusive and market-adjusted definitions reported as robustness. This choice is conditional on evidence about persistence: the simulations in Appendix~\ref{app:robust} favor including the intercept when the underlying alpha persists.

A gradient-boosting classifier provides a complementary assessment of prediction at disclosure. Its 2024 AUC for a 30-day abnormal return above 10\% is \clXgbAUCthree{} (issuer-cluster bootstrap 95\% interval \clXgbAUCci), statistically indistinguishable from random forest and logistic regression. Volatility is its top feature, and volatility alone ranks nearly as well (AUC \clUniTopAUCthree). The evaluation thus helps distinguish predictability of a large positive tail outcome from predictability of directional returns. A separate liquid-universe extension schedules \ExtClusterFlowAnn{} a year of nominal buyer-cluster flow, but its incremental return claim fails the matched-comparison requirements (Section~\ref{sec:extension}).

The paper contributes along three dimensions. First, it decomposes the price-state gradient in disclosure reactions into volatility, the first sessions after publication and subsequent drift, with non-event days, insider sales and earnings announcements as comparisons. Second, it quantifies the sensitivity of Form~4 event studies to price adjustment and benchmark specification. Third, it evaluates classification and trading relevance with issuer-clustered uncertainty, matched model thresholds, costs and capacity constraints. The evidence concerns disclosure responses and measurement; it does not establish classifier superiority or a profitable trading rule.

The closest disclosure studies examine announcement-day reactions to directors' dealings \citep{fidrmuc2006}, Form~4 reactions around the two-business-day reporting rule \citep{brochet2010}, and purchase disclosures conditional on preceding earnings signals \citep{veenman2012}. \citet{lee2019} and \citet{lasfer2024} document insiders' trading around the 52-week reference point. This paper connects that price state to the response at the individual filing, separating the overnight gap, early sessions and later drift. Machine learning on insider filings is also established: \citet{kong2025} use gradient-boosted trees to construct a monthly cross-sectional signal and find the effect strongest in the smallest stocks. The filing-level analysis here complements that portfolio evidence by examining the return components and the uncertainty behind individual predictions.

Sections~\ref{sec:lit}--\ref{sec:design} describe prior work, the data and the empirical protocol. Sections~\ref{sec:results}--\ref{sec:robustness} present prediction results, benchmark decomposition, disclosure reactions, identification tests and robustness. Section~\ref{sec:extension} reports the liquid-universe extension, followed by discussion, limitations and conclusions.

\section{Related literature}\label{sec:lit}

\paragraph{Informativeness of insider purchases.} \citet{lorie1968} and \citet{jaffe1974} established that insiders' reported trades predict returns; \citet{seyhun1986} showed that abnormal returns concentrate in small firms and that trading costs erode outsiders' profits, a point \citet{rozeff1988} sharpened; \citet{lakonishok2001} and \citet{jeng2003} showed that purchases, not sales, are the informative side. \citet{rozeff1998} find that insiders trade as contrarians; \citet{lakonishok2001} that purchases are informative mainly in small firms and when several insiders buy; \citet{cohen2012} separate routine from opportunistic insiders, a distinction Section~\ref{sec:roles} applies; \citet{jeng2003} estimate purchase abnormal returns of about six percent a year; \citet{ke2003}, \citet{piotroski2005} and \citet{jenter2005} tie insider trades to forthcoming accounting news and managers' own view of mispricing. The reaction on the day a trade becomes public is a separate strand: \citet{fidrmuc2006} for directors' dealings in the United Kingdom, \citet{brochet2010} for Form~4 filings before and after the Sarbanes--Oxley two-day rule, and \citet{veenman2012}, who conditions the filing-day reaction to purchase disclosures on the earnings signals that preceded them. Section~\ref{sec:reaction} conditions the same reaction on the pre-event price path.

\paragraph{Reference points and slow diffusion.} \citet{george2004} show that nearness to the 52-week high predicts returns in the direction of momentum and subsumes much of the \citet{jegadeesh1993} effect. That insiders themselves trade against this reference point is documented: \citet{lee2019} read insiders' reluctance to buy near the high as anchoring on the insiders' own part, and \citet{lasfer2024}, on 586,742 insider-days over 1994--2018, read trading at the 52-week extremes as insiders exploiting outside investors' anchoring. The sort variable of this paper and the concentration of insider purchases far below the high are therefore documented. What those papers measure is the return earned from the insider's own transaction date; what Section~\ref{sec:reaction} measures is the market's reaction at the \emph{filing} date, decomposed into the first trading day after the filing, the overnight gap inside it, and the subsequent drift. Here it is distance below the high, not proximity to it, that raises the probability of a large positive abnormal return; Section~\ref{sec:mechanism} sorts on the same variable and separates dispersion from direction, and Section~\ref{sec:reaction} separates the announcement reaction from drift. Slow diffusion of firm-specific information in small, thinly covered stocks \citep{hong1999,hong2000,hou2007} is the natural reason a state variable observed at disclosure could still predict a 30-day abnormal return, and the channel a run-up premium would need; Section~\ref{sec:doublesort} tests it across dollar-volume terciles.

\paragraph{Measurement in event studies.} Mixing retrospectively adjusted per-share data with as-reported amounts can misclassify observations \citep{baber2002,payne2003}; vendor-adjusted price levels need not represent prices on the historical share basis \citep{francis2018}. Benchmark misspecification can also interact with selection on prior performance. \citet{ahern2009} demonstrate this problem using randomly dated non-events, \citet{kothari2007} discuss placebo diagnostics, and \citet{goldsmithpinkham2025} express the role of the intercept under factor-model misspecification. Sections~\ref{sec:artefact} and \ref{sec:mechanism} quantify these effects in the Form~4 sample.

\paragraph{Machine learning, interpretability and inference.} Form~4 applications include predicting the occurrence or size of insider trades \citep{batebi2026}, opportunistic selling \citep{tang2025}, filing delays \citep{huang2025}, unlawful-trade labels \citep{neupane2024} and single-stock prices \citep{chakravorty2025}. \citet{ahn2026} classify narrative tone in filings concurrent with multi-insider buy clusters; \citet{kravet2026} study how acquisition context affects the credibility of purchase signals. \citet{gu2020} and \citet{chen2024} show that tree ensembles and neural networks improve return prediction over linear models. I use XGBoost \citep{chen2016} for its native treatment of missing values and its attribution toolkit, SHAP values from the tree explainer of \citet{lundberg2017} and \citet{lundberg2020}. The evaluation follows the event-study and inference literature: three-factor abnormal returns \citep{fama1993}, the buy-and-hold distinction of \citet{barber1997} and the bid--ask bias of \citet{blume1983}, cross-sectional correlation \citep{kolari2010}, calendar-time portfolios \citep{mandelker1974,fama1998,mitchell2000,kothari2007}, clustered errors \citep{petersen2009,cameron2011}, Newey--West errors \citep{newey1987} and the DeLong test \citep{delong1988}.

\section{Data}\label{sec:data}

\subsection{Sources}\label{sec:sources}

Insider transactions are purchases reported under Form~4 transaction code ``P'' from OpenInsider \citep{openinsider}, a public aggregator of SEC Form~4 filings, whose screener was queried month by month on filing date from January 2018 to December 2024 with the purchase-only filter; it returned \ssRawPurchaseN{} raw lines on \ssRawPurchaseTickers{} ticker symbols with the filing timestamp, transaction date, ticker, insider name and title, price, quantity, post-transaction holdings and value (Appendix~\ref{app:data}). Securities are identified by ticker string, so ``issuers'' are distinct ticker strings; ticker changes, symbol reuse and reverse mergers are not tracked beyond the price-basis checks of Section~\ref{sec:sample}.

\paragraph{What a code-P line is.} Transaction code P covers an ``open market or private purchase'' \citep{sec_form4}, so a line in this sample is a reported purchase, not necessarily an open-market one, and the filing need not show that the buyer used personal funds or acted at discretion. To see what the lines contain, \saN{} retained 2024 filings on \saTickers{} tickers, drawn by a fixed hash within each quarter before any source was opened and without reading any outcome, were reconciled against the original SEC XML. Every accession, issuer, reporting owner, share total and rounded price matches; \saSingleRow{} scrape lines correspond to one XML purchase row and \saAggRows{} aggregate several. On venue, \saExplicitOpen{} filings state an open-market purchase, \saMultiTrade{} carry multiple-trade price footnotes without naming the venue, and \saUnspecified{} are code-P purchases with no venue stated; \saFinancing{} are participations in an issuer financing (a registered direct offering and a private placement), \saDrip{} mixes a dividend-reinvestment purchase with an unspecified one, and \saPlanHeld{} is held through an ESOP or 401(k) plan. Among the same filings, \saIndirect{} report indirect ownership, \saMultiOwner{} name more than one reporting owner and \saPlanAffirmative{} carry an affirmative trading-plan flag. In the 30 days of SEC filings before the purchase, \saCtxPrior{} follow a prior announcement or an organized equity offering, \saCtxRoutine{} show only routine governance filings and \saCtxUnresolved{} remain unresolved; none can be shown to be the first public indication of the trade. The audit is a small, quarter-balanced check of one year, not a prevalence estimate, and the sample is not filtered on it. The results below are therefore results about reported code-P purchases; open-market, personally funded and discretionary buying is the case the mechanism discussion concerns, and the filing establishes it only for a subset.

Daily prices, volume, dividends and split events come from Yahoo Finance through \texttt{yfinance} \citep{yfinance}, pulled in September 2026 with adjustment disabled and corporate actions included, for January 2017 to February 2025; the panel covers \ssPanelTickers{} of the scraped tickers. Yahoo's unadjusted close is still adjusted for every split up to the pull date, so the actual close on day $t$ is reconstructed as the panel close times the product of all split factors dated after $t$. Daily Fama--French factors are from Kenneth French's data library \citep{french_library}. Shares outstanding are point in time where the SEC record allows it: the cover-page count of the latest 10-K or 10-Q filed strictly before the Form~4 filing date, from the SEC XBRL company-facts API \citep{sec_xbrl_companyfacts}, moved to the filing-day share basis and used only if at most 400 days old. This source covers \ssSharesSecPitShare{} of the final sample; the remaining \ssSharesSnapshotShare{} (no SEC observation) and \ssSharesSnapshotStaleShare{} (stale observation) fall back to a January 2026 snapshot from the Yahoo ticker profile, split-adjusted to the filing day, which also supplies sector and industry for the biotechnology indicator.

\subsection{Sample construction}\label{sec:sample}

Table~\ref{tab:funnel} lists every filter in order. The line-level filters are: filing date in 2018--2024 and transaction date from 2018; a reporting lag between 0 and 90 calendar days (\ssFunnelLagDropped{} lines dropped); positive price, quantity and value (\ssFunnelPositiveDropped); a value that agrees with price times quantity within 1\% (\ssFunnelConsistentDropped); a value of at least \$5,000 (\ssFunnelMinValueDropped); and de-duplication of exact repeats, keeping the earliest filing (\ssFunnelDedupDropped). They leave \ssStepSevenN{} lines.

The market-data steps require a Yahoo history covering the filing date (step 8, the survivorship step), 252 daily returns before the event day (\ssFunnelEstWindowDropped{} lines), a full event window, a share count (\ssFunnelSharesDropped), an estimated filing-date capitalization between \$30 million and \$500 million and a 30-trading-day average daily dollar volume above \$200,000 (\ssFunnelAddvDropped). Two price-basis checks follow: the filing-day close over the Form~4 price must lie in $[0.5, 2]$ (\ssFunnelDevBandDropped{} lines, almost all corporate-action mismatches), and the median of that ratio over an issuer-year's same-day filings must lie in $[0.75, 1.33]$ (\ssFunnelPriceBasisDropped{} lines on issuer-years whose panel series is on another basis than the Form~4 prices). Finally, two outlier screens are applied: \ssFunnelCarFloorDropped{} lines whose 30-day abnormal return is below $-100\%$ are excluded, a rule rather than a bound, since a sum of daily abnormal returns has no floor, and so are \ssFunnelJumpFilterDropped{} lines with a one-day price ratio of at least 8 or at most 1/8 inside the event window, the signature of an unrecorded corporate action. Of the \cfFloorDropped{} lines the first screen removes, \cfAlsoJump{} also fail the second; restoring the other \cfRestored{} (the lowest at \cfMinCarPct) leaves the day-one reaction at \cfQone{} in the quintile farthest below the 52-week high against \cfQfive{} nearest it (two-way $t$ = \cfGapTTwo) and the volatility-scaled ratio of Section~\ref{sec:identification} at \cfZRatio. The final sample has \ssN{} lines on \ssNTickers{} tickers (\ssRetention{} of the raw lines), a median transaction value of \ssMedianTxVal, a median capitalization of \ssMedianMcap, a median dollar volume of \ssMedianAddv{} and a median reporting lag of \ssMedianLag{} days.

\begin{table}[tbp]
\centering\tabfoot
\caption{Sample-construction funnel. Each row is one filter applied in order to the OpenInsider purchase lines; ``Dropped'' counts lines removed by that step, ``Remaining'' the lines and ``Tickers'' the distinct ticker strings left after it. $t_0$ is the last trading day on or before the filing date; close\_actual is the filing-day close on the Form~4 share basis; ADDV is 30-day average daily dollar volume. Steps 10b--10d apply the capitalization band to lines whose share count is the SEC point-in-time count, the January 2026 snapshot with no SEC observation, and the snapshot because the SEC observation is stale.}
\label{tab:funnel}
\begin{tabular}{l p{0.6\textwidth} r r r}
\toprule
Step & Filter & Dropped & Remaining & Tickers \\
\midrule
\tabinput{outputs/table_sample_stats_funnel.tex}
\end{tabular}
\end{table}

\paragraph{Survivorship.}\label{sec:survivorship}
Price histories were retrieved in 2026, so issuers whose ticker no longer resolves at Yahoo Finance (delisted, acquired or renamed) are excluded entirely. Of the \ssStepSevenN{} lines that pass the line filters, \ssLostN{} (\ssLostShare) are lost at step 8: \ssLostAbsentN{} on \ssLostAbsentTickers{} tickers absent from the panel and \ssLostEarlyN{} filed before the ticker's first panel row. Table~\ref{tab:byyear} shows the loss rate falling monotonically with the filing year, from \ssLostShareFirstYear{} of 2018 lines to \ssLostShareLastYear{} of 2024 lines, the gradient that attrition through delisting and acquisition produces. The sign of the bias is not determined a priori, because distressed delistings and premium acquisitions both leave the universe, but the training years (\ssLostShareTrainYears{} lost) are more heavily filtered than the test year (\ssLostShareTestYear), so the temporal split embeds a survivorship-driven distribution shift.

\paragraph{Unit of observation.}
The unit is one transaction line as reported on Form~4. A filing or an issuer-day can carry several lines, so the \ssN{} lines span \ssNIssuerDays{} issuer-filing-days and \ssShareSharedDay{} of lines share their filing day, and hence their abnormal return, with another line of the same issuer. Every statistic therefore reports issuer-clustered standard errors by default, and Section~\ref{sec:robustness} repeats the model evaluation on one row per issuer-day. Table~\ref{tab:byyear} gives the composition by year; the positive rate ranges from \ssPosRateYearMin{} in \ssPosRateYearMinYear{} to \ssPosRateYearMax{} in \ssPosRateYearMaxYear.

\begin{table}[tbp]
\centering\tabfoot
\caption{Sample by filing year: lines, issuers (distinct ticker strings), insiders (distinct names), positive rate $\Pr(\CAR > 10\%)$, mean and median $\CAR$, and the share of lines lost at the Yahoo-coverage step (funnel step 8) among lines passing the line filters. The temporal split is 2018--2022 (training), 2023 (threshold) and 2024 (test).}
\label{tab:byyear}
\begin{tabular}{l r r r r r r r}
\toprule
Year & Lines & Issuers & Insiders & Positive rate & Mean CAR & Median CAR & Coverage loss \\
\midrule
\tabinput{outputs/table_sample_stats_byyear.tex}
\end{tabular}
\end{table}

\subsection{Variables}\label{sec:variables}

The classifier uses \nFeaturesFull{} features observable at the close of the filing day (Appendix~\ref{app:vars}). The \emph{base} set contains \nFeaturesBase{} features, defined in Appendix~\ref{app:vars}: an ordinal title score, transaction value, a first-purchase-in-twelve-months indicator, the ratio of the value to the insider's twelve-month average, the disclosure run-up, distance from the 52-week high and low, 30-day volatility, month-to-date return, point-in-time capitalization, 30-day dollar volume and a biotechnology indicator. The \nFeaturesAdded{} added features are role dummies (CEO, CFO, other officer, director, 10\% beneficial owner, parsed from the title by token rules), point-in-time history (prior purchases in the issuer within 365 days, days since the insider's last purchase), cluster and routine indicators (other insiders buying the same issuer in the prior 30 days, their number of lines, and the \citet{cohen2012} routine flag, defined from 2021 on), the reporting lag, the purchase as a share of post-trade holdings, 5- and 21-day returns and an indicator for filings accepted before the 16:00~ET close.

The \emph{disclosure run-up} is the reconstructed filing-day close divided by the reported Form~4 price, minus one, on the same share basis as the trade; a variant on an adjusted close over the nominal price appears only in Section~\ref{sec:artefact}. Distance from the 52-week high is the filing-day close over the maximum daily high of the trailing 252 trading days, minus one, and is non-positive, so ``low values'' are prices far below the high. Several features are missing by construction (the value ratio for \ssShareRatioMissing{} of lines, the routine flag before 2021 for \ssShareRoutineMissing); XGBoost handles missing values natively, the baselines impute the training-split median. Table~\ref{tab:features} reports summary statistics of every feature on the training lines.

\subsection{Abnormal returns and target}\label{sec:car}

For an event with filing day $t$, let $t_0$ be the last trading day on or before $t$. Daily simple returns $R_{i,\tau}$ come from the adjusted close. The factor loadings are estimated by OLS of the stock's daily excess return on the three factors and a constant over $[t_0-252, t_0-1]$, with the stock's returns in that window winsorized at their own 1st and 99th percentiles. The abnormal return and the cumulative abnormal return are
\begin{equation}\label{eq:car}
\begin{split}
\mathrm{AR}_{i,\tau} &= (R_{i,\tau} - R_{f,\tau}) - \big[\hat\beta^{\mathrm{MKT}}_i (R_{\mathrm{MKT},\tau} - R_{f,\tau}) + \hat\beta^{\mathrm{SMB}}_i \mathrm{SMB}_\tau + \hat\beta^{\mathrm{HML}}_i \mathrm{HML}_\tau\big],\\
\CAR &= \sum_{\tau = t_0+1}^{t_0+30} \mathrm{AR}_{i,\tau},
\end{split}
\end{equation}
where the risk-free rate enters on both sides and the estimation-window intercept $\hat\alpha_i$ is estimated but not carried into the expected return. The window starts with the close-to-close return of the first trading day after $t_0$, so no feature uses information from it, and every retained event has exactly 30 event days.

\paragraph{Why the intercept is not carried.} For a stock far below its 52-week high the estimation window is the drawdown itself, so $\hat\alpha_i$ is negative and the benchmark expects the decline to continue: a stock that merely stops falling is credited with a positive abnormal return, and a stock near its high is debited. In this sample the daily intercept has a median of \ssAlphaDailyMedianBp{} basis points and is negative for \ssAlphaNegShare{} of lines, and the implied 30-day carry $-30\hat\alpha_i$ averages \ssAlphaCarryMeanPp{} points over the sample but \mechQoneAlphaCarryPp{} points in the quintile of lines farthest below the high and \mechQfiveAlphaCarryPp{} nearest it (Table~\ref{tab:carry} in Appendix~\ref{app:robust}). The carry is a deterministic function of the pre-event price path, so any sort on a trailing-performance variable inherits it; Section~\ref{sec:mechanism} shows that it accounts for \mechQoneQfiveCarryShareOfVOneGap{} of the quintile gradient in mean abnormal return that the intercept-inclusive benchmark produces, and that firm-days of the same issuers on which no insider bought, sorted on the same variable, show a gradient of \plGapVOnePp{} points under that benchmark (two-way $t$ = \plGapVOneT), \plGapVOneOverEventVOne{} times the events' gradient, against \plGapPrimaryPp{} points (two-way $t$ = \plGapPrimaryT) under the loadings-only one (Table~\ref{tab:placebo}). Equation~\ref{eq:car} therefore uses the loadings only. Retaining the intercept is the market-model convention for short event windows, where a few days of $\hat\alpha_i$ are immaterial. Over 30 days the choice matters, and which choice is right depends on whether the estimation-window intercept persists into the event window. If it does, as when it reflects a premium the factors do not span, subtracting it is what isolates the event, and \citet{goldsmithpinkham2025} include $\hat\alpha$ for that reason; if it reflects a transient drawdown that the sort selects on, carrying it attributes the end of the drawdown to the event. Table~\ref{tab:benchsim} in Appendix~\ref{app:robust} makes this concrete with a known-truth simulation of \smPaths{} paths per scenario: after a transient drawdown, or in a null selected on its prior drawdown, the intercept-inclusive benchmark overstates the event effect by \smTransientIntBias{} and \smSelectedIntBias{} basis points over the window and the loadings-only one by \smTransientLoadBias{} and \smSelectedLoadBias; under a persistent alpha the ranking reverses (\smPersPosIntBias{} against \smPersPosLoadBias{} basis points). Neither benchmark is right in general. The choice here rests on this sample's evidence about persistence: firm-days of the same issuers on which no insider bought, sorted the same way, reproduce \plGapVOneOverEventVOne{} times the events' gradient under the intercept-inclusive benchmark and \plGapPrimaryPp{} of the events' \plEventGapPrimaryPp{} points under the loadings-only one (two-way $t$ = \plGapPrimaryT; Table~\ref{tab:placebo}), which is what a drawdown that does not continue produces and not what a persistent one would. The placebo cannot rule out every process for normal returns, so the intercept-carrying definition is retained throughout as a robustness benchmark rather than discarded. A loadings-only factor benchmark is the counterpart of the reference-portfolio benchmarks of \citet{barber1997}, whose expected return carries no firm-specific intercept, and the choice follows the point of \citet{fama1998} and \citet{kothari2007} that expected-return misspecification, negligible over days, accumulates over longer windows and is most damaging when the sample is selected on prior performance; \citet{lyon1999} document that this misspecification is pervasive in exactly such non-random samples even where the same tests are well specified in random ones. A related but distinct problem is a contaminated estimation window, in which unrelated events inside $[t-252, t-1]$ distort the loadings themselves; \citet{aktas2007} address it with a two-state market model. The concern here is narrower: the intercept is estimated cleanly, and the question is only whether the drawdown it measures continues. The \emph{intercept-inclusive benchmark} is $\mathrm{AR}^{\alpha}_{i,\tau} = \mathrm{AR}_{i,\tau} - \hat\alpha_i$, with $\CAR^{\alpha} = \CAR - 30\hat\alpha_i$. It is reported alongside benchmark-dependent quantities; its label agrees with the primary one on \ssTargetAgreeVOne{} of lines and has a positive rate of \ssPosRateVOne{} against \ssPosRate{} (Table~\ref{tab:car_dist}). The market-adjusted return, which uses no estimation window, is a second robustness benchmark.

\begin{figure}[tbp]
\centering
\includegraphics[width=0.95\textwidth]{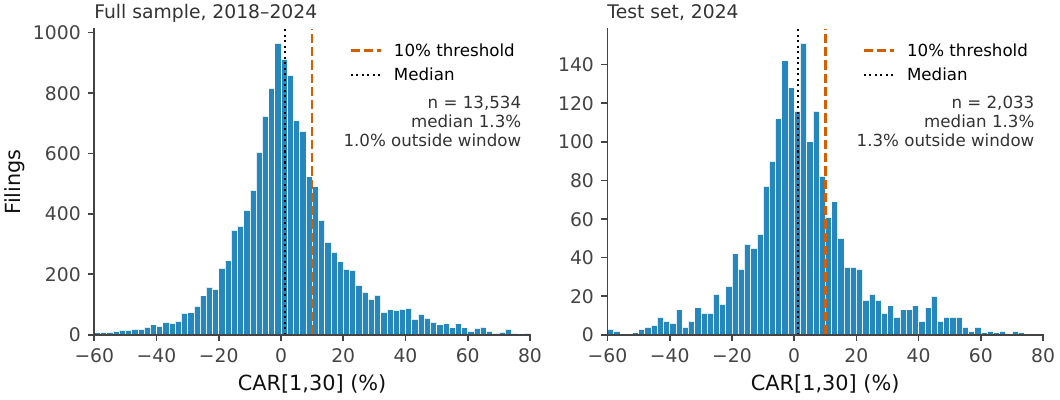}
\caption{Distribution of $\CAR$ in the full sample (left, $N$ = \figCarN) and the 2024 test set (right, $N$ = \figCarTestN), in bins of \figCarBinWidth{} percentage points over $[\figCarRangeLo\%, \figCarRangeHi\%]$; \figCarClipShare{} of full-sample events and \figCarTestClipShare{} of test events lie outside the window. The dashed line is the 10\% threshold and the dotted line the median.}
\label{fig:car_dist}
\end{figure}

The target is $y = \mathbf{1}\{\CAR > 10\%\}$. The 10\% level is fixed a priori as an economically meaningful 30-day outperformance for a microcap position, roughly \ssCarThresholdOverMedian{} times the sample median and \ssCarThresholdInSd{} of a standard deviation, not as a distributional cutoff: it lies at the \ssCarPctileTen{} percentile, so \ssPosRate{} of lines are positive, and a rule labeling the highest tenth would need a threshold near \ssCarPninety. Table~\ref{tab:car_dist} and Figure~\ref{fig:car_dist} describe the distribution: mean \ssCarMean, median \ssCarMedian, standard deviation \ssCarStd{} and extremes of \ssCarMin{} and \ssCarMax. \ssCarNAbsGtOne{} lines (\ssCarShareAbsGtOne) have $|\CAR| > 1$, all positive after the floor; Section~\ref{sec:robustness} reports every headline number with and without them. The 2024 test year has a mean of \ssCarTestMean{} and a positive rate of \ssCarTestPosRate.

\begin{table}[tbp]
\centering\tabtiny
\caption{Distribution of abnormal-return measures. Panel A: observations, availability, mean, median, standard deviation, minimum, maximum, 1st and 99th percentiles and moment skewness. Panel B: share above $+30\%$, share below $-30\%$, share with absolute value above 1, the positive rate, the empirical percentile at which the 10\% threshold falls, and the 90th percentile. BHAR is buy-and-hold; the market-adjusted return subtracts the market excess return only; the timing row starts the window one day later for after-close filings; the last three rows use the intercept-inclusive benchmark, which carries the estimation-window intercept $\hat\alpha$ (Section~\ref{sec:car}). The 60-day window is unavailable near the panel end.}
\label{tab:car_dist}
\shrinkbox{
\begin{tabular}{l r r r r r r r r r r}
\toprule
\multicolumn{11}{l}{\emph{Panel A: moments}} \\
Measure & $N$ & Avail. & Mean & Median & SD & Min & Max & p1 & p99 & Skew \\
\midrule
\tabinput{outputs/table_sample_stats_car_dist.tex}
\end{tabular}}

\medskip
\begin{tabular}{l r r r r r r}
\toprule
\multicolumn{7}{l}{\emph{Panel B: tails and threshold}} \\
Measure & $> +30\%$ & $< -30\%$ & $|x| > 1$ & $\Pr(x > 10\%)$ & Percentile of 10\% & p90 \\
\midrule
\tabinput{outputs/table_sample_stats_car_tails.tex}
\end{tabular}
\end{table}

\paragraph{Acceptance time.}
\ssSharePreOpen{} of lines are accepted before the 09:30~ET open, \ssShareIntraday{} during the session and \ssShareAfterClose{} at or after the 16:00~ET close. The information set is fixed at the close of the filing day, so for after-close filings the first event-day return contains the overnight reaction, which an outsider who observes the filing after the close cannot capture. After-close filings have a mean $\CAR$ of \ssCarPostClose{} against \ssCarPreClose{} (issuer-clustered $t$ of the difference \ssCarPostMinusPreTCl), and the primary model's 2024 AUC is \ssAucPostCloseThree{} on after-close and \ssAucPreCloseThree{} on before-close filings (unpaired, independence-based $z$-test on the DeLong standard errors, $p$ = \ssAucPreMinusPostP).

\FloatBarrier
\section{Empirical design}\label{sec:design}

The sample is split by filing year: 2018--2022 for training (\nTrain{} lines, \nTrainPos{} positives), 2023 for the classification threshold (\nVal{} lines, \valPosRate{} positive) and 2024 as the test year (\nTest{} lines, \testPosRate{} positive), never used for any selection. The primary model is XGBoost with hyperparameters fixed a priori: maximum depth \xgbMaxDepth, learning rate \xgbLearningRate, all other parameters at library defaults (no subsampling, minimum child weight 1, $L_2$ penalty 1, no class re-weighting); no search informs these values (Appendix~\ref{app:model}). Only the number of boosting rounds is data-chosen. Validation-based early stopping on the 2023 AUC (patience \xgbPatience{} out of at most \xgbMaxRounds) would use the 2023 lines twice, for the rounds and for the threshold, and the flat 2023 curve makes the stopping round unstable across otherwise identical fits; that rule is reported as robustness. The protocol is five-fold \texttt{TimeSeriesSplit} cross-validation within 2018--2022 in filing-date order, each fold fitted for up to \xgbMaxRoundsCV{} rounds, the number of rounds set to the first maximum of the mean fold AUC, then one refit on all of 2018--2022 with that many rounds and no early stopping. This selects \xgbRoundsCV{} rounds (mean fold AUC \xgbCvMeanAUC; \xgbRoundsCVBase{} for the base set).

The folds are contiguous, without a purge or embargo: a training line filed within 30 trading days of a fold's first validation filing has an outcome window ending inside the validation block (\robEmbOverlapRowsList{} lines in the five folds), and \robEmbValLateN{} of the 2023 threshold lines (\robEmbValLatePct) have windows resolving in January or February 2024. No 2024 information enters any selection. Purged folds and a threshold on the \robEmbValKeptN{} 2023 lines whose windows end before the first 2024 filing choose \robEmbPurgedRounds{} rounds and give a test AUC of \robEmbPurgedTestAUCthree{} against \robPrimaryAUCthree{} (independence-based DeLong $p$ = \robEmbPurgedVsPrimaryDeLongP), an F1 of \robEmbPurgedTestFone{} against \robEmbPrimTestFone, and a threshold identical to three decimals.

Random forest (\rfTrees{} trees, maximum depth \rfMaxDepth) and $L_2$-regularized logistic regression on standardized features are baselines fitted on the same lines with training-median imputation. Because the positive class is under a third of lines and F1 is the operating criterion, each model's threshold maximizes F1 on the 2023 lines (\clXgbOptThr{} for the primary model, validation F1 \clXgbOptValF); the default 0.5 is reported alongside. Inference on AUCs treats issuers as clusters: the headline uncertainty statement for every AUC and paired difference is a percentile interval, standard error and two-sided $p$-value from a bootstrap that resamples whole issuers (\clBootB{} draws, \clBootClusters{} clusters in 2024, seed \seedBootstrap). The DeLong test \citep{delong1988} assumes independent lines and is anti-conservative here (the cluster-bootstrap standard error of the XGBoost-minus-random-forest difference is \clBootOverDeLongSeXgbRf{} times the DeLong one), so DeLong statistics are reported next to the bootstrap and labeled independence-based, and no difference is called significant on DeLong alone. Mean differences carry issuer-clustered errors \citep{petersen2009,cameron2011}, two-way issuer-and-month clustering where stated, and calendar-time tests with Newey--West errors \citep{newey1987}.

\section{Results}\label{sec:results}

\subsection{Classification performance}\label{sec:classification}

\begin{figure}[tbp]
\centering
\includegraphics[width=0.95\textwidth]{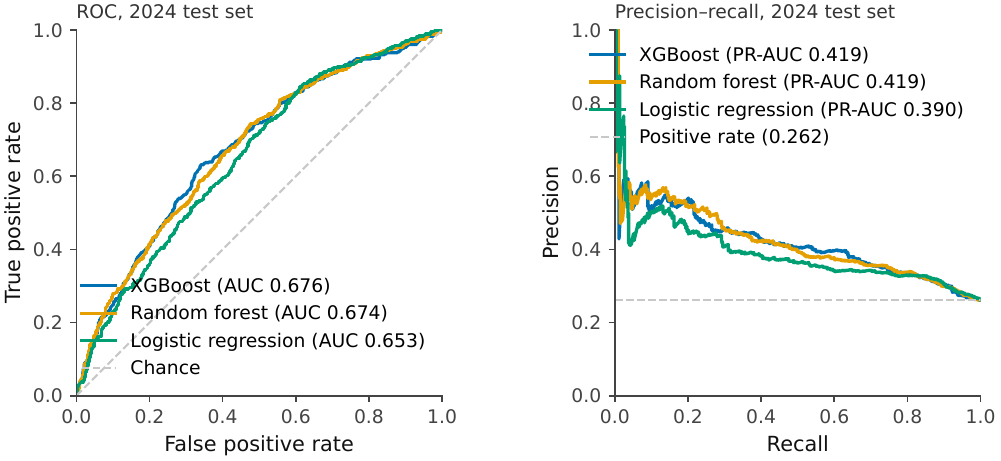}
\caption{ROC (left) and precision--recall (right) curves of the three full-feature models on the 2024 test set ($N$ = \figRocN, \figRocNPos{} positives). AUCs are \figRocXgbAUCthree{} (XGBoost), \figRocRfAUCthree{} (random forest) and \figRocLrAUCthree{} (logistic regression); precision--recall AUCs are \figPrXgbAPthree, \figPrRfAPthree{} and \figPrLrAPthree{} against a positive rate of \figPrBaseRate. Issuer-cluster bootstrap intervals are in Table~\ref{tab:perf}.}
\label{fig:roc}
\end{figure}

Table~\ref{tab:perf} reports 2024 performance for the three learners on both feature sets, Figure~\ref{fig:roc} the ROC and precision--recall curves, and Table~\ref{tab:delong} the paired differences. The primary model attains an AUC of \clXgbAUCthree{} (95\% CI \clXgbAUCci) and a precision--recall AUC of \clXgbPRAUCthree{} against a positive rate of \clFlagAllP; the random forest attains \clRfAUCthree{} (\clRfAUCci) and logistic regression \clLrAUCthree{} (\clLrAUCci). XGBoost minus random forest is \clDiffXgbRf{} (cluster-bootstrap 95\% CI \clDiffXgbRfCi, $p$ = \clBootXgbRfP; independence-based DeLong $p$ = \clDeLongXgbRfP): no accuracy advantage of the boosted model under either test, and the gap is no larger than the forest's own seed-to-seed spread (\clRfSeedRange{} over five seeds, \clRfSeedMin--\clRfSeedMax). XGBoost minus logistic regression is \clDiffXgbLr{} (CI \clDiffXgbLrCi, cluster-bootstrap $p$ = \clBootXgbLrP) and random forest minus logistic regression \clDiffRfLr{} (CI \clDiffRfLrCi, cluster-bootstrap $p$ = \clBootRfLrP): the independence-based DeLong test calls both significant ($p$ = \clDeLongXgbLrP{} and \clDeLongRfLrP), the issuer-clustered test calls neither, and I read both learners as statistically indistinguishable from the linear baseline. XGBoost is retained as the primary model for its treatment of missing values and its attribution toolkit, not for an accuracy advantage.

\begin{table}[tbp]
\centering\tabfoot
\caption{Classification performance on the 2024 test set ($N$ = \nTest{} lines, \nTestPos{} positives). Full = \nFeaturesFull{} features, Base = the \nFeaturesBase{}-feature set listed in Appendix~\ref{app:vars}. AUC with its issuer-cluster bootstrap 95\% interval (\clBootB{} draws), precision--recall AUC, and precision, recall, F1 and flagged lines at the default threshold 0.5 and at each model's own validation-F1-optimal threshold (``optimized'', chosen on 2023). The last block selects rounds by validation-based early stopping.}
\label{tab:perf}
\shrinkbox{
\begin{tabular}{l l r r r r r r r r}
\toprule
Model & Features & AUC & 95\% CI & PR-AUC & Threshold & Precision & Recall & F1 & Flagged \\
\midrule
\tabinput{outputs/table_classification_perf.tex}
\end{tabular}}
\end{table}

\begin{table}[tbp]
\centering\tabfoot
\caption{Paired AUC differences on the 2024 test set. $\Delta$ is the first model's AUC minus the second's; the 95\% interval, standard error and two-sided $p$ come from the issuer-cluster bootstrap (\clBootB{} draws), the headline uncertainty statement; the DeLong $z$ and $p$ assume independent lines and are an independence-based reference. ``CV rounds'' selects rounds within the training years; ``validation-based early stopping'' selects them on 2023.}
\label{tab:delong}
\shrinkbox{
\begin{tabular}{l r r r r r r}
\toprule
Comparison & $\Delta$ AUC & Cluster CI & Cluster SE & Cluster $p$ & DeLong $z$ & DeLong $p$ \\
\midrule
\tabinput{outputs/table_classification_delong.tex}
\end{tabular}}
\end{table}

The \nFeaturesAdded{} added features do not raise accuracy: full minus base is \clDiffXgbFullBase{} (CI \clDiffXgbFullBaseCi, cluster-bootstrap $p$ = \clBootXgbFullBaseP), so their contribution is interpretive. The AUC difference between the primary protocol and validation-based early stopping is \clDiffCvEs{} (CI \clDiffCvEsCi, cluster-bootstrap $p$ = \clBootCvEsP; DeLong $p$ = \clDeLongCvEsP), not significant under clustering. A univariate ranking on the best single feature, \clUniTopFeat, attains \clUniTopAUCthree, and it is the only feature above 0.6 alone (\clUniNAboveSixty{} of \nFeaturesFull); distance from the 52-week high ranks \clUniHighRank{} among the \nFeaturesFull{} features with \clUniHighAUCthree{} (its deficit relative to the full model has an independence-based DeLong $p$ of \clDeLongXgbUniHighP; no cluster interval was computed for this pair). In feature ablations under the primary protocol, dropping the run-up feature raises the test AUC to \clAblNoPdevAUCthree{} (independence-based DeLong $p$ \clAblNoPdevDeLongP), consistent with that feature carrying noise, and dropping distance from the 52-week high leaves it at \clAblNoHighAUCthree{} ($p$ = \clAblNoHighDeLongP), because correlated features substitute for it.

\subsection{Operating points and calibration}\label{sec:operating}

At the default 0.5 threshold the primary model flags \clXgbDefN{} lines with precision \clXgbDefP{} and recall \clXgbDefR{} (F1 \clXgbDefF); at its validation-optimal threshold of \clXgbOptThr{} it flags \clXgbOptN{} lines (\clXgbOptNPct{} of the test set) with precision \clXgbOptP, recall \clXgbOptR{} and F1 \clXgbOptF{}. A classifier that flags every filing attains F1 = \clFlagAllF, because precision then equals the positive rate, and at a common budget of \clMatchedK{} flags F1 spans only \clMatchedFRangeFull{} across the three full-feature learners, so the recall gain over the default is a property of the threshold rule, not of the learner. Across thresholds from \clSwThrLo{} to \clSwThrHi{} F1 is flat between \clSwFMin{} and \clSwFMax{} while precision rises from \clSwPTen{} to \clSwPThirtyFive{} and the number of flags falls from \clSwNTen{} to \clSwNThirtyFive; the operating point is a choice of how many lines to act on. Predictions are concentrated (median \clPredMedian, \clCalLoShare{} below 0.4, maximum \clPredMax). The Brier score is \clBrier{} against \clBrierBase{} for a constant forecast at the base rate (skill \clBrierSkill).

\subsection{What the model uses}\label{sec:importance}

Table~\ref{tab:importance} reports, for every feature, the normalized gain, total gain, split count (weight) and cover of the \impNTreesUsed{} trees used and the mean absolute SHAP value on the 2024 lines. Thirty-day volatility ranks first on gain (\impTopGainPct, \impTopSecondRatioGain{} times the runner-up), total gain (\impTopTotalGainPct) and mean absolute SHAP (\impTopShapPct, \impTopSecondRatioShap{} times the runner-up), and \impTopRankWeight{} on splits. Distance from the 52-week high has rank \mechTgtPrimaryDistRankGain{} on gain (\mechTgtPrimaryDistGain), rank \impFullPctFromFivetwowHighRankWeight{} on splits and rank \mechTgtPrimaryDistRankShap{} on SHAP (\impSecondShapPct). Section~\ref{sec:mechanism} shows that it dominates only under the intercept-carrying label: retrained on that label, the same configuration gives it \mechTgtVOneDistGain{} of gain and \mechTgtVOneDistShap{} of SHAP, first on both bases.

\begin{table}[tbp]
\centering\tabfoot
\caption{Feature attribution in the primary model (XGBoost, \nFeaturesFull{} features, \impNTreesUsed{} rounds). Gain, total gain, weight (share of splits) and cover are normalized shares over the trees used, with ranks in parentheses on the total-gain, weight and cover bases (the Rank column orders by gain); mean $|$SHAP$|$ is the average absolute SHAP value (log-odds) on the 2024 lines and the SHAP share its normalized share with its rank in parentheses. Features never used in a split are listed last.}
\label{tab:importance}
\begin{tabular}{r l r r r r r r}
\toprule
Rank & Feature & Gain & Total gain & Weight & Cover & Mean $|$SHAP$|$ & SHAP share \\
\midrule
\tabinput{outputs/table_importance_full.tex}
\end{tabular}
\end{table}

\begin{figure}[tbp]
\centering
\includegraphics[width=0.72\textwidth]{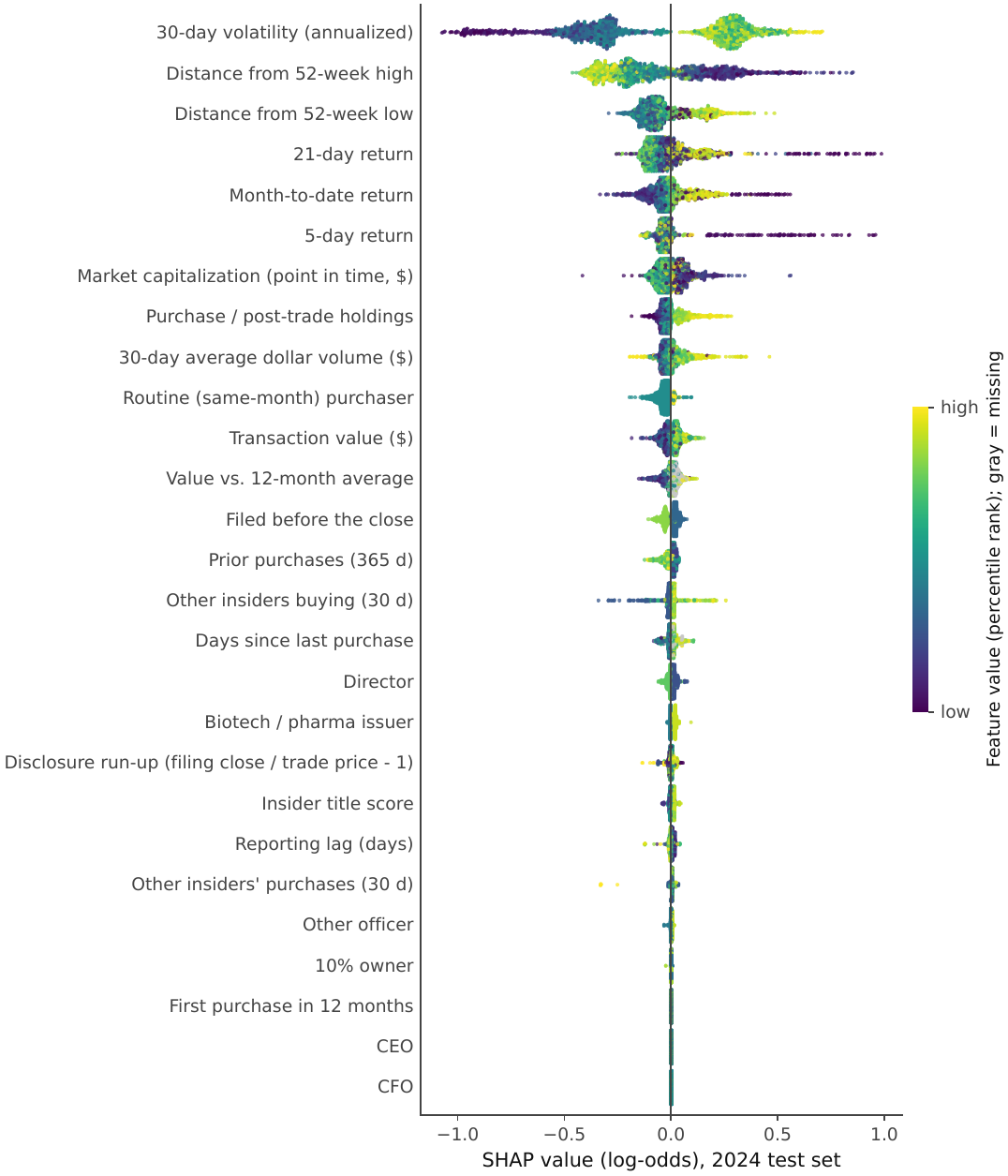}
\caption{SHAP values of the primary model on the 2024 test set ($N$ = \figBeeN{} lines), one point per line and feature, ordered by mean $|$SHAP$|$; color is the feature's percentile rank in the test set (gray = missing). The top feature, \figBeeTopFeature, carries \figBeeTopShare{} of the mean absolute SHAP (mean $|$SHAP$|$ \figBeeTopMeanAbs); \figBeeSecondFeature{} ranks second (\figBeeSecondMeanAbs), and for it low values (prices far below the high) lie on the positive side. The model's base rate corresponds to a probability of \figBeeExpectedProb.}
\label{fig:beeswarm}
\end{figure}

Figure~\ref{fig:beeswarm} shows the SHAP beeswarm. Volatility pushes predictions toward the positive class: the correlation between the feature and its own SHAP value is \impTopCorrPearson{} (CI \impTopCorrPearsonCI), and the mean SHAP is \impTopShapBelowMedian{} below the median volatility of \impTopMedianValue{} and \impTopShapAboveMedian{} above it, which is what a fixed threshold on a summed return produces (Section~\ref{sec:mechanism}). Because distance from the 52-week high is the filing-day close over the trailing 252-day high, minus one, and so non-positive, low values are prices far below the annual high, and these also push predictions toward the positive class: the correlation between the feature and its own SHAP value is \impSignPctFromFivetwowHighPearson, the mean SHAP is \impSignPctFromFivetwowHighShapLow{} below the median feature value and \impSignPctFromFivetwowHighShapHigh{} above it, and decile means fall from \impFtwhDecBottomShap{} in the decile furthest from the high (mean distance \impFtwhDecBottomValue) to \impFtwhDecTopShap{} in the decile nearest it (\impFtwhDecTopValue; rank correlation across deciles \impFtwhDecSpearman). The month-to-date return is U-shaped in decile means (\impMtdDecBottomShap, \impMtdDecMiddleShap{} and \impMtdDecTopShap{} in the lowest, middle and highest decile; end intervals \impMtdDecBottomShapCI{} and \impMtdDecTopShapCI), but the clustered quadratic term is not significant ($t$ = \impMtdQuadT), so the U-shape is suggestive only.

\subsection{The disclosure run-up}\label{sec:runup}

The disclosure run-up is the filing-day close divided by the reported price, minus one. Table~\ref{tab:runup} sorts the sample into five run-up buckets and three finer buckets above 10\%. Lines disclosed at or below the trade price earn a mean $\CAR$ of \ruLowMeanCAR{} with a positive rate of \ruLowPrPos; lines disclosed after a run-up above 10\% earn \ruHighMeanCAR{} with a positive rate of \ruHighPrPos. Neither means nor positive rates are monotone: the 0--3\% bucket has the lowest mean (\ruBZeroThreeMean) and positive rate (\ruBZeroThreePrPos), and within the highest bucket the 10--25\% sub-bucket earns \ruBTenTwentyfiveMean{} while the 50--100\% sub-bucket (\ruBFiftyHundredN{} lines) earns \ruBFiftyHundredMean{} (interval \ruBFiftyHundredCI).

\begin{table}[tbp]
\centering\tabfoot
\caption{Abnormal returns by disclosure run-up bucket. The run-up is the filing-day close on the Form~4 share basis divided by the reported transaction price, minus one. Columns: lines, issuers, mean $\CAR$ with its analytic issuer-clustered 95\% interval, median, mean after winsorizing at the 5th and 95th percentiles within the bucket, positive rate, median 30-day annualized volatility (as a fraction of one), median $|\CAR|$ and median distance from the 52-week high. The indented rows split the highest bucket.}
\label{tab:runup}
\shrinkbox{
\begin{tabular}{l r r r r r r r r r r}
\toprule
Bucket & $N$ & Issuers & Mean CAR & 95\% CI (clustered) & Median & Wins.\ mean & $\Pr(\mathrm{CAR} > 10\%)$ & Med.\ vol. & Med.\ $|$CAR$|$ & Med.\ dist.\ high \\
\midrule
\tabinput{outputs/table_runup_buckets.tex}
\end{tabular}}
\end{table}

The difference between the extreme buckets is \ruDiffPP{} percentage points. A Welch test that treats lines as independent gives $t$ = \ruTWelch{} ($p$ = \ruPWelch); with issuer-clustered errors (\ruNClusters{} clusters) it is $t$ = \ruTCl{} ($p$ = \ruPCl), with two-way issuer-and-month clustering $t$ = \ruTClTwoWay. The issuer-cluster bootstrap interval of the difference is [\ruDiffBootCILo, \ruDiffBootCIHi] points, so the mean gap is statistically indistinguishable from zero. What differs is the probability of a large outcome: $\Pr(\CAR > 10\%)$ is \ruAllPrDiff{} points higher in the highest bucket (clustered $t$ = \ruAllPrTCl). The composition columns say why: lines in the highest bucket sit further below their 52-week high (median \ruHighMedFtwh{} against \ruLowMedFtwh) and are more volatile (median volatility \ruHighMedVol{} against \ruLowMedVol). A higher tail probability with an unchanged mean is what greater dispersion produces. On the 20-day, 60-day and buy-and-hold outcomes the clustered $t$ of the contrast is \ruOutCarTwentyTCl, \ruOutCarSixtyTCl{} and \ruOutBharTCl, and on the intercept-inclusive benchmark \ruOutCarThirtyVOneTCl{} (\ruOutCarThirtyVOneDiff{} points); only the market-adjusted return, which does not remove the size and value exposures of the volatile names, separates the buckets (\ruOutMarDiff{} points, $t$ = \ruOutMarTCl).

\begin{figure}[tbp]
\centering
\includegraphics[width=0.95\textwidth]{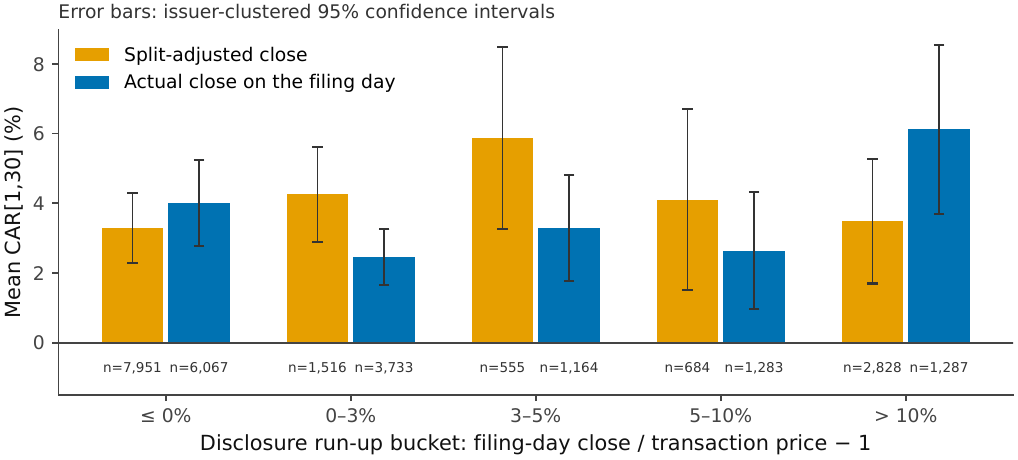}
\caption{Mean $\CAR$ by disclosure run-up bucket under the adjusted-close basis (orange) and the actual filing-day close (blue), with issuer-clustered 95\% intervals and line counts. The highest-minus-lowest gap is \figArtAdjGapGtTenLe{} points on the adjusted basis and \figArtActGapGtTenLe{} on the actual basis; the highest bucket holds \figArtAdjGtTenShare{} of lines under the adjusted basis and \figArtActGtTenShare{} under the actual one.}
\label{fig:artefact}
\end{figure}

\paragraph{The adjusted price basis.}\label{sec:artefact}
Table~\ref{tab:artefact} and Figure~\ref{fig:artefact} compare the adjusted-close and actual-basis run-up variables on the same observations. The table also retains the original sample from the February 2026 version as a reference (Appendix~\ref{app:changes}). The adjusted basis reassigns lines across buckets: \ruArtShareGtHundred{} of lines show a ``run-up'' above 100\% and \ruArtNAtClip{} sit at the clip; the highest bucket holds \figArtAdjGtTenShare{} of lines under the adjusted basis against \figArtActGtTenShare{} under the actual one. Of the lines in the adjusted-basis highest bucket, \ruArtHighShareRatioGtTwo{} have a filing-day close more than twice the reported price, and only \ruArtHighShareActualHigh{} are above 10\% on the actual basis. On the present sample the extreme-bucket mean gap is \ruArtAdjDiff{} points under the adjusted basis (Welch $t$ = \ruArtAdjTWelch, clustered $t$ = \ruArtAdjTCl) and \ruDiffPP{} under the actual basis (clustered $t$ = \ruTCl), both statistically indistinguishable from zero and neither monotone across buckets. The original sample has a gap of \ruArtVOneDiff{} points (Welch $t$ = \ruArtVOneTWelch, clustered $t$ = \ruArtVOneTCl), so price adjustment alone does not explain the difference between samples. What the adjusted basis changes on common observations is the tail-probability contrast: the extreme-bucket difference in $\Pr(\CAR > 10\%)$ is \ruArtAdjPrDiff{} points (clustered $t$ = \ruArtAdjPrTCl), with the share rising in every bucket, against \ruAllPrDiff{} on the actual basis ($t$ = \ruAllPrTCl), where it is not monotone. Reverse-splitting issuers concentrate in the highest adjusted-basis bucket and dividend payers in the lowest. Price basis therefore affects bucket composition and dispersion, while the mean-return comparison also depends on sample construction, including the filters, full event windows, abnormal-return screen, point-in-time capitalization and survivorship layer.

\begin{table}[tbp]
\centering\tabfoot
\caption{The run-up pattern under two price bases and two samples. Adjusted basis: the variable (Yahoo's split- and dividend-adjusted close on the filing day divided by the nominal Form~4 price, minus one, clipped to $[-1, 10]$) recomputed on this sample. Actual basis: the variable on the actual price basis for the same lines. Original sample: the February 2026 sample with its adjusted-basis variable (Appendix~\ref{app:changes}). Each block gives lines, mean $\CAR$ and the share above 10\% by bucket; the last two rows give the highest-minus-lowest difference in percentage points with Welch and issuer-clustered $t$-statistics, and the difference in the share above 10\% with its clustered $t$.}
\label{tab:artefact}
\shrinkbox{
\begin{tabular}{l r r r r r r r r r}
\toprule
 & \multicolumn{3}{c}{Adjusted basis, this sample} & \multicolumn{3}{c}{Actual basis, this sample} & \multicolumn{3}{c}{Original sample, adjusted basis} \\
\cmidrule(lr){2-4}\cmidrule(lr){5-7}\cmidrule(lr){8-10}
Bucket & $N$ & Mean CAR & $\Pr(> 10\%)$ & $N$ & Mean CAR & $\Pr(> 10\%)$ & $N$ & Mean CAR & $\Pr(> 10\%)$ \\
\midrule
\tabinput{outputs/table_runup_artefact.tex}
\end{tabular}}
\end{table}

Two signatures identify the mechanism. The split signature is a per-issuer constant ratio between the adjusted close and the reported price equal to a later reverse-split factor, which moves an issuer's every filing into the highest bucket regardless of its price path. The dividend signature is visible among same-day filings by financial-sector issuers, which pay regular dividends: their median adjusted-basis deviation is \ruDivEighteenFinAdj{} in 2018 and shrinks monotonically to \ruDivTwentyFourFinAdj{} in 2024, because the adjustment accumulates every dividend between the filing and the download date, whereas on the actual basis the median never exceeds \ruDivFinActMaxAbsMedian{} in absolute value. The adjusted basis thus sorts dividend payers into the ``at or below'' bucket (\ruDivFinShareLowAdj{} of financial lines against \ruDivFinShareLowAct{} on the actual basis) and reverse-splitting issuers into the ``run-up'' bucket, and the return difference between buckets becomes a difference between banks and biotechs.

\subsection{Composition and double sort}\label{sec:doublesort}

Table~\ref{tab:subsamples} repeats the extreme contrast within subsamples. By filing year the mean gap is positive in \ruNYearsDiffPos{} of the seven years, with \ruNYearsDiffSig{} significant under clustering; in the 2024 test year it is \ruDiffPPTest{} points ($t$ = \ruTClTest) and the positive-rate difference is \ruSsTestPrDiff{} points ($t$ = \ruPrTClTest). A few subsample contrasts have a clustered $t$ above two, biotechnology (\ruSsBiotechOneDiff{} points, $t$ = \ruSsBiotechOneTCl, against \ruSsBiotechZeroDiff{} elsewhere), director-only lines (\ruSsRoleDirectorOnlyDiff{} points, $t$ = \ruSsRoleDirectorOnlyTCl), 10\% owners (\ruSsRoleTenPctDiff, $t$ = \ruSsRoleTenPctTCl) and repeat purchasers (\ruSsFirstZeroDiff, $t$ = \ruSsFirstZeroTCl), but they are a handful among the many contrasts in Table~\ref{tab:subsamples}, none was pre-specified, and the double sort below absorbs them; and the gap is not monotone in illiquidity (\ruSsAddvLowDiff, \ruSsAddvMidDiff{} and \ruSsAddvHighDiff{} points across dollar-volume terciles), as a slow-diffusion story would require.

\begin{table}[tbp]
\centering\tabtiny
\caption{The extreme run-up contrast ($> 10\%$ minus $\le 0\%$) by subsample: lines in each bucket, mean $\CAR$, difference in percentage points, Welch and issuer-clustered $t$, and the share above 10\% in each bucket. Terciles and bands are formed on the full sample; role groups are mutually exclusive in the order CEO/CFO, other officer, 10\% owner, director only; serial issuers have at least \ruSerialMinLines{} lines; the last row averages issuer means.}
\label{tab:subsamples}
\shrinkbox{
\begin{tabular}{l r r r r r r r r r}
\toprule
Subsample & $N$ low & $N$ high & Mean low & Mean high & Diff (pp) & Welch $t$ & Clustered $t$ & $\Pr(> 10\%)$ low & $\Pr(> 10\%)$ high \\
\midrule
\tabinput{outputs/table_runup_subsamples.tex}
\end{tabular}}
\end{table}

Table~\ref{tab:doublesort} sorts on both the run-up and full-sample quintiles of distance from the 52-week high: the highest-minus-lowest difference is negative in \ruDsNQuintilesNeg{} of the five quintiles and has a clustered $t$ above two in \ruDsNQuintilesSig, the quintile nearest the high (\ruDsQFiveDiff{} points, $t$ = \ruDsQFiveTCl), and the pooled within-quintile coefficient is \ruDsPooledCoefPP{} points ($t$ = \ruDsPooledTCl). Issuer-clustered regressions on the extreme buckets (Panel B; \ruPanelBaseN{} lines) give the same answer: with year fixed effects only, the highest-bucket indicator carries \ruPanelBaseCoef{} points ($t$ = \ruPanelBaseT); adding \ruPanelNDeciles{} decile dummies of distance from the 52-week high turns it to \ruPanelDecCoef{} ($t$ = \ruPanelDecT), and \ruPanelNControls{} further controls (volatility, month-to-date and 21-day returns, log size, log volume, the biotechnology and role indicators, log value and the lag) leave \ruPanelCtrlCoef{} ($t$ = \ruPanelCtrlT). The sort on the run-up relabels the 52-week-high effect: purchases disclosed after a run-up are more dispersed, not more profitable.

\begin{table}[tbp]
\centering\tabfoot
\caption{Double sort and regressions. Panel A compares the $> 10\%$ and $\le 0\%$ run-up buckets within full-sample quintiles of distance from the 52-week high (Q1 furthest below the high; range in brackets): mean $\CAR$ with lines in parentheses, the difference in percentage points and its issuer-clustered $t$. Panel B regresses the outcome on an indicator for the highest bucket over the \ruPanelBaseN{} lines of the two extreme buckets with issuer-clustered errors: (a) year fixed effects; (b) adding \ruPanelNDeciles{} decile dummies of distance from the 52-week high; (c) adding the \ruPanelNControls{} controls listed in the text. Coefficients are in percentage points, log-odds for the logit.}
\label{tab:doublesort}
\begin{tabular}{l r r r r}
\toprule
\multicolumn{5}{l}{\emph{Panel A: extreme run-up contrast within quintiles of distance from the 52-week high}} \\
Quintile & Mean CAR ($N$), $\le 0\%$ & Mean CAR ($N$), $> 10\%$ & Diff (pp) & Clustered $t$ \\
\midrule
\tabinput{outputs/table_runup_doublesort.tex}
\end{tabular}

\medskip
\begin{tabular}{l r r r r r r}
\toprule
\multicolumn{7}{l}{\emph{Panel B: highest-bucket indicator in issuer-clustered regressions}} \\
 & \multicolumn{2}{c}{(a) Year FE} & \multicolumn{2}{c}{(b) + distance deciles} & \multicolumn{2}{c}{(c) + controls} \\
\cmidrule(lr){2-3}\cmidrule(lr){4-5}\cmidrule(lr){6-7}
Outcome & Coef. & $t$ & Coef. & $t$ & Coef. & $t$ \\
\midrule
\tabinput{outputs/table_runup_regressions.tex}
\end{tabular}
\end{table}

\subsection{Insider roles, history and clustering}\label{sec:roles}

Table~\ref{tab:roles_car} reports $\CAR$ by insider role, with overlapping groups. Purchases by chief financial officers earn a mean of \roleCfoMean{} (median \roleCfoMedian, positive rate \roleCfoPos), \roleCfoDiffPp{} points above the rest of the sample with a clustered $t$ of \roleCfoTCl; any officer title earns \roleOfficerAnyMean{} ($t$ = \roleOfficerAnyTCl). Directors without an officer title earn \roleDirOnlyMean{} and 10\% beneficial owners \roleTenPctMean, \roleTenPctDiffPp{} points relative to the rest: the Welch $t$ of \roleTenPctTWelch{} would call this significant, the clustered $t$ of \roleTenPctTCl{} and the bootstrap interval \roleTenPctVsNonDiffCI{} do not, because owner purchases cluster in a few issuers. An issuer-clustered regression of $\CAR$ on the role dummies, a first-purchase indicator, an indicator for other insiders buying in the prior 30 days, log transaction value and year fixed effects (\roleOlsN{} lines, $R^2$ = \roleOlsRsq) gives the CFO dummy \roleOlsCfoCoef{} points ($t$ = \roleOlsCfoT), the 10\%-owner dummy \roleOlsTenPctCoef{} ($t$ = \roleOlsTenPctT), a first purchase \roleOlsFirstCoef{} ($t$ = \roleOlsFirstT), other insiders buying \roleOlsClusterCoef{} ($t$ = \roleOlsClusterT) and log value \roleOlsLogValueCoef{} points per log unit ($t$ = \roleOlsLogValueT); the routine flag of \citet{cohen2012}, available from 2021, carries \roleOlsTwoRoutineCoef{} points ($t$ = \roleOlsTwoRoutineT). On the intercept-inclusive benchmark the role contrasts have the same signs and are slightly larger (CFO \roleCfoDiffPpVOne{} points, $t$ = \roleCfoTClVOne; 10\% owners \roleTenPctDiffPpVOne, $t$ = \roleTenPctTClVOne; routine purchasers \roleRoutineVOneDiffPp, $t$ = \roleRoutineVOneDiffTCl).

\begin{table}[tbp]
\centering\tabfoot
\caption{$\CAR$ by insider role: lines, share, mean, median, positive rate, and the issuer-clustered and Welch $t$-statistics of the group mean against the rest of the sample. Groups overlap when a title carries several roles; ``Director only'' excludes any officer title.}
\label{tab:roles_car}
\begin{tabular}{l r r r r r r r}
\toprule
Group & $N$ & Share & Mean & Median & $\Pr(\mathrm{CAR} > 10\%)$ & Clustered $t$ & Welch $t$ \\
\midrule
\tabinput{outputs/table_roles_car.tex}
\end{tabular}
\end{table}

The univariate patterns behind the regression: routine purchasers (\roleRoutineN{} lines, \roleRoutineShare{} of the \roleRoutineYears{} lines) earn \roleRoutineMean{} against \roleOppMean{} for opportunistic purchases (difference \roleRoutineDiffPp{} points, clustered $t$ = \roleRoutineDiffTCl, bootstrap interval \roleRoutineDiffCI); purchases preceded by another insider's purchase in the same issuer within 30 days earn \roleClusterAnyMean{} against \roleClusterZeroMean{} (clustered $t$ = \roleClusterAnyVsRefTCl), consistent with \citet{lakonishok2001}; first purchases in twelve months earn \roleFirstMean{} against \roleRepeatMean{} for repeat purchasers ($t$ = \roleFirstDiffTCl). In the classifier these ``who is buying'' features carry \roleImpWhoShap{} of SHAP in total (role dummies and title score \roleImpRoleShap, history \roleImpHistoryShap, cluster and routine \roleImpClusterShap).

\subsection{Economic significance}\label{sec:economics}

\paragraph{Per-trade returns.}
Each 2024 line flagged at the threshold of \econThreshold{} is a \$1 position held for 30 trading days with the realized $\CAR$ as its return. The \econNFlag{} flagged lines (\econShareFlag{} of the test set, \econTickersFlag{} issuers) earn a mean of \econMeanFlag{} against \econMeanRej{} for the \econNRej{} rejected lines and \econMeanAll{} for all 2024 lines. The \econDiffFlagRejPp-point difference has a Welch $t$ of \econDiffTWelch, an issuer-clustered $t$ of \econDiffTCl{} (\econDiffNClusters{} clusters) and a two-way $t$ of \econDiffTTwoWay. Flagged lines beat rejected lines in \econMonthsFlagGtRej{} of \econMonthsTotal{} months and in medians (\econMedianFlag{} against \econMedianRej) and hit rates (\econHitFlag{} against \econHitRej). The flagged mean is right-tail-driven: the twenty largest outcomes supply \econTopTwentyShare{} of the summed flagged $\CAR$, and excluding the \econNFlagGtFifty{} trades above 50\% lowers the flagged mean to \econMeanFlagExFifty{} (the same trim lowers the rejected mean to \econMeanRejExFifty{} and the all-2024 mean to \econMeanAllExFifty), so the ordering survives the trim while the level does not. On the intercept-inclusive benchmark the flagged and rejected means are \econMeanFlagVOne{} and \econMeanRejVOne{} (difference \econDiffFlagRejVOnePp{} points, clustered $t$ = \econDiffVOneTCl). The gap is further qualified by price level. Flagged lines are concentrated in low-priced stocks (\mechSubTwoShareFlagged{} below \$2 against \mechSubTwoShareRejected{} of rejected lines), whose close-to-close returns are biased upward by bid--ask bounce \citep{blume1983}; the buy-and-hold variant, which compounds instead of summing, keeps the gap at \robBharTradeDiffPP{} points ($t$ = \robBharTradeTCl, Table~\ref{tab:tails_model}), but the per-trade gap and the flagged-portfolio alphas are not re-estimated under a price floor (the announcement reaction and the quintile portfolios of Section~\ref{sec:reaction} are), and Table~\ref{tab:floor} shows that the run-up contrast vanishes at a \$2 floor while the AUC is unchanged.

\paragraph{Calendar-time portfolios.}
Table~\ref{tab:calendar} gives the calendar-time evidence \citep{mandelker1974,fama1998,mitchell2000}. The all-filings portfolio has a monthly alpha of \econCtAllAlphaMonthly{} ($t$ = \econCtAllAlphaT) with a size loading of \econCtAllBetaSmb. The 2024 flagged portfolio has a monthly alpha of \econCtFlagAlphaMonthly{} ($t$ = \econCtFlagAlphaT, $p$ = \econCtFlagAlphaP), the rejected portfolio \econCtRejAlphaMonthly{} ($t$ = \econCtRejAlphaT), and the flagged-minus-rejected portfolio \econCtFlagMinusRejAlphaMonthly{} with $t$ = \econCtFlagMinusRejAlphaT{} ($p$ = \econCtFlagMinusRejAlphaP), short of the 5\% level over a single year; on the intercept-inclusive benchmark the flagged-minus-rejected mean abnormal return is \econCtFlagMinusRejArMonthlyVOne{} a month ($t$ = \econCtFlagMinusRejArTVOne). Section~\ref{sec:reaction} shows that the all-filings alpha falls to \rxCtAllEwSkipAlphaMonthly{} ($t$ = \rxCtAllEwSkipAlphaT) once the first event day is skipped.

\begin{table}[tbp]
\centering\tabfoot
\caption{Calendar-time portfolios: daily equal-weighted portfolios of the positions open on each trading day (each event held for 30 trading days from the day after the event day), regressed on the three factors with Newey--West errors (\econNwLags{} lags). Columns: trading days, mean open positions, monthly alpha (daily alpha times 21) with $t$, loadings, and mean daily abnormal return times 21 with its $t$. The 2024 portfolios hold positions into February 2025.}
\label{tab:calendar}
\shrinkbox{
\begin{tabular}{l r r r r r r r r r}
\toprule
Portfolio & Days & Positions & $\alpha$ (monthly) & $t$ & $\beta_{\mathrm{MKT}}$ & $\beta_{\mathrm{SMB}}$ & $\beta_{\mathrm{HML}}$ & Mean AR (monthly) & $t$ \\
\midrule
\tabinput{outputs/table_economics_calendar.tex}
\end{tabular}}
\end{table}

\paragraph{Costs and capacity.}
Table~\ref{tab:cost} subtracts round-trip costs of 1\% to 5\% from every gross figure, with issuer-cluster bootstrap intervals of the gross means. Net of 3\%, a charge in the range that \citet{lesmond2004} and \citet{novymarx2016} document for small illiquid stocks, the flagged mean falls from \econGrossFlag{} (interval [\econGrossCILoFlag, \econGrossCIHiFlag]) to \econNetFlagThree{} and its interval, [\econNetCILoFlagThree, \econNetCIHiFlagThree], includes zero, as does the buy-every-filing interval ([\econNetCILoBuyAllThree, \econNetCIHiBuyAllThree] around \econNetBuyAllThree); at 5\% the flagged mean is \econNetFlagFive{} ([\econNetCILoFlagFive, \econNetCIHiFlagFive]). The break-even round-trip cost is \econBreakEvenFlag{} for flagged lines and \econBreakEvenBuyAll{} for all 2024 lines. The charge is consistent with the paper's own spread measure: \citet{corwin2012} high--low spreads over the \econCsWindowDays{} trading days before the filing have a median of \econCsMedianAll{} across all events, \econCsMedianFlag{} for flagged 2024 lines and \econCsMedianRej{} for rejected ones, and net of the median flagged spread alone the flagged mean is \econNetFlagCsMedian. Capacity is small: a \econOrderSize{} order exceeds 5\% of trailing daily dollar volume for \econShareFiftyKOverFivePctAll{} of sample events and 10\% for \econShareFiftyKOverTenPctAll{} (\econShareFiftyKOverTenPctFlag{} of the flagged 2024 lines). The gross differences between flagged and rejected lines are real in the 2024 data, but at realistic microcap execution costs, with capacity limited to small orders in the least liquid names, the signal is economically marginal. The full-sample mean of \ssCarMean{} per 30 days (about \econFullAnnualised{} annualized; \econFullMedianAnnualised{} for the median) is several times the roughly six percent a year that \citet{jeng2003} estimate for portfolios of insider purchases; survivorship, the concentration of large outcomes in sub-\$5 stocks and the summed-return construction account for the gap.

\begin{table}[tbp]
\centering\tabfoot
\caption{Transaction-cost sensitivity. Gross mean $\CAR$ with its issuer-cluster bootstrap 95\% interval, and the mean net of a round-trip cost of 1\%, 2\%, 3\% and 5\% subtracted once per position, for the full sample, the run-up buckets, all 2024 lines (buy-all), the model-flagged and rejected 2024 lines and the 2024 highest run-up bucket.}
\label{tab:cost}
\shrinkbox{
\begin{tabular}{l r r r r r r r}
\toprule
Group & $N$ & Gross mean & 95\% CI & Net 1\% & Net 2\% & Net 3\% & Net 5\% \\
\midrule
\tabinput{outputs/table_economics_cost.tex}
\end{tabular}}
\end{table}

\FloatBarrier
\section{Is the 52-week-high effect mechanical?}\label{sec:mechanism}

A fixed 10\% threshold on a summed 30-day return is easier to clear for a volatile stock, and stocks far below their 52-week high are volatile. Table~\ref{tab:quintiles} sorts the sample into quintiles of distance from the high on the primary benchmark. From the quintile furthest below the high (mean distance \mechQoneMeanDist) to the quintile nearest it (\mechQfiveMeanDist), the share of events above $+10\%$ falls from \mechQonePrPos{} to \mechQfivePrPos, a ratio of \mechQoneQfivePrPosRatio, and the share below $-10\%$ falls with it, from \mechQonePrNeg{} to \mechQfivePrNeg{} (difference \mechQoneQfivePrNegDiffPp{} points, clustered $t$ = \mechQoneQfivePrNegTCl), while the standard deviation of $\CAR$ falls from \mechQoneSdCAR{} to \mechQfiveSdCAR{} and mean volatility from \mechQoneMeanVol{} to \mechQfiveMeanVol. Both tails widen far from the high, which is what dispersion does. Mean $\CAR$ also falls, from \mechQoneMeanCAR{} to \mechQfiveMeanCAR{} (difference \mechQoneQfiveMeanCARDiffPp{} points, clustered $t$ = \mechQoneQfiveMeanCARTCl), but the share of events with a volatility-normalized abnormal return $z = \CAR / (\sigma_{30}\sqrt{30/252})$ above one is nearly flat, \mechQonePrZ{} against \mechQfivePrZ{} (difference \mechQoneQfivePrZDiffPp{} points, $t$ = \mechQoneQfivePrZTCl). Measured in units of each stock's own volatility, a purchase far below the high is not much more likely to be followed by a large positive outcome than a purchase near it; the difference in means is real, and Section~\ref{sec:reaction} locates about half of it in the first trading day after disclosure (\rxDiffArTonePp{} of \rxDiffCarThirtyPp{} points) and the rest in a drift that is far less precisely estimated.

\begin{table}[tbp]
\centering\tabfoot
\caption{Outcomes by quintile of distance from the 52-week high (Q1 furthest below the high): lines, mean distance, share of events above $+10\%$ and below $-10\%$, mean, median and standard deviation of $\CAR$, mean 30-day annualized volatility, share of events with a volatility-normalized abnormal return $z > 1$, mean $z$, and share with a buy-and-hold abnormal return above 10\%. Volatility is annualized and expressed as a fraction of one.}
\label{tab:quintiles}
\shrinkbox{
\begin{tabular}{l r r r r r r r r r r r}
\toprule
Quintile & $N$ & Mean dist. & $\Pr(\mathrm{CAR} > 10\%)$ & $\Pr(\mathrm{CAR} < -10\%)$ & Mean CAR & Median & SD & Mean vol. & $\Pr(z > 1)$ & Mean $z$ & $\Pr(\mathrm{BHAR} > 10\%)$ \\
\midrule
\tabinput{outputs/table_mechanism_quintiles.tex}
\end{tabular}}
\end{table}

Distance from the high and 30-day volatility are strongly related (Pearson \mechDistVolPearson, Spearman \mechRhoVol), the one pair among the \mechNPairsHigh{} feature pairs with an absolute correlation above 0.5 that matters for interpretation, most of the others being related by construction. The Spearman correlation of distance from the high with the label is \mechRhoTarget, with an indicator for $|\CAR| > 10\%$ it is \mechRhoAbsTarget, and with the normalized label $\mathbf{1}\{z > 1\}$ it is \mechRhoTargetZ, so normalizing by pre-event volatility removes about \mechRhoDropShare{} of the rank association and retains \mechRhoRetainedShare. Issuer-clustered logits give the same answer: the distance variable predicts the label with a coefficient of \mechLogitClsDistCoef{} ($t$ = \mechLogitClsDistTCl), the normalized label with \mechLogitZDistCoef{} ($t$ = \mechLogitZDistTCl), and \mechLogitZDistVolCoef{} ($t$ = \mechLogitZDistVolTCl) once log volatility is added, so a directional component that is not volatility survives once volatility is held fixed, but it is a small part of what the classifier learns from the variable.

On the normalized label the primary configuration attains a test AUC of \mechTgtZAUCthree{} (independence-based DeLong SE \mechTgtZAUCSE) against \mechTgtPrimaryAUCthree{} on the primary label, and the gain share of distance from the high falls from \mechTgtPrimaryDistGain{} to \mechTgtZDistGain{} and its SHAP share from \mechTgtPrimaryDistShap{} to \mechTgtZDistShap. The price level tells the same story: lines with a filing-day price below \$1 earn a mean $\CAR$ of \mechPriceSubOneMeanCAR{} and lines between \$1 and \$2 earn \mechPriceOneTwoMeanCAR, against \mechPriceAboveTenMeanCAR{} above \$10; the positive rate is \mechSubTwoPrPos{} below \$2 and \mechAboveTenPrPos{} above \$10; sub-\$2 lines are \mechSubTwoShareAll{} of the sample but \mechSubTwoShareFlagged{} of the model's 2024 flags and \mechSubTwoShareRejected{} of its rejections. Deciles of predicted probability on the 2024 lines give a mean $\CAR$ rising from \mechDecOneMeanCAR{} in the lowest decile to \mechDecTenMeanCAR{} in the highest (difference \mechDecTopBottomDiffPp{} points, clustered $t$ = \mechDecTopBottomTCl, bootstrap interval \mechDecTopBottomCIpp), with mean volatility rising from \mechDecOneMeanVol{} to \mechDecTenMeanVol{} across the same deciles.

\paragraph{The benchmark carry.} Table~\ref{tab:carry} in Appendix~\ref{app:robust} repeats the sort on the intercept-inclusive benchmark and reports the estimation-window intercept by quintile. The daily intercept averages \mechQoneAlphaDailyBp{} basis points in the quintile farthest below the high and \mechQfiveAlphaDailyBp{} nearest it, so the intercept-inclusive benchmark adds \mechQoneAlphaCarryPp{} points to the 30-day abnormal return of the former and \mechQfiveAlphaCarryPp{} points to that of the latter. Under that benchmark mean $\CAR$ falls from \mechQoneMeanCARVOne{} to \mechQfiveMeanCARVOne, a gradient of \mechQoneQfiveMeanCARVOneDiffPp{} points (clustered $t$ = \mechQoneQfiveMeanCARVOneTCl) of which the carry, \mechQoneQfiveAlphaCarryDiffPp{} points ($t$ = \mechQoneQfiveAlphaCarryTCl), is \mechQoneQfiveCarryShareOfVOneGap; the positive-rate gradient is \mechQoneQfivePrPosVOneDiffPp{} points against \mechQoneQfivePrPosDiffPp{} on the primary benchmark. The carry is a property of the design, not of insider purchases: it is fixed by the estimation window before the event occurs, so any firm-day sorted on the same variable should show it, and the placebo below shows that it does. The carry also makes distance from the 52-week high the dominant feature of a classifier trained on the intercept-carrying label. Retraining the primary configuration on the intercept-inclusive label gives a test AUC of \robVOneRefitAUCthree{} (interval \robVOneRefitAUCci) against \robPrimaryAUCthree{} on the primary label, with the distance variable's gain share at \robVOneRefitDistGain{} and its SHAP share at \robVOneRefitDistShap, first on both bases, against \mechTgtPrimaryDistGain{} and \mechTgtPrimaryDistShap{} on the primary label; the primary model's probabilities score \robVOneRefitPrimaryAUCthree{} against the intercept-inclusive label.

\paragraph{A placebo without purchases.} If the gradient under the intercept-inclusive benchmark belongs to the benchmark rather than to the purchase, firm-days on which no insider bought, sorted the same way, must show it too, and must not show the gradient that the loadings-only benchmark leaves. Table~\ref{tab:placebo} draws \plDrawsPerYear{} random trading days per issuer and calendar year from \plYearFirst{} to \plYearLast{} for every issuer of the sample, among days that satisfy the sample's own market-data requirements and event-window gates (252 daily returns before the day, 30 trading days after it, a 30-day average dollar volume above \$200,000, and the two gates of Table~\ref{tab:funnel}) and that lie more than \plExclDays{} trading days from the filing day of every purchase line of the issuer in the raw scrape, before any sample filter (\plNRawLines{} lines; \plNRowsExcluded{} of the issuers' \plNRowsPanel{} panel days are excluded). This leaves \plNPlacebo{} placebo days on \plNTickers{} issuers, \plEventsPerPlacebo{} per event line. For each day the loadings and the intercept are estimated on $[t-252, t-1]$, the three 30-day outcomes are computed on the code path that produced the events' outcomes, and the day is assigned to the quintile cutoffs of Table~\ref{tab:quintiles}; in Q1 the placebo days resemble the events in distance from the high (\plQoneCeboMeanDist{} against \plQoneEventMeanDist) and in price level (\plQoneCeboShareSubFive{} below \$5 against \plQoneEventShareSubFive). Under the intercept-inclusive benchmark the placebo days reproduce the gradient: their mean $\CAR^{\alpha}$ falls from \plQoneCeboVOne{} in Q1 to \plQfiveCeboVOne{} in Q5, a gap of \plGapVOnePp{} points (two-way issuer-and-month $t$ = \plGapVOneT, clustered $t$ = \plGapVOneTCl), \plGapVOneOverEventVOne{} times the events' \plEventGapVOnePp{} points, and the carry $-30\hat\alpha$ of the placebo days, \plQoneCeboAlphaCarryPp{} points in Q1 and \plQfiveCeboAlphaCarryPp{} in Q5 (gap \plGapCarryPp{} points, $t$ = \plGapCarryT), is \plGapCarryShareOfVOne{} of that gap. On the primary benchmark the placebo gap is \plGapPrimaryPp{} points (two-way $t$ = \plGapPrimaryT) against \plEventGapPrimaryPp{} for the events (two-way $t$ = \plEventGapPrimaryT), and on the market-adjusted return \plGapMarPp{} ($t$ = \plGapMarT) against \plEventGapMarPp{} ($t$ = \plEventGapMarT). In Q1, events earn \plQoneEventPrimary{} on the primary benchmark against \plQoneCeboPrimary{} on the placebo days (difference \plQoneDiffPrimaryPp{} points, two-way $t$ = \plQoneDiffPrimaryT; placebo-to-event ratio \plRatioQonePrimary), whereas under the intercept-inclusive benchmark the placebo days earn \plQoneCeboVOne{} and the ratio is \plRatioQoneVOne. The placebo Q1 mean on the primary benchmark is itself right-tail-driven: winsorized at the placebo distribution's 1st and 99th percentiles it is \plWinQoneCeboPrimary{} and its median is \plMedQoneCeboPrimary, against \plWinQoneEventPrimary{} and \plMedQoneEventPrimary{} for the events, and the winsorized placebo gap is \plWinGapPrimary{} points on the primary benchmark ($t$ = \plWinGapPrimaryT) and \plWinGapVOne{} on the intercept-inclusive one ($t$ = \plWinGapVOneT). A random day has no disclosure to react to: the placebo abnormal return on the day after the placebo day is \plQoneCeboArToneMean{} in Q1 ($t$ = \plQoneCeboArToneTTwo) against \rxQoneArToneMean{} for the events (Section~\ref{sec:reaction}). The gradient the intercept-inclusive benchmark produces therefore needs no purchase; the gradient the loadings-only benchmark leaves does. The placebo days are drawn from the same survivor-only issuers as the events, so the comparison controls for the benchmark and the price path, not for survivorship.

\begin{table}[tbp]
\centering\tabfoot
\caption{The no-purchase placebo by quintile of distance from the 52-week high (Q1 furthest below the high). Placebo days are \plDrawsPerYear{} random trading days per issuer and calendar year, \plYearFirst--\plYearLast, for every issuer of the sample, drawn from days that meet the sample's market-data requirements and event-window gates and lie more than \plExclDays{} trading days from the filing day of any purchase line of the issuer in the raw scrape (\plNPlacebo{} days on \plNTickers{} issuers); each day's loadings, intercept and 30-day outcomes are computed as for the events, and the day is assigned to the event sample's quintile cutoffs. Columns: event lines, placebo days, the placebo days' mean carry $-30\hat\alpha$ in percentage points, and the mean 30-day abnormal return of the events and of the placebo days on the primary benchmark (loadings only), the intercept-inclusive benchmark ($\hat\alpha$ carried) and the market-adjusted return. Lower rows: the Q1 $-$ Q5 gap in percentage points with its two-way issuer-and-month clustered $t$, for the events and the placebo days separately (the carry column gives the placebo carry gap); the Q1 event-minus-placebo difference with its two-way $t$; and the ratio of the placebo to the event Q1 mean.}
\label{tab:placebo}
\shrinkbox{
\begin{tabular}{l r r r r r r r r r}
\toprule
 & & & & \multicolumn{2}{c}{Primary benchmark} & \multicolumn{2}{c}{Intercept included} & \multicolumn{2}{c}{Market-adjusted} \\
\cmidrule(lr){5-6}\cmidrule(lr){7-8}\cmidrule(lr){9-10}
Quintile & $N$ events & $N$ placebo & Placebo carry (pp) & Events & Placebo & Events & Placebo & Events & Placebo \\
\midrule
\tabinput{outputs/table_placebo_quintiles.tex}
\end{tabular}}
\end{table}

\section{The announcement reaction and the post-disclosure drift}\label{sec:reaction}

The gradient in mean abnormal return under the loadings-only benchmark could be a reaction at disclosure or a drift after it. To separate the two I decompose $\CAR$ into the abnormal return on the first event day, $\mathrm{AR}_{t_0+1}$, and the drift over event days 2 to 30, and add the abnormal return on the filing day itself, $\mathrm{AR}_{t_0}$, which lies outside the event window and contains post-disclosure trading only for the \rxSharePreClose{} of lines accepted before the 16:00~ET close. For the \rxSharePostClose{} accepted at or after the close, $\mathrm{AR}_{t_0+1}$ contains the overnight gap from the $t_0$ close to the next open, which I also report raw, together with the raw intraday return of day $t_0+1$. Every cell in Tables~\ref{tab:reaction} and \ref{tab:reaction_timing} carries an issuer-clustered $t$, the text adds the two-way issuer-and-filing-month clustered $t$ where the two differ, and quintiles are those of Table~\ref{tab:quintiles}.

\paragraph{The first day.} Table~\ref{tab:reaction} shows that the reaction is concentrated on the first trading day after disclosure and rises steeply with distance from the high. $\mathrm{AR}_{t_0+1}$ is \rxQoneArToneMean{} in the quintile farthest below the high (clustered $t$ = \rxQoneArToneTCl) and \rxQfiveArToneMean{} nearest it ($t$ = \rxQfiveArToneTCl), a ratio of \rxQoneQfiveArToneRatio{} and a difference of \rxDiffArTonePp{} points (clustered $t$ = \rxDiffArToneTCl, two-way $t$ = \rxDiffArToneTTwo); the means fall from Q1 through Q4 and are nearly equal in Q4 and Q5, and the two-way $t$ exceeds two in \rxNQuintilesArToneSigTwo{} of the five quintiles. The pattern does not depend on the quintile cuts: lines at least 50\% below the high (\rxBBelowFiftyN{} lines) earn \rxBBelowFiftyArToneMean{} on the first day against \rxBAboveFiveArToneMean{} for lines within 5\% of it (difference \rxBDiffArTonePp{} points, two-way $t$ = \rxBDiffArToneTTwo). For lines accepted before the close the filing day's own return contains post-disclosure trading: $\mathrm{AR}_{t_0}$ is \rxQoneArTzeroMean{} in Q1 (\rxQoneArTzeroN{} lines, $t$ = \rxQoneArTzeroTCl) against \rxQfiveArTzeroMean{} in Q5 (difference \rxDiffArTzeroPp{} points, two-way $t$ = \rxDiffArTzeroTTwo). The means are line-weighted; collapsed to one row per issuer-day (\rxIssuerDayN{} rows), the first-day reaction is \rxIssuerDayQoneArToneMean{} in Q1 against \rxIssuerDayQfiveArToneMean{} in Q5 (difference \rxIssuerDayDiffArTonePp{} points, clustered $t$ = \rxIssuerDayDiffArToneTCl).

\begin{table}[tbp]
\centering\tabfoot
\caption{The announcement reaction by quintile of distance from the 52-week high (Q1 furthest below the high; primary benchmark). Columns: lines, mean distance, the mean abnormal return on the filing day for lines accepted before the 16:00~ET close (\rxAllArTzeroN{} lines in total), the mean abnormal return on the first event day with its issuer-clustered $t$, the mean drift over event days 2--30 with its $t$, the mean drift over days 2--60 (available for \rxShareDriftSixtyAvail{} of lines), mean $\CAR$ with its $t$, and the share of mean $\CAR$ realized on the first day. The last row is the Q1 minus Q5 difference in percentage points with its clustered $t$; two-way issuer-and-month clustered $t$-statistics are given in the text.}
\label{tab:reaction}
\shrinkbox{
\begin{tabular}{l r r r r r r r r r r r}
\toprule
Quintile & $N$ & Mean dist. & AR$_{t_0}$ (pre-close) & AR$_{t_0+1}$ & $t$ & Drift 2--30 & $t$ & Drift 2--60 & CAR & $t$ & Day-1 share \\
\midrule
\tabinput{outputs/table_reaction_quintiles.tex}
\end{tabular}}
\end{table}

\paragraph{Filing time and the overnight gap.} Table~\ref{tab:reaction_timing} splits the extreme quintiles by acceptance time. For the \rxPostQoneN{} Q1 lines accepted at or after the close, the filing-day abnormal return is \rxPostQoneArTzeroMean{} ($t$ = \rxPostQoneArTzeroTCl) and the next day's is \rxPostQoneArToneMean{} ($t$ = \rxPostQoneArToneTCl), of which the raw overnight gap is \rxPostQoneGapMean{} ($t$ = \rxPostQoneGapTCl) and the intraday return \rxPostQoneIntradayMean{} ($t$ = \rxPostQoneIntradayTCl); against Q5 the gap difference is \rxPostDiffGapPp{} points (two-way $t$ = \rxPostDiffGapTTwo). For the \rxPreQoneN{} Q1 lines accepted before the close, the reaction is split between the filing day (\rxPreQoneArTzeroMean, $t$ = \rxPreQoneArTzeroTCl) and the next (\rxPreQoneArToneMean, $t$ = \rxPreQoneArToneTCl), a two-day total of \rxPreQoneTwoDayMean{} ($t$ = \rxPreQoneTwoDayTCl) that differs from the after-close one-day reaction by \rxPreQoneTwoDayMinusPostDayOnePp{} points. The reaction therefore occurs when the filing becomes public, whichever day that is, and an outsider who observes an after-close filing can act only at the next open, after the gap has been realized.

\begin{table}[tbp]
\centering\tabfoot
\caption{The reaction by acceptance time: all lines and the extreme quintiles (primary benchmark). Columns: lines, the abnormal return on the filing day and on the first event day, the raw overnight gap (open of $t_0+1$ over close of $t_0$, minus one) and the raw intraday return of $t_0+1$, each with its issuer-clustered $t$, the mean drift over days 2--30 and mean $\CAR$. Q1 $-$ Q5 rows are differences in percentage points with their clustered $t$.}
\label{tab:reaction_timing}
\shrinkbox{
\begin{tabular}{l r r r r r r r r r r r}
\toprule
Group & $N$ & AR$_{t_0}$ & $t$ & AR$_{t_0+1}$ & $t$ & Gap & $t$ & Intraday & $t$ & Drift 2--30 & CAR \\
\midrule
\tabinput{outputs/table_reaction_timing.tex}
\end{tabular}}
\end{table}

\paragraph{Drift.} The drift over event days 2 to 30 is \rxQoneDriftThirtyMean{} in Q1 (clustered $t$ = \rxQoneDriftThirtyTCl, two-way $t$ = \rxQoneDriftThirtyTTwo) and \rxQfiveDriftThirtyMean{} in Q5 ($t$ = \rxQfiveDriftThirtyTCl); the difference is \rxDiffDriftThirtyPp{} points with a clustered $t$ of \rxDiffDriftThirtyTCl{} and a two-way $t$ of \rxDiffDriftThirtyTTwo, and over days 2 to 60 it is \rxDiffDriftSixtyPp{} points (two-way $t$ = \rxDiffDriftSixtyTTwo). The first day supplies \rxQoneDayOneShare{} of Q1's mean $\CAR$ of \rxQoneCarThirtyMean{} and \rxAllDayOneShare{} of the full-sample mean. The drift has a two-way $t$ above two in \rxNQuintilesDriftSigTwo{} of the five quintiles, and its cross-section is not the reaction's: within Q1 the first-day reaction is nearly the same for officers, 10\% owners and directors (\rxQoneRoleOfficerArToneMean, \rxQoneRoleTenPctArToneMean{} and \rxQoneRoleDirectorArToneMean; officer minus director \rxQoneRoleOfficerMinusDirectorArTonePp{} points, $t$ = \rxQoneRoleOfficerMinusDirectorArToneTCl), whereas the drift is \rxQoneRoleOfficerDriftThirtyMean, \rxQoneRoleTenPctDriftThirtyMean{} and \rxQoneRoleDirectorDriftThirtyMean{} ($t$ = \rxQoneRoleOfficerDriftThirtyTCl, \rxQoneRoleTenPctDriftThirtyTCl{} and \rxQoneRoleDirectorDriftThirtyTCl). On the intercept-inclusive benchmark the first-day reaction is almost unchanged, \rxVoneQoneArToneMean{} in Q1 and a Q1--Q5 difference \rxDiffArToneVoneMinusPrimaryPp{} points larger than on the primary benchmark, but the Q1 drift becomes \rxVoneQoneDriftThirtyMean{} ($t$ = \rxVoneQoneDriftThirtyTCl) and the Q1--Q5 drift difference \rxVoneDiffDriftThirtyPp{} points: of the \mechQoneQfiveAlphaCarryDiffPp-point difference in carry between the extreme quintiles (Table~\ref{tab:carry}), \rxDiffDriftThirtyVoneMinusPrimaryPp{} points sit in the drift. More than half of the drift measured under the intercept-inclusive benchmark is the benchmark itself. The drift that remains is concentrated in time: the \rxQoneCovidN{} Q1 lines filed in March and April 2020 have a mean drift of \rxQoneDriftThirtyCovidMean{} ($t$ = \rxQoneDriftThirtyCovidTCl) and the other \rxQoneExCovidN{} Q1 lines \rxQoneDriftThirtyExCovidMean{} ($t$ = \rxQoneDriftThirtyExCovidTCl, two-way $t$ = \rxQoneDriftThirtyExCovidTTwo), whereas the first-day reaction is \rxQoneArToneExCovidMean{} outside those two months and \rxQoneArToneCovidMean{} inside them.

\paragraph{Price floors.} The sort on distance from the high is also a sort on price level: \rxFloorFiveQoneShareOfQone{} of Q1 lines but \rxFloorFiveQfiveShareOfQfive{} of Q5 lines have a filing-day close of at least \$5, and the close-to-close returns of low-priced stocks carry bid--ask bounce \citep{blume1983}. Table~\ref{tab:reaction_floor} repeats the decomposition within the \rxFloorFiveN{} lines priced at \$5 or more (\rxFloorFiveShare{} of the sample, \rxFloorFiveNTickers{} issuers) and within the \rxFloorFiveLiqN{} lines priced at \$5 or more with a 30-day average dollar volume of at least \$1 million (\rxFloorFiveLiqShare, \rxFloorFiveLiqNTickers{} issuers), keeping the full-sample quintile cuts. The reaction survives both floors. At the \$5 floor $\mathrm{AR}_{t_0+1}$ is \rxFloorFiveQoneArToneMean{} in Q1 (\rxFloorFiveQoneN{} lines; clustered $t$ = \rxFloorFiveQoneArToneTCl, two-way $t$ = \rxFloorFiveQoneArToneTTwo) against \rxFloorFiveQfiveArToneMean{} in Q5, a difference of \rxFloorFiveDiffArTonePp{} points (clustered $t$ = \rxFloorFiveDiffArToneTCl, two-way $t$ = \rxFloorFiveDiffArToneTTwo) and a ratio of \rxFloorFiveQoneQfiveArToneRatio; in the liquid subset it is \rxFloorFiveLiqQoneArToneMean{} (\rxFloorFiveLiqQoneN{} lines) against \rxFloorFiveLiqQfiveArToneMean{} (difference \rxFloorFiveLiqDiffArTonePp{} points, clustered $t$ = \rxFloorFiveLiqDiffArToneTCl, two-way $t$ = \rxFloorFiveLiqDiffArToneTTwo), of which the raw overnight gap from the $t_0$ close to the next open averages \rxFloorFiveLiqQoneGapMean{} in Q1 ($t$ = \rxFloorFiveLiqQoneGapTCl) against \rxFloorFiveLiqQfiveGapMean{} in Q5. The drift is not distinguishable from zero in either subset: over days 2 to 30 it is \rxFloorFiveQoneDriftThirtyMean{} in Q1 at the \$5 floor (clustered $t$ = \rxFloorFiveQoneDriftThirtyTCl, two-way $t$ = \rxFloorFiveQoneDriftThirtyTTwo) and \rxFloorFiveLiqQoneDriftThirtyMean{} in the liquid subset (clustered $t$ = \rxFloorFiveLiqQoneDriftThirtyTCl, two-way $t$ = \rxFloorFiveLiqQoneDriftThirtyTTwo), the Q1--Q5 drift differences are \rxFloorFiveDiffDriftThirtyPp{} and \rxFloorFiveLiqDiffDriftThirtyPp{} points (two-way $t$ = \rxFloorFiveDiffDriftThirtyTTwo{} and \rxFloorFiveLiqDiffDriftThirtyTTwo), and the first day supplies \rxFloorFiveQoneDayOneShare{} and \rxFloorFiveLiqQoneDayOneShare{} of the Q1 mean $\CAR$ against \rxQoneDayOneShare{} in the full sample. Whatever post-disclosure drift the full sample retains sits in the sub-\$5 names.

\begin{table}[tbp]
\centering\tabfoot
\caption{The announcement reaction and the drift under price floors (primary benchmark). Lines with a filing-day close of at least \$5, and lines with a close of at least \$5 and a 30-day average dollar volume of at least \$1 million, in the full-sample quintiles of distance from the 52-week high of Table~\ref{tab:quintiles} (Q1 furthest below the high). Columns: lines, the mean abnormal return on the first event day with its issuer-clustered and its two-way issuer-and-month clustered $t$, the mean drift over event days 2--30 with both $t$-statistics, mean $\CAR$ and the share of mean $\CAR$ realized on the first day; the Q1 $-$ Q5 rows are differences in percentage points with their $t$-statistics. The line below each block gives the subset's Q1 calendar-time portfolio (equal-weighted, positions held for event days 1--30 or 2--31, three factors and momentum, Newey--West $t$ with \rxCtNwLags{} lags), as in Table~\ref{tab:reaction_calendar}.}
\label{tab:reaction_floor}
\begingroup
\let\spanhead\multicolumn
\renewcommand{\multicolumn}[3]{\spanhead{#1}{@{}p{0.85\textwidth}@{}}{#3}}
\shrinkbox{
\begin{tabular}{l r r r r r r r r r}
\toprule
 & & \spanhead{3}{c}{AR$_{t_0+1}$} & \spanhead{3}{c}{Drift 2--30} & & \\
\cmidrule(lr){3-5}\cmidrule(lr){6-8}
Group & $N$ & Mean & Clustered $t$ & Two-way $t$ & Mean & Clustered $t$ & Two-way $t$ & CAR & Day-1 share \\
\midrule
\tabinput{outputs/table_reaction_floor.tex}
\end{tabular}}
\endgroup
\end{table}

\paragraph{Calendar time.} Table~\ref{tab:reaction_calendar} holds each event for 30 trading days from $t_0+1$ or from $t_0+2$ (skipping the first day), in equal- and value-weighted portfolios regressed on the three factors and momentum with Newey--West errors. The Q1 portfolio earns \rxCtQoneEwBaseAlphaMonthly{} a month equal-weighted ($t$ = \rxCtQoneEwBaseAlphaT) and \rxCtQoneVwBaseAlphaMonthly{} value-weighted ($t$ = \rxCtQoneVwBaseAlphaT) with the first day, and \rxCtQoneEwSkipAlphaMonthly{} ($t$ = \rxCtQoneEwSkipAlphaT) and \rxCtQoneVwSkipAlphaMonthly{} ($t$ = \rxCtQoneVwSkipAlphaT) without it, \rxCtQoneEwSkipShareOfBase{} and \rxCtQoneVwSkipShareOfBase{} of the with-day-one alpha. Within the price-floor subsets of Table~\ref{tab:reaction_floor} the Q1 portfolio earns, equal-weighted, \rxFloorFiveCtQoneEwBaseAlphaMonthly{} a month with the first day ($t$ = \rxFloorFiveCtQoneEwBaseAlphaT) and \rxFloorFiveCtQoneEwSkipAlphaMonthly{} without it ($t$ = \rxFloorFiveCtQoneEwSkipAlphaT) at the \$5 floor, and \rxFloorFiveLiqCtQoneEwBaseAlphaMonthly{} ($t$ = \rxFloorFiveLiqCtQoneEwBaseAlphaT) and \rxFloorFiveLiqCtQoneEwSkipAlphaMonthly{} ($t$ = \rxFloorFiveLiqCtQoneEwSkipAlphaT) in the liquid subset, where the value-weighted alphas are \rxFloorFiveLiqCtQoneVwBaseAlphaMonthly{} ($t$ = \rxFloorFiveLiqCtQoneVwBaseAlphaT) with the first day and \rxFloorFiveLiqCtQoneVwSkipAlphaMonthly{} ($t$ = \rxFloorFiveLiqCtQoneVwSkipAlphaT) without it. The Q1-minus-Q5 portfolio has an alpha of \rxCtLsVwBaseAlphaMonthly{} value-weighted ($t$ = \rxCtLsVwBaseAlphaT) and \rxCtLsEwBaseAlphaMonthly{} equal-weighted ($t$ = \rxCtLsEwBaseAlphaT) with the first day and \rxCtLsVwSkipAlphaMonthly{} ($t$ = \rxCtLsVwSkipAlphaT) and \rxCtLsEwSkipAlphaMonthly{} ($t$ = \rxCtLsEwSkipAlphaT) without it. The all-filings portfolio's alpha of \rxCtAllEwBaseAlphaMonthly{} ($t$ = \rxCtAllEwBaseAlphaT), which on the three factors alone is the \econCtAllAlphaMonthly{} of Table~\ref{tab:calendar}, falls to \rxCtAllEwSkipAlphaMonthly{} ($t$ = \rxCtAllEwSkipAlphaT) when the first day is skipped. Two properties of the calendar-time test qualify its reading. It weights trading days equally, so events that cluster in a few months carry less weight than in the event-time means of Table~\ref{tab:reaction}, which lowers its power against a drift concentrated in time \citep{kothari2007}, and the Q1 drift is concentrated in two months; and skipping the first day removes the intraday part of that day as well as the overnight gap, although only the gap is untradable, so the skip-day-one alpha is a lower bound on what an entry at the next open could collect (in after-close Q1 lines the intraday return of day one is \rxPostQoneIntradayMean, $t$ = \rxPostQoneIntradayTCl, against a gap of \rxPostQoneGapMean).

\begin{table}[tbp]
\centering\tabtiny
\caption{Calendar-time portfolios by quintile of distance from the 52-week high: daily portfolios of the positions open on each trading day, equal-weighted (EW) or weighted by estimated capitalization at entry (VW), holding each event for event days 1--30 or 2--31 (skipping the first day), regressed on the three factors and momentum with Newey--West errors (\rxCtNwLags{} lags). Columns: trading days with an open position, mean open positions, monthly alpha (daily alpha times \rxCtMonthDays) with its $t$, loadings, and the three-factor alpha with its $t$. The all-filings equal-weighted 1--30 three-factor alpha equals the all-filings row of Table~\ref{tab:calendar}.}
\label{tab:reaction_calendar}
\shrinkbox{
\begin{tabular}{l l l r r r r r r r r r r}
\toprule
Portfolio & Weight & Holding & Days & Positions & $\alpha$ (monthly) & $t$ & $\beta_{\mathrm{MKT}}$ & $\beta_{\mathrm{SMB}}$ & $\beta_{\mathrm{HML}}$ & $\beta_{\mathrm{UMD}}$ & FF3 $\alpha$ & $t$ \\
\midrule
\tabinput{outputs/table_reaction_calendar.tex}
\end{tabular}}
\end{table}

\paragraph{Interpretation.} The first-day reaction to an insider purchase in a beaten-down stock is \rxQoneQfiveArToneRatio{} times the reaction to one near the high, and \rxQoneDayOneShare{} of the quintile's mean $\CAR$ is realized that day. That is a ratio of returns, not of information: Section~\ref{sec:identification} shows that most of it is the volatility of the stocks involved. This is consistent with the contrarian-insider information channel of \citet{rozeff1998} and \citet{jenter2005} being impounded at disclosure. It does not by itself exclude the slow correction of an anchoring bias in the sense of \citet{george2004}, which would require drift in the direction of the reaction: the point estimate of the Q1 drift, \rxQoneDriftThirtyMean{} over 29 days, is as large as the first-day reaction of \rxQoneArToneMean{} and supplies \rxQoneDriftShare{} of the quintile's mean $\CAR$, but it is far less precisely estimated (two-way $t$ = \rxQoneDriftThirtyTTwo{} against \rxQoneArToneTTwo{} for the reaction, \rxDiffDriftThirtyTTwo{} against \rxDiffArToneTTwo{} for the Q1--Q5 difference), it is concentrated in two months of 2020 whereas the reaction is not, and in calendar time the Q1 portfolio that skips the first day earns \rxCtQoneEwSkipAlphaMonthly{} a month ($t$ = \rxCtQoneEwSkipAlphaT), statistically indistinguishable from zero. Nor is the reaction confined to one session: Q1 earns a further \dpTtwoQone{} on the second session after publication (two-way $t$ = \dpTtwoQoneTTwo; Q1 $-$ Q5 \dpTtwoGapPp{} points, $t$ = \dpTtwoGapTTwo) and \dpTthreeQone{} on the third ($t$ = \dpTthreeQoneTTwo), so that \dpQoneShareTthree{} of the quintile's mean $\CAR$ is realized by the third session, after which no window is precisely estimated (Section~\ref{sec:identification}). The reaction over the first sessions is established; the drift is not. What remains after the first day sits in survivor-only, mostly sub-\$5 names and cannot be separated from survivorship with this panel (Section~\ref{sec:limitations}); in the lines priced at \$5 or more with a dollar volume of at least \$1 million, the names an outsider can trade, the first-day reaction is \rxFloorFiveLiqQoneArToneMean{} in Q1 and the drift \rxFloorFiveLiqQoneDriftThirtyMean{} (two-way $t$ = \rxFloorFiveLiqQoneDriftThirtyTTwo), and the calendar-time portfolio that skips the first day earns \rxFloorFiveLiqCtQoneEwSkipAlphaMonthly{} a month ($t$ = \rxFloorFiveLiqCtQoneEwSkipAlphaT).

\section{What the reaction is and is not}\label{sec:identification}

The day-one gap of Table~\ref{tab:reaction} invites four objections, each of which would reduce it
to something other than information in the disclosure: that it is volatility, that it is a
confound, that it is survivorship, and that beaten-down stocks reprice any news more. Five tests,
T1 to T5, had the reading of each outcome fixed before they were run (Section~\ref{sec:limitations}
states the scope of that pre-specification). One of them fails, and the claim is narrowed
accordingly. The further comparisons in this section are post hoc and are labelled as such, here and
in Table~\ref{tab:identification}.

Two volatility measures are used. Within the purchase sample, $\sigma$ is the sample's own 30-day
volatility feature converted to a daily unit, $\sigma = \texttt{volatility\_30d}/\sqrt{252}$, and
$z = \mathrm{AR}_{t_0+1}/\sigma$ is the standardized reaction. Across disclosure types, where that
feature does not exist, $\sigma$ is the standard deviation of the stock's daily returns over
$[t_0-29, t_0]$ computed from the price panel with one definition for every event; a variant ending
at $t_0-1$, so that the filing day's own return does not enter, gives the same answers and is
reported where it matters.

\paragraph{T1: volatility (pass; narrows the claim).} Stocks far below their 52-week high are more
volatile: the median daily volatility over the 30 days before the event is \dxSigmaRatio{} times
higher in Q1 than in Q5, so any news moves them more. Part of the raw ratio is therefore scale.
Standardized by each stock's own pre-event volatility, the day-one reaction is \dxZQone{} standard
deviations in Q1 against \dxZQfive{} in Q5, a ratio of \dxZRatio{} (difference \dxZDiff, two-way
$t$ = \dxZDiffTTwo; issuer-cluster bootstrap interval for the ratio [\drMeanZLo, \drMeanZHi]). The
sign of the reaction, which volatility cannot inflate, moves the same way: the market reacts
positively to \dxPosQone{} of Q1 disclosures and \dxPosQfive{} of Q5 disclosures (difference
\dxPosDiffPp{} points, clustered $t$ = \dxPosDiffT, two-way $t$ = \drPosTTwo). T1 fixed one
estimator, the ratio of mean $z$; the size of the per-risk gradient depends on which is used, and
the alternatives are reported post hoc. The ratio is \drMedianZRatio{} on medians
([\drMedianZLo, \drMedianZHi]), \drWinsOneRatio{} and \drWinsFiveRatio{} with $z$ winsorized at
1\% and 5\%, \drMeanArMeanSigRatio{} as the ratio of mean reaction to mean volatility, \drAcZRatio{}
on after-close filings, \drExTzeroZRatio{} with the volatility window ending the day before $t_0$,
and \drIssuerDayZRatio{} with lines collapsed to issuer-days. Holding volatility fixed rather than
dividing by it gives the smallest figures: reweighting the near-high quintile to the far quintile's
volatility-decile distribution, the standardized ratio is \drMatchedZRatio{} and the raw ratio
\drMatchedRawRatio; a regression of $z$ on the quintile dummies with volatility-decile fixed effects
puts the Q1 $-$ Q5 difference at \drFeZDiff{} standard deviations (two-way $t$ = \drFeZTTwo) and
the raw difference at \drFeRawDiffPp{} points ($t$ = \drFeRawTTwo); and within terciles of
volatility the standardized ratio is \drTercMidZRatio{} in the middle tercile and \drTercHiZRatio{}
in the upper one (two-way $t$ = \drTercMidZTTwo{} and \drTercHiZTTwo; the lower tercile holds only
\drTercLoQoneN{} Q1 lines). A slope through the origin, which weights each disclosure by its
variance, shows no gradient (\drOriginSlopeQone{} against \drOriginSlopeQfive). The statement the
paper makes is therefore that the gradient at fixed volatility is positive and precisely estimated
and that its size depends on the estimator: between \drMatchedZRatio{} and \drMedianZRatio{} on
those reported here other than the variance-weighted slope, which the most volatile stocks dominate
and which shows none;
``about twice'' is the pre-specified estimator, and part of it is the volatility composition of the
quintiles rather than a difference at fixed volatility.

\paragraph{T2: confounds (fail).} Distance from the high is correlated with size, price level,
liquidity, trade size, insider role, reporting lag, recent return and filing time. Regressing the
day-one abnormal return on quintile dummies and all of these, with year effects and two-way
clustering by issuer and month, leaves a Q1 coefficient of \dxCtrlQoneCoefPp{} points (two-way $t$
= \dxCtrlQoneTTwo) against \dxRawGapPp{} points without controls. The pre-specified threshold was a
two-way $|t|$ of 3; the estimate does not meet it. Adding the controls one at a time (post hoc)
shows where the gap goes: log volatility alone takes it to \dxVolOnlyGapPp{} points ($t$ =
\dxVolOnlyGapTTwo) and price level alone to \dxPriceOnlyGapPp, while size, liquidity, trade size,
filing time, reporting lag and the prior month's return each move it little. With the full control
set the standardized reaction keeps \dxCtrlZQoneCoef{} standard deviations (two-way $t$ =
\dxCtrlZQoneTTwo), which clears the bar, and the probability of a positive reaction keeps
\dxPosCtrlPp{} points (two-way $t$ = \dxPosCtrlT), which does not; neither outcome was named in T2,
so both are post hoc. The raw four-to-five-fold gap is mostly that beaten-down microcaps are
volatile and cheap. Whether volatility should be treated as a confound or as the channel is not
something this regression can decide: if a purchase resolves uncertainty about the firm, the value
of that resolution rises with how much uncertainty there is, and volatility measures exactly that.

\paragraph{T3: survivorship (pass).} The panel keeps only tickers that still resolve in 2026, and
the loss rate falls from \ssLostShareFirstYear{} of 2018 lines to \ssLostShareLastYear{} of 2024
lines. A day-one reaction is measured inside a single session, so later delisting can move it only
by changing which firms are in the sample. It does not appear to: the Q1 $-$ Q5 day-one gap is
\dxSurvHiGapPp{} points in the high-attrition years 2018--2020 (two-way $t$ = \dxSurvHiT) and
\dxSurvLoGapPp{} points in the low-attrition years 2023--2024 ($t$ = \dxSurvLoT), and it is positive
in every year, with a minimum of \dxSurvMinYearGapPp{} points. The pre-specified reading (same sign,
low-attrition gap at least half the high-attrition gap) is met. Stated as a difference, the
low-attrition gap exceeds the high-attrition one by \dcSurvDiffPp{} points with a 95\% interval of
[\dcSurvDiffLoPp, \dcSurvDiffHiPp] (two-way $t$ = \dcSurvDiffTTwo), so the test excludes a change
of more than about half the gap in either direction, and no more than that.

\paragraph{T4: any good news, or this news? (pass on the primary rule; inconclusive against the
strongest news).} A random day has no news in it, so the placebo of Section~\ref{sec:mechanism}
cannot rule out that beaten-down stocks simply reprice any good news more. Earnings announcements
by the same issuers test that directly: \dxEarnNEightK{} original 8-K filings under Item~2.02 in
2018--2024, signed by a seasonal-random-walk surprise in quarterly EPS from the issuers' XBRL
filings (\dxEarnNSue{} with a usable surprise), scored on the same code path and benchmark and
assigned to the purchase sample's quintile cutoffs. Panel~B of Table~\ref{tab:identification}
restricts every disclosure type to filings accepted after the close, so that the abnormal return on
$t_0+1$ is the first session after publication for all of them. For positive surprises the day-one
reaction is \dxEarnAcQone{} in Q1 and \dxEarnAcQfive{} in Q5, a ratio of \dxEarnAcRatio{} with a
95\% issuer bootstrap interval of [\dxEarnRatioLo, \dxEarnRatioHi]; the purchase ratio on the same
timing, \dxPurAcRatio, lies outside it, which was the pre-specified criterion for a
purchase-specific effect. Beaten-down stocks do not react more to earnings news as such: the
first-session spread between positive and negative surprises is \dcEarnSpreadQonePp{} points in Q1
against \dcEarnSpreadQfivePp{} in Q5 (difference \dcEarnSpreadDiffPp{} points, two-way $t$ =
\dcEarnSpreadDiffTTwo). The criterion is not the whole story. Restricted to the top tercile of
surprises, earnings news does reprice more in beaten-down stocks (\dxEarnTopAcQone{} against
\dxEarnTopAcQfive, ratio \dxEarnTopRatio, interval [\dxEarnTopRatioLo, \dxEarnTopRatioHi]), and on
raw returns the purchase ratio falls inside that interval: a purchase far below the high is not
distinguishable from a strong earnings beat. The comparison is also the least powerful in the
paper. The after-close positive-surprise sample has \dcEarnGoodAcQoneN{} events by
\dcEarnGoodQoneIssuers{} issuers in Q1 (\dcEarnTopAcQoneN{} events by \dcEarnTopQoneIssuers{}
issuers in the top tercile) against \dcPurQoneN{} purchases by \dcPurQoneIssuers{} issuers, because
issuers far below their high rarely report positive surprises. Two construction checks (post hoc)
leave the results unchanged: \dcEarnNonFirst{} of the \dcEarnEvents{} events are Item~2.02 filings
that share a fiscal quarter with an earlier one and inherit its surprise, and restricting each
quarter to its first filing gives a positive-surprise ratio of \dcEarnGoodFirstRatio{} and a
top-tercile gap of \dcEarnTopFirstGapPp{} points ($t$ = \dcEarnTopFirstGapTTwo).

The two disclosure types separate partly, not fully, once volatility is accounted for. Standardizing
every disclosure by its own pre-event volatility --- a comparison specified after the pre-specified
results were known --- the Q1/Q5 ratio is \dxZPurRatio{} for purchases (difference \dxZPurGap{}
standard deviations, two-way $t$ = \dxZPurGapTTwo), \dxZEarnTopRatio{} for top-tercile earnings
surprises ($t$ = \dxZEarnTopGapTTwo, interval [\dxZEarnTopLo, \dxZEarnTopHi]) and
\dxZEarnGoodRatio{} for all positive surprises ($t$ = \dxZEarnGoodGapTTwo). Tested as a difference
in gaps in one pooled regression --- purchases and the comparator together, with a purchase
indicator and its interaction with Q1, two-way clustered --- the standardized purchase gradient
exceeds that of all positive surprises by \dcDigGoodZ{} standard deviations ($t$ = \dcDigGoodZTTwo;
on raw returns \dcDigGoodRawPp{} points, $t$ = \dcDigGoodRawTTwo), but that of the strongest
surprises by \dcDigTopZ{} ($t$ = \dcDigTopZTTwo; raw $t$ = \dcDigTopRawTTwo). A joint issuer
bootstrap of the two ratios puts the difference between the standardized purchase and top-tercile
earnings ratios at [\dcBootTopZRatioDiffLo, \dcBootTopZRatioDiffHi] and between the raw ones at
[\dcBootTopRatioDiffLo, \dcBootTopRatioDiffHi]. Against positive surprises as a class, then, the
purchase gradient is larger on raw and on standardized returns; against the strongest surprises it
is larger in standardized terms but not precisely so on \dcEarnTopAcQoneN{} events, and the raw
earnings gradient there is the scale of the stock while the purchase gradient is not.

\paragraph{T5: the mirror (pass).} Insider sales by the same issuers, \dxSaleEvents{} filings over
the same period of which \dxSaleAcN{} pass the eligibility rules after the close, show no gradient:
the day-one reaction is \dxSaleAcQone{} in Q1 (\dcSaleQoneN{} events) and \dxSaleAcQfive{} in Q5
(\dcSaleQfiveN), a difference of \dxSaleAcGapPp{} points (two-way $t$ = \dxSaleAcGapTTwo) against
\dxPurAcGapPp{} points for purchases on the same timing. The ratio of the sale gradient to the
purchase gradient is $k$ = \dxKSale, with a 95\% issuer bootstrap interval of [\dxKSaleLo,
\dxKSaleHi]; the pre-specified reading fixed an upper bound below 0.5 as favouring the
costly-signal account, and it is met. Standardized by volatility the sale gradient is equally flat
(\dxZSaleGap{} standard deviations, $t$ = \dxZSaleGapTTwo), and the difference in gaps against the
sign-flipped sale gradient is \dcDigSaleZ{} standard deviations ($t$ = \dcDigSaleZTTwo; raw
\dcDigSaleRawPp{} points, $t$ = \dcDigSaleRawTTwo). Two qualifications. The inference is weaker than
the test: purchases are more informative than sales in general \citep{lakonishok2001,jeng2003}, so
a flatter sale gradient is what several accounts predict and the test is a necessary rather than a
sufficient condition for the costly-signal reading. And the sale sample is less clean than the
purchase sample: it did not pass through the purchase pipeline's price-basis and capitalization
filters, and on an approximate point-in-time capitalization \dcSaleAboveFiveHundredShare{} of the
after-close sale events (\dcSaleAboveFiveHundredShareQfive{} in Q5 against
\dcSaleAboveFiveHundredShareQone{} in Q1) sit above \$500 million, so many sellers are larger firms
than any buyer in the sample. Restricting sales to the price-basis band, a ticker-year basis check
and the \$30--\$500 million band (post hoc) leaves \dcSaleCapN{} events, on which the gap is
\dcSaleCapGapPp{} points ($t$ = \dcSaleCapGapTTwo) and $k$ = \dcSaleCapK{} [\dcSaleCapKLo,
\dcSaleCapKHi].

\paragraph{Concurrent news (post hoc).} The pre-specified tests attribute the reaction to the
Form~4 by its timing alone. Using the SEC submissions records fetched for T4, I flag every purchase
whose issuer filed an 8-K of any item, amendments included, between the start of the session before
$t_0$ and the close of the $t_0+1$ session, or an Item~2.02 8-K whose first reacting session falls
within the five sessions before $t_0$ or on $t_0+1$. \dnNoCikLines{} lines (\dnNoCikTickers{}
issuers) have no CIK record and \dnDeadCikLines{} more (\dnDeadCikTickers{} issuers) map to a CIK
that filed no 8-K within a year of the purchase, a probable mapping failure; both groups are
excluded from the screened set rather than counted as news-free, and issuers reporting on Form 6-K
and news that reaches the market outside an SEC filing are not screened at all. Of the
\dnCheckedN{} remaining lines, \dnFlagShare{} are flagged (\dnFlagShareQone{} in Q1,
\dnFlagShareQfive{} in Q5): \dnBroadShare{} by an 8-K inside the two-session window and
\dnErnShare{} by a nearby earnings release. Concurrent news adds to the reaction --- flagged Q1
purchases earn \dnFlaggedQone{} against \dnUnflaggedQone{} for unflagged ones (difference
\dnQoneFlagDiffPp{} points, clustered $t$ = \dnQoneFlagDiffT) --- but does not create the gradient.
On the \dnUnflaggedN{} news-free purchases the day-one reaction is \dnUnflaggedQone{} in Q1 against
\dnUnflaggedQfive{} in Q5 (gap \dnUnflaggedGapPp{} points, two-way $t$ = \dnUnflaggedGapTTwo;
standardized ratio \dnUnflaggedZRatio, $t$ = \dnUnflaggedZGapTTwo), and \dnUnflaggedAcQone{}
against \dnUnflaggedAcQfive{} for after-close filings (gap $t$ = \dnUnflaggedAcGapTTwo). The T2
regression on the news-free set leaves \dnCtrlUnflaggedQoneCoefPp{} points on raw returns ($t$ =
\dnCtrlUnflaggedQoneTTwo) and \dnCtrlUnflaggedZQoneCoef{} standard deviations on the standardized
outcome ($t$ = \dnCtrlUnflaggedZQoneTTwo): on news-free purchases the per-risk gradient no longer
clears the bar T2 set, while the raw gradient without controls does.

\paragraph{After the first session (post hoc).} The pre-specified outcome is the first session.
Recomputing the abnormal-return path from the panel with the paper's own benchmark code, which
reproduces the stored $\mathrm{AR}_{t_0+1}$ and drift exactly, Q1 earns \dpTtwoQone{} on $t_0+2$
(clustered $t$ = \dpTtwoQoneTCl, two-way $t$ = \dpTtwoQoneTTwo; Q1 $-$ Q5 \dpTtwoGapPp{} points,
$t$ = \dpTtwoGapTTwo) and \dpTthreeQone{} on $t_0+3$ ($t$ = \dpTthreeQoneTTwo; gap $t$ =
\dpTthreeGapTTwo), a further \dpTtwoTthreeQoneSumPp{} points over the two sessions after the first,
and by $t_0+3$ \dpQoneShareTthree{} of the quintile's mean $\CAR$ is realized. Over $t_0+2$ to
$t_0+10$ the Q1 mean is \dpTtwoTenQone{} (two-way $t$ = \dpTtwoTenQoneTTwo) and the gap
\dpTtwoTenGapPp{} points ($t$ = \dpTtwoTenGapTTwo); no later window is precisely estimated. The
reaction is two to three sessions rather than one, and a one-session description would overstate
it.

\paragraph{What the reaction is.} The day-one reaction to an insider purchase far below the 52-week
high is mostly the scale of a volatile, cheap stock, which is why its raw size does not survive a
full set of controls. What remains is a gradient per unit of the stock's own risk: positive and
precisely estimated at fixed volatility, between \drMatchedZRatio{} and \drMedianZRatio{} times
depending on the estimator (a variance-weighted slope shows none), positive within each tercile of
volatility and precisely so in the two that hold most far-quintile lines, and larger than the
corresponding gradient for insider sales and for positive earnings surprises as a class by the same
firms, though not separable from that of the strongest earnings surprises and no longer clearing
the full-control bar on news-free purchases. This is consistent with a costly-signal account, in
which an open-market purchase of a stock that has already fallen commits the insider's own money
against the market's recent verdict, selling it does not, and an earnings beat is not a choice, an
account that applies to the discretionary open-market buys among the code-P lines
(Section~\ref{sec:sources}); it is also consistent
with a purchase resolving more uncertainty where there is more of it. An anchoring account in which
any news moves the price relative to the reference point predicts a sale gradient of the same kind,
and there is none; the strongest earnings news is the case the evidence does not separate.
\begin{table}[tbp]
\centering\tabfoot
\caption{What the day-one reaction is and is not. Panel A: the insider purchase reaction under the
three pre-specified objections that need no new data (T1--T3), followed by post-hoc rows: the same
gradient on purchases with no concurrent 8-K filing by the issuer, and the abnormal return on the
sessions after the first. Panel B: the reaction on the first session after publication for three
kinds of disclosure by the same issuers, 2018--2024, scored on the same code path and benchmark and
assigned to the purchase sample's quintile cutoffs of distance from the 52-week high; after-close
filings only (acceptance at or after 16:00~ET), so that the abnormal return on $t_0+1$ is
unambiguously the first session after publication. Earnings: original 8-K filings with Item~2.02
whose seasonal-random-walk surprise is positive. Sales: code-S sales without option exercise
(OpenInsider \texttt{S - Sale}), one event per issuer and filing timestamp; the restricted row
applies the purchase sample's price-basis and capitalization bands. $N$ is the number of events
and, in parentheses, the number in Q1. Q1 $-$ Q5 gaps carry two-way issuer-and-month clustered
$t$-statistics; sd = standard deviations of the stock's own pre-event volatility. The last column of
Panel A gives the reading fixed before the test was run, or ``post hoc'' for comparisons added
after the pre-specified tests. Below Panel B: the pre-specified
ratio and $k$ criteria, and the post-hoc standardized ratios and difference-in-gaps tests.}
\label{tab:identification}
\shrinkbox{
\begin{tabular}{l r r r r l}
\toprule
\multicolumn{6}{l}{\emph{Panel A: insider purchases, all timings}} \\
 & Q1 & Q5 & Q1 $-$ Q5 & two-way $t$ & Reading \\
 & & & (pp or sd) & & \\
\midrule
Raw day-one abnormal return & \rxQoneArToneMean & \rxQfiveArToneMean & \dxRawGapPp & \dxRawGapTTwo & headline \\
Standardized by own volatility (sd) & \dxZQone & \dxZQfive & \dxZDiff & \dxZDiffTTwo & T1 pass, narrows \\
Share of positive reactions (pp) & \dxPosQone & \dxPosQfive & \dxPosDiffPp & \drPosTTwo & T1 pass \\
Raw, full controls (pp) & & & \dxCtrlQoneCoefPp & \dxCtrlQoneTTwo & \textbf{T2 fail} ($|t|<3$) \\
Standardized, full controls (sd) & & & \dxCtrlZQoneCoef & \dxCtrlZQoneTTwo & post hoc \\
Positive share, full controls (pp) & & & \dxPosCtrlPp & \dxPosCtrlT & post hoc \\
Raw, 2018--2020 (high attrition) & & & \dxSurvHiGapPp & \dxSurvHiT & T3 \\
Raw, 2023--2024 (low attrition) & & & \dxSurvLoGapPp & \dxSurvLoT & T3 pass \\
\quad difference, 2023--24 minus 2018--20 & & & \dcSurvDiffPp & \dcSurvDiffTTwo & post hoc; 95\% [\dcSurvDiffLoPp, \dcSurvDiffHiPp] \\
Raw, no concurrent 8-K ($N$ = \dnUnflaggedN) & \dnUnflaggedQone & \dnUnflaggedQfive & \dnUnflaggedGapPp & \dnUnflaggedGapTTwo & post hoc \\
Standardized, no concurrent 8-K (sd) & \dnUnflaggedZQone & \dnUnflaggedZQfive & \dnUnflaggedZGap & \dnUnflaggedZGapTTwo & post hoc \\
Standardized, no concurrent 8-K, full controls (sd) & & & \dnCtrlUnflaggedZQoneCoef & \dnCtrlUnflaggedZQoneTTwo & post hoc \\
Raw, second session $t_0+2$ & \dpTtwoQone & \dpTtwoQfive & \dpTtwoGapPp & \dpTtwoGapTTwo & post hoc \\
Raw, third session $t_0+3$ & \dpTthreeQone & \dpTthreeQfive & \dpTthreeGapPp & \dpTthreeGapTTwo & post hoc \\
Raw, sessions $t_0+2$ to $t_0+10$ & \dpTtwoTenQone & \dpTtwoTenQfive & \dpTtwoTenGapPp & \dpTtwoTenGapTTwo & post hoc \\
\midrule
\multicolumn{6}{p{0.96\textwidth}}{\emph{Per-risk Q1/Q5 ratio by estimator (post hoc):} mean $z$ \drMeanZRatio{} [\drMeanZLo, \drMeanZHi]; median \drMedianZRatio; winsorized at 1\%/5\% \drWinsOneRatio/\drWinsFiveRatio; mean reaction over mean volatility \drMeanArMeanSigRatio; after-close filings \drAcZRatio; volatility window ending $t_0-1$ \drExTzeroZRatio; issuer-days \drIssuerDayZRatio; Q5 reweighted to Q1's volatility deciles \drMatchedZRatio{} (raw \drMatchedRawRatio); within volatility terciles \drTercMidZRatio{} (middle), \drTercHiZRatio{} (upper); variance-weighted slope through the origin \drOriginSlopeRatio. Volatility-decile fixed effects: Q1 $-$ Q5 = \drFeZDiff{} sd ($t$ = \drFeZTTwo), \drFeRawDiffPp{} pp ($t$ = \drFeRawTTwo).} \\
\midrule
\multicolumn{6}{l}{\emph{Panel B: first session after publication, after-close filings}} \\
 & $N$ ($N_{\mathrm{Q1}}$) & Q1 & Q5 & Q1 $-$ Q5 (pp) & two-way $t$ \\
\midrule
Insider purchases & \dxPurAcN{} (\dcPurQoneN) & \dxPurAcQone & \dxPurAcQfive & \dxPurAcGapPp & \dxPurAcGapTTwo \\
Positive earnings surprises & \dxEarnAcN{} (\dcEarnGoodAcQoneN) & \dxEarnAcQone & \dxEarnAcQfive & \dxEarnAcGapPp & \dxEarnAcGapTTwo \\
\quad top surprise tercile & \dxEarnTopAcN{} (\dcEarnTopAcQoneN) & \dxEarnTopAcQone & \dxEarnTopAcQfive & \dxEarnTopAcGapPp & \dxEarnTopAcGapTTwo \\
\quad negative surprises & \dxEarnBadAcN{} (\dcEarnBadAcQoneN) & \dxEarnBadAcQone & \dxEarnBadAcQfive & \dxEarnBadAcGapPp & \dxEarnBadAcGapTTwo \\
Insider sales & \dxSaleAcN{} (\dcSaleQoneN) & \dxSaleAcQone & \dxSaleAcQfive & \dxSaleAcGapPp & \dxSaleAcGapTTwo \\
\quad restricted to purchase-sample bands (post hoc) & \dcSaleCapN{} (\dcSaleCapQoneN) & \dcSaleCapQone & \dcSaleCapQfive & \dcSaleCapGapPp & \dcSaleCapGapTTwo \\
\midrule
\multicolumn{6}{p{0.96\textwidth}}{\emph{Pre-specified criteria.} Q1/Q5 raw ratio: purchases \dxPurAcRatio; positive earnings \dxEarnAcRatio{} [\dxEarnRatioLo, \dxEarnRatioHi], purchase ratio outside (T4 pass); top tercile \dxEarnTopRatio{} [\dxEarnTopRatioLo, \dxEarnTopRatioHi], purchase ratio inside. Sale-to-purchase gradient $k$: \dxKSale{} [\dxKSaleLo, \dxKSaleHi], upper bound below 0.5 (T5 pass); restricted sample \dcSaleCapK{} [\dcSaleCapKLo, \dcSaleCapKHi].} \\
\multicolumn{6}{p{0.96\textwidth}}{\emph{Post hoc.} Standardized Q1/Q5: purchases \dxZPurRatio, positive earnings \dxZEarnGoodRatio, top tercile \dxZEarnTopRatio, sales \dxZSaleRatio. Difference in gaps, purchases minus comparator, two-way $t$ (standardized; raw): positive earnings \dcDigGoodZTTwo; \dcDigGoodRawTTwo, top tercile \dcDigTopZTTwo; \dcDigTopRawTTwo, sales sign-flipped \dcDigSaleZTTwo; \dcDigSaleRawTTwo. Joint bootstrap of the standardized ratio difference, purchases minus top tercile: [\dcBootTopZRatioDiffLo, \dcBootTopZRatioDiffHi].} \\
\bottomrule
\end{tabular}}
\end{table}
\section{Robustness}\label{sec:robustness}

\paragraph{Rolling origin, regimes and issuers.}
Table~\ref{tab:rolling} re-runs the protocol with each year from 2020 to 2024 as the test year, training on years up to two before it and thresholding on the year before. Test AUCs range from \robRollAUCminThree{} (\robRollAUCminYear, trained on \robRollNTrainTwenty{} lines) to \robRollAUCmaxThree{} (\robRollAUCmaxYear), with a mean of \robRollAUCmeanThree{} and a standard deviation of \robRollAUCsd; the random forest matches or exceeds XGBoost in every origin before 2024 and trails it by \clDiffXgbRf{} in 2024 (Table~\ref{tab:rolling}), consistent with Section~\ref{sec:classification}. Within 2024 the quarterly AUC lies between \robQAUCminThree{} and \robQAUCmaxThree; it is \robVolLowAUCthree{} below and \robVolHighAUCthree{} above the median stock volatility, where the positive rate is \robVolLowPosRate{} against \robVolHighPosRate; and it is \robMktVolLowAUCthree{} against \robMktVolHighAUCthree{} on either side of the median trailing \robMktVolWindow-day realized market volatility computed from the market factor. Issuer memorization is limited: \robShareTestSeenTrainVal{} of 2024 lines are on issuers present in the training or threshold years; removing every 2024 issuer from those years (\robBlockNDropRows{} lines on \robBlockNDropTickers{} issuers) and refitting gives a test AUC of \robBlockAllAUCthree{} against \robPrimaryAUCthree{} (independence-based DeLong $p$ = \robBlockDeLongP), so issuer recognition is worth about \robBlockVsCtrlIssAUC{} to \robBlockVsCtrlRowAUC{} of AUC relative to size-matched random drops. Collapsing to one row per issuer-day with the maximum predicted probability gives an AUC of \robFirmDayAUCthree{} (interval \robFirmDayAUCci) on \robFirmDayN{} issuer-days.

\begin{table}[tbp]
\centering\tabfoot
\caption{Rolling-origin evaluation. For each test year $Y$ the model is trained on lines filed up to $Y-2$ with rounds chosen by the cross-validation protocol inside those years, the threshold is chosen on $Y-1$ and the model is evaluated on $Y$; the last row is the primary fit. Columns: training years, threshold year, training and test lines, test positive rate, rounds, XGBoost test AUC with its issuer-cluster bootstrap interval, random forest and logistic regression test AUCs, and XGBoost F1 at the threshold.}
\label{tab:rolling}
\shrinkbox{
\begin{tabular}{l l l r r r r r r r r r}
\toprule
Test year & Train years & Threshold year & $N$ train & $N$ test & Pos.\ rate & Rounds & AUC & 95\% CI & RF AUC & LR AUC & F1 \\
\midrule
\tabinput{outputs/table_robustness_rolling.tex}
\end{tabular}}
\end{table}

\begin{figure}[tbp]
\centering
\includegraphics[width=0.85\textwidth]{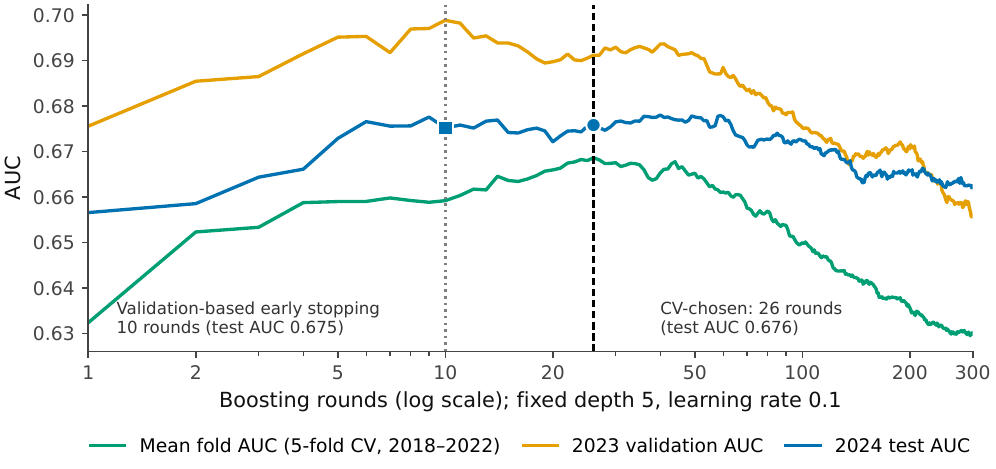}
\caption{AUC by number of boosting rounds for the a priori configuration (depth \xgbMaxDepth, learning rate \xgbLearningRate; log scale up to \figRcMaxRounds{} rounds): mean fold AUC of the five-fold cross-validation on time-ordered folds inside 2018--2022, 2023 AUC and 2024 test AUC. The dashed line marks the cross-validation choice of \figRcCvRounds{} rounds (test AUC \figRcTestAUCatCv) and the dotted line the validation-based early-stopping choice of \figRcEsRounds{} rounds (test AUC \figRcTestAUCatEs); the test curve peaks at \figRcTestAUCMax{} at round \figRcTestAUCArgmax{} and spans \figRcTestAUCRangeTenOn{} from round 10 on.}
\label{fig:rounds}
\end{figure}

\paragraph{Boosting rounds, random seeds and early stopping.}
Figure~\ref{fig:rounds} plots the mean fold, 2023 and 2024 AUC of the a priori configuration by number of rounds up to \figRcMaxRounds. The 2024 curve is flat from round 10 on: it ranges over \robTestCurveRangeTenToHundred{} between 10 and \xgbMaxRounds{} rounds and peaks at \robTestAUCbyRoundsMax{} at round \robTestAUCbyRoundsArgmax, against \robPrimaryAUCthree{} at the cross-validation choice of \figRcCvRounds{} rounds, \robEsFullTestAUCthree{} at the validation-based early-stopping choice of \figRcEsRounds{} (Appendix~\ref{app:model}), \robXgbAUCnoESthree{} at a fixed \xgbMaxRounds{} and \robXgbAUCmaxRounds{} at \xgbMaxRoundsCV. Across five seeds the cross-validation choice stays at \clSeedZeroRounds{} rounds and the XGBoost test AUC at \clSeedZeroXgbAUC, the fit being deterministic without subsampling, while the random forest ranges from \clRfSeedMin{} to \clRfSeedMax.

\paragraph{Tails, benchmarks and timing.}
Appendix~\ref{app:robust} repeats the headline quantities when the \robNBigCAR{} lines with $|\CAR| > 1$ (on \robNBigCARTickers{} issuers, all positive) are excluded, when $\CAR$ is winsorized at the full-sample 1st and 99th percentiles, on the buy-and-hold abnormal return, and on the timing variant. Across the five variants the flagged-minus-rejected gap lies between \robVariantTradeDiffMinPP{} and \robVariantTradeDiffMaxPP{} points with clustered $t$ between \robVariantTradeTClMin{} and \robVariantTradeTClMax, while the extreme run-up contrast lies between \robVariantContrastMinPP{} and \robVariantContrastMaxPP{} points with clustered $t$ between \robVariantContrastTClMin{} and \robVariantContrastTClMax, at most marginal. The timing variant, which removes the overnight reaction for after-close filings, lowers the flagged-minus-rejected gap to \robTimingTradeDiffPP{} points ($t$ = \robTimingTradeTCl), so part of the per-trade difference is the untradable first-day gap of Section~\ref{sec:reaction}. On the intercept-inclusive benchmark (last row of Table~\ref{tab:tails_model}) the primary model's probabilities score \robVOneAUCthree{} against the intercept-inclusive label and the flagged-minus-rejected gap is \robVOneTradeDiffPP{} points ($t$ = \robVOneTradeTCl); a refit on the buy-and-hold label gives \robBharRefitAUCthree{} (interval \robBharRefitAUCci) and on the timing label \robTimingRefitAUCthree{} (\robTimingRefitAUCci). Price floors of \$1, \$2 and \$5 are tabulated in Appendix~\ref{app:robust}.

\section{A liquid-universe extension: opportunity without a validated increment}\label{sec:extension}

The sample of this paper is microcap, survivor-only and thin: the median line
trades \ssMedianAddv{} a day, so a position large enough to matter cannot be
filled, and Section~\ref{sec:economics} reports per-trade returns rather than a
portfolio. That is a property of the universe, not of the disclosure. This
section reports a separate, larger-capitalization extension built to ask a
different question -- how many dollars a screened Form~4 rule can actually put to
work, and whether the resulting book beats a matched comparison -- and it
answers the first question and fails the second.

The extension is a distinct dataset, universe, sample period and research
protocol. Nothing in it is used to support, extend or qualify the microcap
results of Sections~\ref{sec:results}--\ref{sec:robustness}, and the two must not
be read as one sample. It is reported because tradability is part of what these disclosures are
worth, and because its design constraints show what such a study requires.

\paragraph{What is already known about buyer clusters.} The construct this extension
uses is established. \citet{kang2018} define cluster trading as several insiders of the same
firm trading in the same direction on the same or consecutive days, find that more than 40\% of
insider trades are clustered on 457{,}539 purchases and 1{,}001{,}188 sales over 1986--2016, and
document that clustered purchases predict higher returns than isolated ones; \citet{alldredge2019}
establish the same clustering behaviour on daily data in a published sample. \citet{ahn2026}
studies the same Form~4 cluster events on a survivorship-free 2015--2025 panel of 1{,}932{,}506
accessions and 38{,}261 cluster events, and reports that a five-day long-only cross-section of
them earns an annualised net Sharpe ratio of 1.77 at a 25~bp round-trip cost before any
classification overlay.\footnote{Design and results of \citet{ahn2026} as reported in its abstract.} Those samples are two to four orders of magnitude larger than the cohort below and document
that cluster purchases earn positive raw returns. This section does not revisit
that question, and its positive dollars are not evidence on it.

The question this extension asks is not whether those returns exist but what happens to them when a fixed amount of money has to be put to work, and that question has already been answered once for Form~4 filings: \citet{oenschlager2025} report that the returns to trading on insider filings vanish and turn negative once the tradable dollar amount per signal is capped at a reasonable size, and that they are negatively correlated with liquidity. The extension is consistent with that conclusion on a different universe and sample period and adds the accounting behind it: the schedulable flow and inventory, the funding requirement, the matched-comparison support, and the behaviour of the excess when the cost assumption doubles. Neither \citet{kang2018}, whose working paper contains no treatment of transaction costs, capacity or implementable flow, nor \citet{ahn2026}, whose abstract reports a single 25~bp cost assumption with no capacity or matched-control analysis, reports those quantities for the cluster construct specifically. The answer here is negative.

\paragraph{Design.} The filing cohort runs from March 2025 to June 2026 and is
drawn from the same SEC quarterly archives, screened to direct common-equity
purchases of at least \$100{,}000 by a single officer or director acting as a
natural person, excluding affirmative trading plans and explicit
issuer-financing or compensation. The screen runs on the SEC's bulk Form~4
data. The original XML filing was checked for \ExtPersonalXmlVerified{} of the
\ExtPersonalPositions{} accepted personal positions, so \ExtPersonalXmlUnverified{}
(\ExtPersonalXmlUnverifiedFlowShare{} of that family's scheduled flow) rest on the bulk
data alone, and for all \ExtClusterPositions{} cluster positions, of which
\ExtClusterNarrativeConflict{} (\ExtClusterNarrativeConflictFlow{} of flow) are coded as
purchases in a filing whose narrative describes a sale; they are kept under the frozen
coding rule. ``Personal purchase'' is the screen's definition, not a filing-by-filing
confirmation. Three nested families are reported: all such
personal purchases; the subset with a prior distinct qualifying buyer within
thirty days (the \emph{cluster} family); and the first arrival of a second
distinct buyer (the \emph{second-buyer} family), which is the one design of the
three that isolates an incremental-information hypothesis. Entry is the first
session strictly after the filing date, exit is twenty sessions later, positions
are whole tickets capped at \$250{,}000 against a \$5 million nominal budget, and
no two positions in one issuer overlap. Every rule, source correction, comparison
assignment and capital schedule was frozen before the first event return was
extracted; the freeze times and hashes are in the research record.

\paragraph{Opportunity.} Table~\ref{tab:extension} gives the result. The liquid
rule enters materially more capital than the microcap pool: the personal family
schedules \ExtPersonalFlowAnn{} a year across \ExtPersonalPositions{} positions,
the cluster family \ExtClusterFlowAnn{} across \ExtClusterPositions, and the
second-buyer family \ExtSecondFlowAnn{} across \ExtSecondPositions. On daily
adjusted prices and a 25~bp round-trip cost the full sixteen-month cohort
returns \ExtPersonalNetTwentyFive, \ExtClusterNetTwentyFive{} and
\ExtSecondNetTwentyFive{} respectively, through the last scheduled exit. These
are cohort proxy dollars, not annual alpha and not filled profit: they carry no
opening-auction depth, no arrival quote, no fill probability and no measured
slippage.

\paragraph{Why the positive dollars do not establish an edge.} Three findings
stand against reading those totals as evidence of a tradable signal, and each is
prespecified rather than discovered.

First, every comparison fails its support gate. Matching each position to
same-day, same-sector controls under fixed size, liquidity and prior-return
calipers covers \ExtPersonalCoverage{} of scheduled personal notional and
\ExtClusterCoverage{} of cluster notional, against a frozen 80\% requirement,
and the strict source- and news-matched second-buyer comparison -- the only
design here that addresses the increment -- covers \ExtSecondCoverage, nine
positions. The covered-subset net excess is \ExtPersonalExcess{} for the personal
family and \ExtClusterExcess{} for the cluster family at 25~bp; the
second-buyer primary comparison is \ExtSecondExcess. Low coverage is not a
technicality: an unmatched majority of the book means the excess describes a
subset selected by data availability.

Second, what survives costs is small and the intervals contain zero. The cluster
covered-subset excess falls from \ExtClusterExcess{} at 25~bp to
\ExtClusterExcessFifty{} at 50~bp, so a doubling of the cost assumption
removes almost all of it. Twenty-session moving-block resampling puts the cluster
excess in \ExtClusterCiLow{} to \ExtClusterCiHigh{} and the second-buyer estimate
in \ExtSecondCiLow{} to \ExtSecondCiHigh. The second-buyer point estimate is
negative, but with nine matched positions it establishes neither an effect nor
its absence; it is an underpowered comparison, and reporting it as evidence of
zero would be as wrong as reporting the raw total as evidence of profit.

Third, the profits are concentrated in one entry cohort. Entries in
\ExtClusterTopQuarter{} supply \ExtClusterTopQuarterShare{} of the cluster
family's net proxy profit and \ExtSecondTopQuarterShare{} of the second-buyer
family's; removing those cohorts, holding every other booked position fixed,
leaves \ExtClusterExTopQuarter{} and \ExtSecondExTopQuarter. A result that
depends on one quarter out of five is a statement about that quarter until a
longer cohort says otherwise. This is a descriptive check on unchanged positions,
not an alternative backtest and not a calendar filter.

\paragraph{Capital, not just return.} The frozen schedules are also not free
books. The personal and cluster schedules reach a most negative marked cash
balance requiring \ExtPersonalExtraCash{} and \ExtClusterExtraCash{} beyond the
nominal \$5 million, so they are not self-financed at that size; maximum
drawdowns at 25~bp are \ExtPersonalMaxDD{} and \ExtClusterMaxDD. Read as a
business rather than a backtest, the observed annualized flow means
\$500{,}000 of profit before fixed expense requires about
\ExtPersonalEdgeFiveHundred, \ExtClusterEdgeFiveHundred{} or
\ExtSecondEdgeFiveHundred{} net basis points for the three families. None of
those edges is demonstrated here. Enlarging the account does not help: the
binding constraint is the number of qualifying disclosures, not the budget.

\paragraph{What the extension contributes.} It shows that the economic question
and the statistical question separate. A screened insider-purchase rule in liquid
names schedules tens of millions of dollars a year of nominal flow. Under the same
ticket rule (1\% of 30-day dollar volume, capped at \$250{,}000 per position) the
microcap sample of this paper schedules \mfClusterFlowAnnM{} a year on its nearest
analogue of the cluster screen and \mfAllFlowAnnM{} on every purchase line, but at a
median ticket of \mfAllMedianTicket{} spread over \mfAllStockDaysPerYear{} stock-days a
year, so what separates the two universes is the size of a position, not the nominal
total. Nominal flow is the part most easily established. The incremental claim, that these
particular disclosures beat a comparable stock bought the same day, is the part
that the data do not support at the coverage, cost and concentration that the
frozen design requires. The extension is retained as a bounded research lead with
explicit external requirements outstanding: permanent security identifiers with
delisting-complete total returns, historical quote and official auction records,
the strategy's own receipt and deadline clocks, and authentic fill and cost
records. Until those exist, the liquid family is best described as an
opportunity of measurable size and an advantage of unmeasured sign.

\begin{table}[tbp]
\centering\tabfoot
\caption{The liquid-universe extension, March 2025--June 2026 filing cohort.
Positions and flow are the frozen \$250{,}000-ticket, \$5 million nominal
schedule; net proxy P\&L is the full sixteen-month cohort on daily adjusted
prices through the last scheduled exit at a 25~bp round-trip cost, not
annual alpha and not filled profit. Coverage is the share of scheduled notional
with a complete matched comparison against a frozen 80\% requirement; the
covered-subset excess and its twenty-session moving-block interval describe that
subset only. The second-buyer column uses the strict source- and news-matched
comparison, the only design of the three that addresses an incremental claim.
The last row removes the largest entry-quarter cohort holding all other booked
positions fixed.}
\label{tab:extension}
\shrinkbox{
\begin{tabular}{l r r r}
\toprule
 & Personal & Cluster & Second buyer \\
\midrule
Positions & \ExtPersonalPositions & \ExtClusterPositions & \ExtSecondPositions \\
Annualized scheduled flow & \ExtPersonalFlowAnn & \ExtClusterFlowAnn & \ExtSecondFlowAnn \\
Average nominal inventory & \ExtPersonalAvgInventory & \ExtClusterAvgInventory & \ExtSecondAvgInventory \\
Net proxy P\&L, 25~bp & \ExtPersonalNetTwentyFive & \ExtClusterNetTwentyFive & \ExtSecondNetTwentyFive \\
Maximum drawdown, 25~bp & \ExtPersonalMaxDD & \ExtClusterMaxDD & \ExtSecondMaxDD \\
Additional cash required & \ExtPersonalExtraCash & \ExtClusterExtraCash & \ExtSecondExtraCash \\
\midrule
Matched positions & \ExtPersonalMatched & \ExtClusterMatched & \ExtSecondMatched \\
Scheduled-notional coverage & \ExtPersonalCoverage & \ExtClusterCoverage & \ExtSecondCoverage \\
Support gate (80\%) & \ExtPersonalGatePass & \ExtClusterGatePass & \ExtSecondGatePass \\
Covered-subset excess, 25~bp & \ExtPersonalExcess & \ExtClusterExcess & \ExtSecondExcess \\
Covered-subset excess, 50~bp & \ExtPersonalExcessFifty & \ExtClusterExcessFifty & \ExtSecondExcessFifty \\
95\% block interval, low & \ExtPersonalCiLow & \ExtClusterCiLow & \ExtSecondCiLow \\
95\% block interval, high & \ExtPersonalCiHigh & \ExtClusterCiHigh & \ExtSecondCiHigh \\
\midrule
Largest entry quarter & \ExtPersonalTopQuarter & \ExtClusterTopQuarter & \ExtSecondTopQuarter \\
Its share of net P\&L & \ExtPersonalTopQuarterShare & \ExtClusterTopQuarterShare & \ExtSecondTopQuarterShare \\
Net P\&L excluding it & \ExtPersonalExTopQuarter & \ExtClusterExTopQuarter & \ExtSecondExTopQuarter \\
Net edge for \$500k/year (bp) & \ExtPersonalEdgeFiveHundred & \ExtClusterEdgeFiveHundred & \ExtSecondEdgeFiveHundred \\
\bottomrule
\end{tabular}}
\end{table}

\section{Discussion}\label{sec:discussion}

Distance below the 52-week high is informative for three reasons that the paper separates. First, an intercept-inclusive benchmark credits it with \mechQoneQfiveCarryShareOfVOneGap{} of a quintile gradient that no purchase is needed to produce; that part belongs to the benchmark. Second, about \mechRhoDropShare{} of its remaining rank association with the label is removed by volatility normalization: stocks far below their annual high are volatile, and a fixed threshold on a summed return is a tail event whose probability rises with volatility in both directions, which is also why volatility is the classifier's top feature. Third, of the directional part that survives, a \mechQoneQfiveMeanCARDiffPp-point gradient in mean $\CAR$, \rxDiffArTonePp{} points are realized on the trading day after the filing day, for after-close filings mostly in the overnight gap, with a two-way $t$ of \rxDiffArToneTTwo; the remaining \rxDiffDriftThirtyPp{} points are a drift with a two-way $t$ of \rxDiffDriftThirtyTTwo{} that a calendar-time portfolio does not collect and that is not distinguishable from zero in the lines priced at \$5 or more (Table~\ref{tab:reaction_floor}). Section~\ref{sec:identification} then shows that most of the day-one gradient is the volatility of the beaten-down stocks, that the gradient per unit of that volatility is positive and precisely estimated but of estimator-dependent size, that it is larger than the corresponding gradients for insider sales and positive earnings surprises as a class though not precisely separable from the strongest surprises, and that the reaction runs over the first two to three sessions. It is consistent with insiders buying as contrarians after large declines \citep{rozeff1998,jenter2005,piotroski2005} and with the market reacting more to such purchases; the precisely measured part is not evidence of slow correction of underpricing in stocks that few analysts follow \citep{hong2000}, which would show up as drift after the first sessions. The sign is the opposite of the momentum use of the same variable in \citet{george2004}: there, proximity to the high predicts continuation in the broad cross-section; here, in a sample conditioned on an insider having bought, distance below the high predicts a larger one-day reaction.

Adjusted daily price series are the default output of most retail and academic feeds, and Form~4 prices are reported as paid. Any run-up, discount or price-improvement variable that divides one by the other inherits every subsequent split and dividend, and because reverse splits concentrate in distressed issuers and dividends in banks, the contamination is correlated with industry, price level and subsequent returns rather than random. Researchers using such feeds should reconstruct the transaction-day share basis, or measure the run-up on adjusted prices at both ends, and check for the two signatures of Section~\ref{sec:artefact}. The benchmark lesson is a procedure rather than a rule: in an event study that sorts on trailing performance, or trains a model on trailing-performance features, whether to carry the estimation-window intercept depends on whether the pre-event drawdown persists, the choice can move the sorted gradient by more than the effect of interest (Table~\ref{tab:benchsim}), and a placebo of firm-days without an event, sorted the same way, measures that persistence in the sample at hand: here it reproduces \plGapVOneOverEventVOne{} times the events' gradient under the intercept-inclusive benchmark and \plGapPrimaryPp{} of their \plEventGapPrimaryPp{} points under the loadings-only one (Table~\ref{tab:placebo}).

On implementation, the model's flags in 2024 are a genuine screen: flagged and rejected lines differ by \econDiffFlagRejPp{} points per trade and by \econCtFlagMinusRejAlphaMonthly{} per month in calendar time, in \econMonthsFlagGtRej{} of \econMonthsTotal{} months and in medians and hit rates. But the screen selects into the least liquid, lowest-priced and widest-spread names, its flagged mean is right-tail-driven, its calendar-time alpha over rejected lines does not reach the 5\% level over one year, part of the per-trade gap is a first-day reaction an outsider cannot trade, and its capacity is small; Section~\ref{sec:limitations} lists what the evidence cannot settle. The paper's contribution is therefore less a trading signal than a characterization of where the information in microcap insider purchases sits, in the reaction over the first sessions after disclosure, in the price state and in a few insider types, and of how easily a spurious signal can be manufactured from it by a price feed or a benchmark.

\section{Limitations}\label{sec:limitations}

Five limits apply to the paper's central finding specifically. First, the decision rules of the five tests of Section~\ref{sec:identification} were fixed before the tests were run but after the day-one pattern was known, and they were not filed with a registry; they rule out choosing a reading after the fact, not the choice of which questions to test. Second, the raw day-one gap does not survive the full control set at the pre-specified threshold (two-way $t$ = \dxCtrlQoneTTwo); what survives is the volatility-scaled gradient, whose size depends on the estimator (\drMatchedZRatio{} to \drMedianZRatio, and none on a variance-weighted slope) and which no longer clears the same bar on the purchases free of concurrent 8-K news, and whether volatility is a confound or the channel through which a purchase resolves uncertainty is not something this design can settle. Third, the comparison of volatility-scaled gradients across purchases, sales and earnings, the difference-in-gaps tests, the concurrent-news screen, the alternative per-risk estimators and the path after day one are all post hoc; the pre-specified earnings test uses raw returns and is passed on all positive surprises but not on the strongest ones, where the raw purchase ratio lies inside the earnings interval and the standardized difference is imprecise on \dcEarnTopAcQoneN{} events; and the news screen covers 8-K filers only, so issuers without a CIK record, issuers whose CIK filed no 8-K, foreign private issuers reporting on Form 6-K and news that reaches the market outside an SEC filing are not screened. Fourth, the earnings comparison covers \dxEarnNSue{} of the \dxEarnNEightK{} Item~2.02 filings, because three-month fourth-quarter EPS is often absent from XBRL; a consensus-based surprise measure, which this study does not have, would cover them all. Fifth, the costly-signal reading concerns discretionary open-market buys with the insider's own money, which Form~4 code P does not identify: in the source audit of Section~\ref{sec:sources}, \saExplicitOpen{} of \saN{} filings state an open-market venue, \saFinancing{} are participations in an issuer financing and \saIndirect{} report indirect ownership.

The remaining limits apply to the sample and the design as a whole. First, survivorship: the sample is conditioned on a ticker having a price history in 2026, which removes \ssLostShare{} of the lines that pass the line filters with a loss rate declining from \ssLostShareFirstYear{} in 2018 to \ssLostShareLastYear{} in 2024; the direction of the bias is not determined and the training years are more affected than the test year. Second, share counts: \ssSharesSecPitShare{} of lines use a point-in-time SEC count, but the remainder use a 2026 snapshot split-adjusted to the filing day, which misstates capitalization for issuers that have since issued or repurchased shares. Third, identifiers: issuers are ticker strings, so symbol reuse and renames are handled only by the price-basis checks, and insiders are name strings. Fourth, the outcome: $\CAR$ is close-to-close, so for the \ssShareAfterClose{} of filings accepted after the close the first event-day return contains an overnight gap an outsider cannot trade, Section~\ref{sec:reaction} shows that for beaten-down stocks most of the reaction is that gap, and the timing variant lowers the per-trade gap accordingly; $\CAR$ is summed rather than compounded, and the buy-and-hold benchmark gives lower AUCs and a smaller run-up contrast; the flagged-minus-rejected gap is not re-estimated under a price floor. The announcement reaction is attributed to the filing by its timing alone, on the filing day for filings accepted before the close and overnight for those accepted after, with the concurrent-news screen of Section~\ref{sec:identification}, which covers 8-K filers only, as the sole control for other news; the overnight gap is measured to the official opening print, which is thin in these names and is not a fill an outsider obtains; the first-day return, like the 30-day one, is observed only for issuers that survive to 2026; and the sort on distance from the high is also a sort on price level: the reaction survives a \$5 price floor and a \$1 million dollar-volume floor (Table~\ref{tab:reaction_floor}), but the liquid subset holds \rxFloorFiveLiqQoneShareOfQone{} of the Q1 lines. Dropping the estimation-window intercept removes the carry component documented in this sample but still measures a stock in a drawdown against its loadings-only three-factor return; the placebo of Section~\ref{sec:mechanism} draws its days from the same survivor-only issuers as the events, so it controls for the benchmark and not for survivorship; and the post-day-one drift that remains in sub-\$5 names cannot be separated from survivorship with a survivor-only panel. Fifth, a single test year: every out-of-sample number is for 2024, the rolling-origin AUCs of \robRollAUCminThree{} to \robRollAUCmaxThree{} show how much a year can move it, and the 2023 lines choose the threshold, so 2023 statistics are in-sample selections. Sixth, no execution model: costs are fixed charges and spreads are \citet{corwin2012} estimates from daily highs and lows, with no price impact or fill assumptions. Seventh, the scrape begins in January 2018, so history features are left-censored in the first sample year and the routine flag exists only from 2021.

\section{Conclusion}\label{sec:conclusion}

Reported insider purchases in microcap stocks far below their 52-week high are followed by a larger disclosure reaction: \rxQoneArToneMean{} on the first trading day after filing, against \rxQfiveArToneMean{} near the high. Most of the raw gap reflects volatility and does not meet the pre-specified threshold after controls. The volatility-scaled gradient is positive on the primary estimator, but its magnitude depends on the estimator, a variance-weighted slope shows none, and the controlled gap falls below that threshold among purchases free of concurrent 8-K news. Comparisons with sales and earnings announcements narrow the interpretation without establishing a causal channel.

The response extends over the first two to three sessions; subsequent drift is imprecise, and calendar-time portfolios earn no significant alpha after skipping the first day. For after-close filings, much of the initial response occurs overnight. The results locate the clearest information in the early disclosure response, while leaving the size and tradability of later returns unresolved.

Price adjustment and benchmark construction materially affect the measured patterns. Consistent share bases prevent run-up bucket misclassification, and no-purchase placebos help assess whether an estimation-window intercept should persist into expected returns. The gradient-boosting classifier remains a useful summary of prediction at disclosure (test AUC \clXgbAUCthree), with no demonstrated advantage over random forest or logistic regression. Neither its microcap implementation nor the separate liquid-universe extension establishes a profitable incremental trading rule. The contribution is a decomposition of disclosure reactions and their measurement in the observed microcap sample.

\bibliography{references}

\appendix

\section{Variable definitions}\label{app:vars}

Table~\ref{tab:features} reports summary statistics of the features on the training lines and Table~\ref{tab:vardef} defines every feature and outcome. Price-based quantities are measured at the close of $t_0$, the last trading day on or before the filing date. ``Adjusted close'' is the panel's split- and dividend-adjusted close; ``actual close'' is the panel close on the Form~4 share basis; daily highs and lows are split-adjusted only. History and cluster features use the full scrape of purchase lines and only lines filed strictly before the current filing; the insider key is the casefolded name.

\begin{table}[tbp]
\centering\tabtiny
\caption{Summary statistics of the \nFeaturesFull{} features on the training lines (2018--2022): non-missing observations, mean, standard deviation, 1st, 10th, 50th, 90th and 99th percentiles, and the share of missing values. Dollar amounts are abbreviated.}
\label{tab:features}
\begin{tabular}{l r r r r r r r r r}
\toprule
Feature & $N$ & Mean & SD & p1 & p10 & p50 & p90 & p99 & Missing \\
\midrule
\tabinput{outputs/table_sample_stats_features_train.tex}
\end{tabular}
\end{table}

\begin{table}[tbp]
\centering\tabtiny
\caption{Variable definitions. Base = included in both feature sets; Added = included only in the full specification.}
\label{tab:vardef}
\begin{tabular}{l l p{0.68\textwidth}}
\toprule
Variable & Set & Definition \\
\midrule
\texttt{insider\_title\_score} & Base & Substring rule on the reported title: CEO or chief executive = 5, CFO or chief financial = 4, COO or chief operating = 3, ``Dir'' or director = 2, otherwise 1. \\
\texttt{transaction\_value} & Base & Reported transaction value in dollars. \\
\texttt{is\_first\_purchase\_12m} & Base & 1 if the insider has no earlier purchase line in the issuer with a transaction date within the prior 365 days. \\
\texttt{value\_vs\_avg\_ratio} & Base & Transaction value divided by the mean value of the insider's purchase lines in the issuer over the prior 365 days; missing when there is none. \\
\texttt{price\_deviation} & Base & Disclosure run-up: actual close at $t_0$ divided by the reported transaction price, minus one. \\
\texttt{pct\_from\_52w\_high} & Base & Actual close at $t_0$ divided by the maximum daily high over the 252 trading days ending at $t_0$, minus one (non-positive). \\
\texttt{pct\_from\_52w\_low} & Base & Actual close at $t_0$ divided by the minimum daily low over the same window, minus one (non-negative). \\
\texttt{volatility\_30d} & Base & Standard deviation of daily log returns of the adjusted close over the 30 trading days ending at $t_0$, annualized with $\sqrt{252}$. \\
\texttt{return\_mtd} & Base & Close at $t_0$ divided by the last close of the previous calendar month, minus one. \\
\texttt{market\_cap\_pit} & Base & Actual close at $t_0$ times point-in-time shares outstanding (Section~\ref{sec:sources}). \\
\texttt{addv\_30d} & Base & Mean of daily volume times close over the 30 trading days ending at $t_0$ (at least 20 observations). \\
\texttt{is\_biotech} & Base & 1 if the Yahoo sector or industry string contains ``biotechnology'', ``pharmaceutical'', ``drug'' or ``biotech''. \\
\texttt{is\_ceo}, \texttt{is\_cfo} & Added & Token rules on the title: the token CEO or a ``chief executive'' pattern; CFO or a ``chief financial'' pattern. \\
\texttt{is\_other\_officer} & Added & Any other officer token (president, EVP, SVP, VP, COO, chief, officer, treasurer, secretary, controller, general counsel, managing director and similar) after removing the CEO and CFO patterns. \\
\texttt{is\_director} & Added & Director, chair or board tokens; ``Dir of'' a function is not a board seat, ``Director of'' a subsidiary, bank or parent is. \\
\texttt{is\_10pct\_owner} & Added & 1 if the title contains ``10\%''. \\
\texttt{n\_prior\_purchases\_365d} & Added & Number of the insider's earlier purchase lines in the issuer with a transaction date within the prior 365 days. \\
\texttt{days\_since\_last\_purchase} & Added & Days from the insider's most recent earlier purchase in the issuer to the current transaction date; missing when there is none. \\
\texttt{n\_other\_insiders\_30d} & Added & Distinct other insiders with a purchase line in the issuer filed within the 30 days before the filing. \\
\texttt{n\_purchases\_30d} & Added & Number of those other insiders' purchase lines. \\
\texttt{cmp\_routine} & Added & 1 if the insider bought the issuer in the same calendar month in each of the three prior years \citep{cohen2012}; defined for transaction years from 2021 on. \\
\texttt{filing\_lag\_days} & Added & Filing date minus transaction date, calendar days. \\
\texttt{pct\_of\_holdings} & Added & Shares purchased divided by post-transaction holdings. \\
\texttt{return\_5d}, \texttt{return\_21d} & Added & Adjusted close at $t_0$ divided by the adjusted close 5 and 21 trading days earlier, minus one. \\
\texttt{pre\_close} & Added & 1 if the filing was accepted before 16:00~ET. \\
\midrule
$\CAR$ & Outcome & Sum of daily abnormal returns over the 30 trading days after $t_0$ (equation~\ref{eq:car}): excess return minus the three loadings times the factors, loadings from OLS with a constant on the 252 trading days before $t_0$ with the stock's returns winsorized at their 1st and 99th percentiles; the intercept is not part of the benchmark. \\
$\CAR^{\alpha}$ & Outcome & As $\CAR$ with the estimation-window intercept $\hat\alpha$ subtracted from every daily abnormal return; $\CAR - \CAR^{\alpha} = 30\hat\alpha$ exactly. \\
$\mathrm{AR}_{t_0}$, $\mathrm{AR}_{t_0+1}$, drift & Outcome & The daily abnormal return of equation~\ref{eq:car} on the filing day (post-disclosure only for filings accepted before the close), on the first event day, and summed over event days 2--30 or 2--60; $\mathrm{AR}_{t_0+1}$ plus the 2--30 drift equals $\CAR$. Overnight gap: open of $t_0+1$ over close of $t_0$, minus one; intraday: close over open of $t_0+1$, minus one (raw returns). \\
$\mathrm{BHAR}_{[1,30]}$ & Outcome & Compounded stock return minus the compounded benchmark return (risk-free rate plus the loadings times the factors) over the same 30 days. \\
Market-adjusted return & Outcome & Sum over the 30 days of the stock's excess return minus the market excess return. \\
$\CAR$, timing & Outcome & As $\CAR$, with the event day moved to the next trading day for filings accepted at or after 16:00~ET. \\
$z$, \texttt{target\_z} & Outcome & $\CAR$ divided by \texttt{volatility\_30d} times $\sqrt{30/252}$; \texttt{target\_z} = $\mathbf{1}\{z > 1\}$. \\
\bottomrule
\end{tabular}
\end{table}

\section{Model settings and protocol}\label{app:model}

The primary model is \texttt{xgboost} \xgbVersion{} with \texttt{max\_depth} = \xgbMaxDepth, \texttt{learning\_rate} = \xgbLearningRate, the binary logistic objective, the AUC evaluation metric, seed \seedModel, a single thread and every other parameter at its library default; missing values are handled natively. Rounds are chosen by \texttt{scikit-learn} \sklearnVersion{} \texttt{TimeSeriesSplit} with \xgbCvFolds{} folds over the training lines in filing-date order, each fold fitted for \xgbMaxRoundsCV{} rounds with the validation AUC recorded after every round; the primary model uses \xgbRoundsCV{} rounds (mean fold AUC \xgbCvMeanAUC) and the base model \xgbRoundsCVBase. Under validation-based early stopping (patience \xgbPatience{} on the 2023 AUC) the full model uses \xgbRoundsES{} rounds, attains a 2024 AUC of \robEsFullTestAUCthree{} and an F1 of \robEsFullTestFone{} at its threshold of \robEsFullThr.

The random forest uses \rfTrees{} trees of maximum depth \rfMaxDepth{} and the logistic regression a standard scaler with at most \lrMaxIter{} iterations; both impute missing values with the training-split median. Statistics use \texttt{statsmodels} \statsmodelsVersion, \texttt{numpy} \numpyVersion{} and \texttt{pandas} \pandasVersion{} under Python \pythonVersion; bootstraps use seed \seedBootstrap{} and \clBootB{} draws.

A hyperparameter grid of \robGridNConfigs{} fixed-round configurations of depth, learning rate and rounds, run with the same five time-ordered folds and no early stopping, describes the surface and did not inform the a priori choice: that configuration ranks \robGridAprioriRank{} of \robGridAprioriRankOf{} on mean fold AUC, the best configuration (depth \robGridBestDepth, learning rate \robGridBestLR, \robGridBestRounds{} rounds) attains a test AUC of \robGridBestTestAUCthree{} (independence-based DeLong $p$ against the primary \robGridBestVsPrimaryDeLongP), and test AUCs across the grid lie between \robGridTestAUCmin{} and \robGridTestAUCmax.

\section{Additional robustness}\label{app:robust}

Table~\ref{tab:tails_model} gives the tail, benchmark and timing variants of Section~\ref{sec:robustness}, Table~\ref{tab:carry} the benchmark carry by quintile of Section~\ref{sec:mechanism}, Table~\ref{tab:floor} the price floors and Table~\ref{tab:benchsim} the benchmark simulation of Section~\ref{sec:car}.

\begin{table}[tbp]
\centering\tabfoot
\caption{Tail, benchmark and timing variants for the primary model's 2024 evaluation: test lines, positive rate under the variant label, AUC of the primary probabilities against that label with its issuer-cluster bootstrap interval, mean outcome of flagged and rejected lines, their difference in percentage points and its issuer-clustered $t$. Variants: baseline $\CAR$; excluding lines with $|\CAR| > 1$; $\CAR$ winsorized at the full-sample 1st and 99th percentiles; the buy-and-hold abnormal return; the timing-adjusted $\CAR$; and the intercept-inclusive benchmark with the estimation-window intercept carried (Section~\ref{sec:car}).}
\label{tab:tails_model}
\shrinkbox{
\begin{tabular}{l r r r r r r r r}
\toprule
Variant & $N$ 2024 & Pos.\ rate & AUC & 95\% CI & Flagged mean & Rejected mean & Diff (pp) & Clustered $t$ \\
\midrule
\tabinput{outputs/table_robustness_tails_model.tex}
\end{tabular}}
\end{table}

\begin{table}[tbp]
\centering\tabfoot
\caption{The benchmark carry by quintile of distance from the 52-week high (Q1 furthest below the high). Columns: lines, mean distance, the mean estimation-window intercept $\hat\alpha$ in basis points per day, the implied 30-day carry $-30\hat\alpha$ in percentage points, mean $\CAR$ and $\CAR^{\alpha}$, median $\CAR^{\alpha}$, and the indicated tail probabilities. The superscript $\alpha$ denotes the intercept-inclusive benchmark, so the carry equals $\CAR^{\alpha}-\CAR$. The Q1 $-$ Q5 gradient is \mechQoneQfiveMeanCARDiffPp{} points on the primary benchmark and \mechQoneQfiveMeanCARVOneDiffPp{} with the intercept included; the difference, \mechQoneQfiveAlphaCarryDiffPp{} points (clustered $t$ = \mechQoneQfiveAlphaCarryTCl), is the carry.}
\label{tab:carry}
\shrinkbox{
\begin{tabular}{l r r r r r r r r r r}
\toprule
Quintile & $N$ & Mean dist. & $\hat\alpha$ (bp/day) & Carry (pp) & Mean CAR & Mean CAR$^{\alpha}$ & Median$^{\alpha}$ & $\Pr(\mathrm{CAR} > 10\%)$ & $\Pr(\mathrm{CAR}^{\alpha} > 10\%)$ & $\Pr(\mathrm{CAR}^{\alpha} < -10\%)$ \\
\midrule
\tabinput{outputs/table_mechanism_quintiles_v1.tex}
\end{tabular}}
\end{table}

\begin{table}[tbp]
\centering\tabfoot
\caption{Price-floor robustness. For each floor on the filing-day price: lines and share retained, 2024 lines and positive rate, the primary model's AUC on the retained 2024 lines and the AUC of a refit on the floored sample, and the extreme run-up contrast on the floored sample (lines and mean $\CAR$ in the $\le 0\%$ and $> 10\%$ buckets, difference in percentage points, issuer-clustered and Welch $t$, share above 10\% in each bucket and the clustered $t$ of that difference).}
\label{tab:floor}
\shrinkbox{
\begin{tabular}{l r r r r r r r r r r r r r r r r}
\toprule
Floor & $N$ & Share & $N$ 2024 & Pos.\ 2024 & AUC & AUC refit & $N$ low & $N$ high & Mean low & Mean high & Diff (pp) & Cl.\ $t$ & Welch $t$ & Pr low & Pr high & Pr cl.\ $t$ \\
\midrule
\tabinput{outputs/table_mechanism_floor.tex}
\end{tabular}}
\end{table}

\begin{table}[tbp]
\centering\tabfoot
\caption{Known-truth benchmark simulation. Each scenario simulates \smPaths{} stock paths from a one-factor model (market beta 1.2, daily residual volatility 3\%) with a 252-day estimation window and a 20-day event window, and estimates the cumulative abnormal return over the event window with the fitted intercept carried or with the loadings only. The target is the event innovation, the realized event-window return minus the scenario's true conditional normal return. Scenarios: no alpha; a transient drawdown (a mean of $-10$ basis points a day over the last 60 estimation days that does not continue); a high-volatility null (residual volatility 8\% over the last 60 estimation days and the event window) keeping the \smPaths{} paths with the lowest estimation-window residual sum out of \smSelectedDraws; a persistent alpha of $\pm 10$ basis points a day in both windows; and the positive persistent alpha plus a true event increment of 20 basis points a day. Bias is the mean error against the target and RMSE its root mean squared error, both in basis points over the 20 days. The simulation is stylized: it shows which benchmark is biased under which process, not the process that generated this sample, which the no-purchase placebo of Section~\ref{sec:mechanism} speaks to.}
\label{tab:benchsim}
\begin{tabular}{l r r r r}
\toprule
 & \multicolumn{2}{c}{Intercept carried} & \multicolumn{2}{c}{Loadings only} \\
Scenario & Bias & RMSE & Bias & RMSE \\
\midrule
No alpha & \smNullIntBias & \smNullIntRmse & \smNullLoadBias & \smNullLoadRmse \\
Transient drawdown & \smTransientIntBias & \smTransientIntRmse & \smTransientLoadBias & \smTransientLoadRmse \\
High-volatility null, selected on prior drawdown & \smSelectedIntBias & \smSelectedIntRmse & \smSelectedLoadBias & \smSelectedLoadRmse \\
Persistent positive alpha & \smPersPosIntBias & \smPersPosIntRmse & \smPersPosLoadBias & \smPersPosLoadRmse \\
Persistent negative alpha & \smPersNegIntBias & \smPersNegIntRmse & \smPersNegLoadBias & \smPersNegLoadRmse \\
Persistent positive alpha plus event increment & \smPersEventIntBias & \smPersEventIntRmse & \smPersEventLoadBias & \smPersEventLoadRmse \\
\bottomrule
\end{tabular}
\end{table}

\section{Data and code availability}\label{app:data}

\paragraph{Form~4 lines.} The OpenInsider screener \citep{openinsider} was queried once per calendar month of filing date from January 2018 to December 2024 with the purchase-only filter (\texttt{xp=1}, \texttt{xs=0}), all other filters blank and the result count at its maximum, in January 2026.

\paragraph{Prices, corporate actions, shares and factors.} Daily open, high, low, close, adjusted close, volume, dividends and split events for every scraped ticker were pulled through \texttt{yfinance} \citep{yfinance} in September 2026 with \texttt{auto\_adjust=False} and \texttt{actions=True} for January 2017 to February 2025 and cached; the pipeline runs offline from the cache. Cover-page share counts (\texttt{dei:Entity\-Common\-Stock\-Shares\-Outstanding}, with \texttt{us-gaap:Common\-Stock\-Shares\-Outstanding} as a fallback) come from the SEC XBRL company-facts API \citep{sec_xbrl_companyfacts}, pulled in September 2026 for every ticker with a CIK. The January 2026 snapshot of shares outstanding, sector and industry comes from the Yahoo ticker profile. Daily factors are from Kenneth French's data library \citep{french_library}.

\paragraph{Source audit and simulation.} The source audit of Section~\ref{sec:sources} retrieved the original SEC XML of the \saN{} audited filings and the SEC filings of the preceding 30 days in September 2026, with the sample drawn and its labels frozen before any outcome was read; the benchmark simulation of Table~\ref{tab:benchsim} is a fixed-seed script. Both, with their manifests and hashes, are in the research record.

\paragraph{Identification-section data.} Insider sales (Section~\ref{sec:identification}) come from the same OpenInsider screener queried per ticker with the sale-only filter for filing dates 2018--2024, keeping code-S sales (open market or private) without option exercise. 8-K filing histories with acceptance timestamps and item codes come from the SEC submissions API (\texttt{data.sec.gov/submissions}, recent and older pages), and quarterly EPS (\texttt{us-gaap:EarningsPerShareDiluted}, basic where diluted is absent; \dcEpsDilutedShare{} of matched events use the diluted tag) from the XBRL company-facts API \citep{sec_xbrl_companyfacts}, both pulled in September 2026 with a declared user agent for the \dnCikTickers{} sample issuers with a CIK. The EPS value used for a quarter is the earliest-filed XBRL value for that period, which is the 10-Q or 10-K filed after the 8-K for \dcEpsAfterShare{} of events and on the same day for \dcEpsSameDayShare; the surprise it produces is ordinarily the figure announced in the 8-K itself. The liquid-universe extension (Section~\ref{sec:extension}) uses the SEC quarterly Form~4 archives for March 2025 to June 2026 and daily prices from the same source as the main sample; its frozen rules, hashes and outcome files are in the research record.

\paragraph{Code.} A deterministic offline builder writes the analysis sample and the funnel; a model module fits every model once; one analysis module per section emits its macros and table bodies; a figure module asserts every plotted quantity against the fit; and an orchestrator writes the macro file read by this document, so that every number in the text, tables and captions is generated. Code and processed sample are available from the author.

\section{Changes from the first version}\label{app:changes}

This appendix summarizes how this version differs from the February 2026 version (v1) in its description of the data, its methods and its results. Section references are to this version.

\begin{itemize}\setlength{\itemsep}{0pt}\setlength{\parskip}{0pt}
    \item \textbf{Data source.} The transactions are SEC Form~4 purchase filings as compiled by the OpenInsider aggregator, and prices come from Yahoo Finance through \texttt{yfinance}; this replaces v1's description of filings parsed directly from EDGAR and prices from an institutional vendor. This version also reconciles a sample of the filings against the original EDGAR documents and takes share counts from SEC XBRL data (Section~\ref{sec:sources}).
    \item \textbf{Identifiers.} Securities are identified by ticker string, with symbol reuse screened by the price-basis checks (Sections~\ref{sec:sources} and \ref{sec:sample}); this replaces v1's description of a mapping to permanent identifiers.
    \item \textbf{Filters.} The 90-day reporting-lag cap and the \$5,000 minimum value described in v1 are applied, together with positivity, value-consistency and de-duplication filters (Table~\ref{tab:funnel}).
    \item \textbf{Capitalization.} Capitalization uses the actual filing-day close and, for \ssSharesSecPitShare{} of lines, SEC cover-page counts filed before the filing date, with a split-adjusted 2026 snapshot for the rest (Section~\ref{sec:sources}); v1 used an adjusted close and a January 2026 snapshot.
    \item \textbf{Volume window.} The dollar-volume filter uses 30 trading days, as the definition states, in place of the 90-calendar-day window of v1's code (Table~\ref{tab:funnel}).
    \item \textbf{Event windows.} Every event has exactly 30 event days; v1's price sample ended on 31 December 2024, which truncated the windows of late-2024 filings (Section~\ref{sec:car}).
    \item \textbf{Survivorship and attrition.} The exclusion of issuers without a current price history and the estimation-window requirement, not reported in v1, are quantified (Section~\ref{sec:survivorship}, Table~\ref{tab:funnel}).
    \item \textbf{Information set.} Features are measured at the close of the filing day; for the \ssShareAfterClose{} of lines accepted at or after that close the first event-day return contains an overnight reaction an outsider cannot trade, and it is reported separately (Sections~\ref{sec:car} and \ref{sec:reaction}), replacing v1's description of that return as tradable.
    \item \textbf{Abnormal-return formula and screens.} Equation~\ref{eq:car} states the formula as computed, with the risk-free rate on both sides and the estimation window specified; lines with a 30-day abnormal return below $-100\%$, which v1 retained, are excluded by an explicit outlier rule whose sensitivity is reported (Sections~\ref{sec:sample} and \ref{sec:car}).
    \item \textbf{Abnormal-return benchmark.} The benchmark uses the factor loadings only. The intercept-carrying benchmark of v1 adds a mechanical carry $-30\hat\alpha$ of \mechQoneAlphaCarryPp{} points in the quintile farthest below the high and \mechQfiveAlphaCarryPp{} nearest it, \mechQoneQfiveCarryShareOfVOneGap{} of the \mechQoneQfiveMeanCARVOneDiffPp-point quintile gradient under that benchmark, and a placebo of firm-days without a purchase reproduces \plGapVOneOverEventVOne{} times that gradient (Table~\ref{tab:placebo}). It is reported as robustness throughout (Sections~\ref{sec:car} and \ref{sec:mechanism}, Appendix~\ref{app:robust}), and the reading of the 52-week-high variable follows Sections~\ref{sec:importance} and \ref{sec:reaction}.
    \item \textbf{Threshold.} The 10\% threshold is justified as an economic level; it lies at the \ssCarPctileTen{} percentile of the distribution rather than near the top decile, as v1 described (Section~\ref{sec:car}).
    \item \textbf{Model protocol.} Hyperparameters are fixed a priori and stated, and the number of boosting rounds is chosen by cross-validation inside 2018--2022; v1's rule, early stopping on the 2023 set that also chooses the threshold, is reported as robustness (Section~\ref{sec:design}, Appendix~\ref{app:model}). This replaces v1's description of cross-validated hyperparameter tuning.
    \item \textbf{Threshold and figures.} Every number and figure comes from the single fitted model and the threshold is printed as applied (\clXgbOptThr), replacing v1's rounded threshold and separately fitted ROC figure (Appendix~\ref{app:data}).
    \item \textbf{Classifier comparison.} With matched thresholds and issuer-clustered inference the three learners are statistically indistinguishable (Section~\ref{sec:classification}); this replaces v1's statement that non-linear classifiers outperform linear baselines.
    \item \textbf{Calibration.} The distribution of predicted probabilities (median \clPredMedian, \clCalLoShare{} below 0.4) and the Brier score are reported (Section~\ref{sec:operating}), replacing v1's description of calibration.
    \item \textbf{Feature importance.} Importance is reported on five bases with ranks and with the sign of each effect (Table~\ref{tab:importance}, Section~\ref{sec:importance}); v1 reported a single normalized gain share on the intercept-carrying label.
    \item \textbf{Prior work.} v1's priority statement is withdrawn: \citet{kong2025} train a gradient-boosted tree on Form~4 trades and report the effect strongest in the smallest stocks. Section~\ref{sec:intro} positions the paper relative to the machine-learning and disclosure-reaction literature.
    \item \textbf{Run-up premium (withdrawn).} v1 reported that purchases disclosed after a price run-up above 10\% earn the highest mean abnormal return. The premium does not survive the rebuilt sample under either price basis: v1's variable recomputed on this sample gives a mean gap of \ruArtAdjDiff{} points (clustered $t$ = \ruArtAdjTCl) and the actual-basis variable \ruDiffPP{} points (clustered $t$ = \ruTCl); neither pattern is monotone, the gap is absorbed by distance from the 52-week high, and it does not replicate in 2024 (Sections~\ref{sec:runup} and \ref{sec:doublesort}). The claim, its robustness statements and the slow-incorporation interpretation built on it are withdrawn.
    \item \textbf{Costs.} Round-trip costs of 1\% to 5\% are subtracted with bootstrap intervals, with measured spreads, break-even costs and capacity (Section~\ref{sec:economics}); this replaces v1's cost discussion, which concerned the withdrawn run-up variable.
    \item \textbf{Robustness.} 20- and 60-day windows, buy-and-hold returns and realized-volatility regimes are reported (Sections~\ref{sec:runup} and \ref{sec:robustness}), in place of v1's statements on alternative windows and implied-volatility regimes.
    \item \textbf{Insider identity features.} History and cluster features are computed point in time from the full scrape rather than the filtered sample, and 10\% owners are a separate role (Sections~\ref{sec:variables} and \ref{sec:roles}).
    \item \textbf{Front matter.} The date line gives the posting dates of both versions.
\end{itemize}

\end{document}

%% file: outputs/metrics.tex
\newcommand{\nTrain}{9,044}
\newcommand{\nVal}{2,457}
\newcommand{\nTest}{2,033}
\newcommand{\nTrainPos}{2,467}

\newcommand{\nTestPos}{532}

\newcommand{\valPosRate}{27.3\%}
\newcommand{\testPosRate}{26.2\%}

\newcommand{\nFeaturesFull}{27}
\newcommand{\nFeaturesBase}{12}
\newcommand{\nFeaturesAdded}{15}
\newcommand{\xgbMaxDepth}{5}
\newcommand{\xgbLearningRate}{0.1}
\newcommand{\xgbCvFolds}{5}
\newcommand{\xgbMaxRoundsCV}{300}
\newcommand{\xgbRoundsCV}{26}
\newcommand{\xgbRoundsCVBase}{12}
\newcommand{\xgbCvMeanAUC}{0.669}

\newcommand{\xgbMaxRounds}{100}
\newcommand{\xgbPatience}{10}
\newcommand{\xgbRoundsES}{10}

\newcommand{\rfTrees}{100}
\newcommand{\rfMaxDepth}{10}
\newcommand{\lrMaxIter}{1,000}
\newcommand{\seedModel}{42}
\newcommand{\seedBootstrap}{0}
\newcommand{\pythonVersion}{3.11.13}
\newcommand{\xgbVersion}{3.0.5}
\newcommand{\sklearnVersion}{1.7.2}

\newcommand{\statsmodelsVersion}{0.14.5}
\newcommand{\numpyVersion}{2.3.3}
\newcommand{\pandasVersion}{2.3.2}

\newcommand{\ssN}{13,534}
\newcommand{\ssNTickers}{1,192}

\newcommand{\ssNIssuerDays}{10,858}

\newcommand{\ssShareSharedDay}{32.1\%}

\newcommand{\ssPosRate}{27.1\%}
\newcommand{\ssYearsRange}{2018--2024}

\newcommand{\ssMedianTxVal}{\$45,400}

\newcommand{\ssMedianMcap}{\$209.6M}
\newcommand{\ssMedianAddv}{\$910k}
\newcommand{\ssMedianLag}{2}

\newcommand{\ssShareRatioMissing}{39.2\%}
\newcommand{\ssShareRoutineMissing}{35.7\%}

\newcommand{\ssLostAbsentN}{25,496}
\newcommand{\ssLostAbsentTickers}{2,376}
\newcommand{\ssLostEarlyN}{826}
\newcommand{\ssLostN}{26,322}
\newcommand{\ssLostShare}{34.7\%}
\newcommand{\ssStepSevenN}{75,843}

\newcommand{\ssRawPurchaseN}{101,093}
\newcommand{\ssRawPurchaseTickers}{6,390}
\newcommand{\ssRetention}{13.4\%}

\newcommand{\ssLostShareFirstYear}{47.7\%}
\newcommand{\ssLostShareLastYear}{14.4\%}
\newcommand{\ssLostShareTrainYears}{39.8\%}
\newcommand{\ssLostShareTestYear}{14.4\%}

\newcommand{\ssPosRateYearMin}{21.5\%}
\newcommand{\ssPosRateYearMax}{35.6\%}
\newcommand{\ssPosRateYearMinYear}{2018}
\newcommand{\ssPosRateYearMaxYear}{2020}

\newcommand{\ssSharePreOpen}{9.8\%}

\newcommand{\ssShareIntraday}{30.2\%}

\newcommand{\ssShareAfterClose}{60.0\%}

\newcommand{\ssCarPreClose}{2.0\%}

\newcommand{\ssAucPreCloseThree}{0.671}

\newcommand{\ssCarPostClose}{4.7\%}

\newcommand{\ssAucPostCloseThree}{0.675}

\newcommand{\ssCarPostMinusPreTCl}{4.25}

\newcommand{\ssAucPreMinusPostP}{0.878}

\newcommand{\ssCarMean}{3.6\%}
\newcommand{\ssCarMedian}{1.3\%}
\newcommand{\ssCarStd}{21.5\%}
\newcommand{\ssCarMin}{$-$94\%}
\newcommand{\ssCarMax}{474\%}

\newcommand{\ssCarPninety}{26.4\%}

\newcommand{\ssCarShareAbsGtOne}{0.51\%}

\newcommand{\ssCarPctileTen}{73rd}

\newcommand{\ssCarTestMean}{3.3\%}

\newcommand{\ssCarTestPosRate}{26.2\%}

\newcommand{\ssCarThresholdOverMedian}{7.6}
\newcommand{\ssCarThresholdInSd}{0.46}
\newcommand{\ssCarNAbsGtOne}{69}

\newcommand{\ssPosRateVOne}{29.7\%}

\newcommand{\ssTargetAgreeVOne}{92.2\%}

\newcommand{\ssAlphaCarryMeanPp}{0.92}

\newcommand{\ssAlphaDailyMedianBp}{$-$3.2}

\newcommand{\ssAlphaNegShare}{60.3\%}

\newcommand{\ssFunnelLagDropped}{1,727}

\newcommand{\ssFunnelPositiveDropped}{1,859}

\newcommand{\ssFunnelConsistentDropped}{3,459}

\newcommand{\ssFunnelMinValueDropped}{17,046}

\newcommand{\ssFunnelDedupDropped}{614}

\newcommand{\ssFunnelEstWindowDropped}{4,235}

\newcommand{\ssFunnelSharesDropped}{81}

\newcommand{\ssFunnelAddvDropped}{6,134}

\newcommand{\ssFunnelDevBandDropped}{203}

\newcommand{\ssFunnelPriceBasisDropped}{192}

\newcommand{\ssFunnelCarFloorDropped}{9}

\newcommand{\ssFunnelJumpFilterDropped}{4}

\newcommand{\ssPanelTickers}{3,710}

\newcommand{\ssSharesSecPitShare}{91.1\%}

\newcommand{\ssSharesSnapshotShare}{6.1\%}

\newcommand{\ssSharesSnapshotStaleShare}{2.9\%}

\newcommand{\clBootB}{1,000}
\newcommand{\clBootClusters}{479}
\newcommand{\clFlagAllF}{0.415}
\newcommand{\clFlagAllP}{26.2\%}

\newcommand{\clXgbAUCthree}{0.676}

\newcommand{\clXgbAUCci}{0.628--0.714}

\newcommand{\clXgbPRAUCthree}{0.419}

\newcommand{\clXgbOptThr}{0.184}
\newcommand{\clXgbOptValF}{0.503}

\newcommand{\clXgbDefP}{0.510}
\newcommand{\clXgbDefR}{0.094}
\newcommand{\clXgbDefF}{0.159}
\newcommand{\clXgbDefN}{98}

\newcommand{\clXgbOptP}{0.338}
\newcommand{\clXgbOptR}{0.765}
\newcommand{\clXgbOptF}{0.469}
\newcommand{\clXgbOptN}{1,205}
\newcommand{\clXgbOptNPct}{59.3\%}

\newcommand{\clRfAUCthree}{0.674}

\newcommand{\clRfAUCci}{0.621--0.718}

\newcommand{\clLrAUCthree}{0.653}

\newcommand{\clLrAUCci}{0.600--0.700}

\newcommand{\clDiffXgbRf}{0.002}

\newcommand{\clDiffXgbRfCi}{[$-$0.012, 0.017]}

\newcommand{\clBootXgbRfP}{0.795}

\newcommand{\clDeLongXgbRfP}{0.671}
\newcommand{\clDiffXgbLr}{0.023}

\newcommand{\clDiffXgbLrCi}{[$-$0.003, 0.051]}

\newcommand{\clBootXgbLrP}{0.113}

\newcommand{\clDeLongXgbLrP}{0.015}
\newcommand{\clDiffRfLr}{0.021}

\newcommand{\clDiffRfLrCi}{[$-$0.004, 0.047]}

\newcommand{\clBootRfLrP}{0.103}

\newcommand{\clDeLongRfLrP}{0.015}

\newcommand{\clDiffXgbFullBase}{$-$0.006}

\newcommand{\clDiffXgbFullBaseCi}{[$-$0.021, 0.008]}

\newcommand{\clBootXgbFullBaseP}{0.429}

\newcommand{\clDiffCvEs}{0.001}

\newcommand{\clDiffCvEsCi}{[$-$0.012, 0.012]}

\newcommand{\clBootCvEsP}{0.919}

\newcommand{\clDeLongCvEsP}{0.859}

\newcommand{\clBootOverDeLongSeXgbRf}{1.64}
\newcommand{\clMatchedK}{1,205}

\newcommand{\clMatchedFRangeFull}{0.010}
\newcommand{\clSwNTen}{1,817}
\newcommand{\clSwPTen}{0.275}

\newcommand{\clSwNThirtyFive}{438}
\newcommand{\clSwPThirtyFive}{0.443}

\newcommand{\clSwThrLo}{0.10}
\newcommand{\clSwThrHi}{0.35}
\newcommand{\clSwFMin}{0.400}

\newcommand{\clSwFMax}{0.477}

\newcommand{\clCalLoShare}{87.3\%}

\newcommand{\clBrier}{0.180}

\newcommand{\clBrierBase}{0.193}

\newcommand{\clBrierSkill}{6.8\%}

\newcommand{\clPredMedian}{0.226}

\newcommand{\clPredMax}{0.917}

\newcommand{\clSeedZeroRounds}{26}

\newcommand{\clSeedZeroXgbAUC}{0.676}

\newcommand{\clRfSeedMin}{0.670}
\newcommand{\clRfSeedMax}{0.677}
\newcommand{\clRfSeedRange}{0.007}

\newcommand{\clUniHighAUCthree}{0.596}
\newcommand{\clUniHighRank}{3}

\newcommand{\clUniTopFeat}{\texttt{volatility\_30d}}

\newcommand{\clUniTopAUCthree}{0.669}
\newcommand{\clUniNAboveSixty}{1}

\newcommand{\clDeLongXgbUniHighP}{$<$0.001}

\newcommand{\clAblNoPdevAUCthree}{0.685}

\newcommand{\clAblNoPdevDeLongP}{0.002}

\newcommand{\clAblNoHighAUCthree}{0.678}

\newcommand{\clAblNoHighDeLongP}{0.753}

\newcommand{\impFullPctFromFivetwowHighRankWeight}{1}

\newcommand{\impSignPctFromFivetwowHighPearson}{$-$0.82}

\newcommand{\impSignPctFromFivetwowHighShapLow}{0.114}
\newcommand{\impSignPctFromFivetwowHighShapHigh}{$-$0.225}

\newcommand{\impNTreesUsed}{26}

\newcommand{\impTopGainPct}{18.7\%}

\newcommand{\impTopTotalGainPct}{28.5\%}

\newcommand{\impTopRankWeight}{5th}

\newcommand{\impTopShapPct}{30.3\%}

\newcommand{\impSecondShapPct}{17.6\%}
\newcommand{\impTopSecondRatioGain}{2.2}

\newcommand{\impTopSecondRatioShap}{1.7}

\newcommand{\impTopCorrPearson}{0.67}

\newcommand{\impTopMedianValue}{58.6\%}
\newcommand{\impTopShapBelowMedian}{$-$0.45}
\newcommand{\impTopShapAboveMedian}{+0.27}

\newcommand{\impTopCorrPearsonCI}{[0.62, 0.72]}

\newcommand{\impFtwhDecBottomShap}{+0.27}
\newcommand{\impFtwhDecTopShap}{$-$0.24}

\newcommand{\impFtwhDecSpearman}{$-$0.94}

\newcommand{\impFtwhDecBottomValue}{$-$81.1\%}
\newcommand{\impFtwhDecTopValue}{$-$3.9\%}

\newcommand{\impMtdDecBottomShap}{+0.06}
\newcommand{\impMtdDecTopShap}{+0.13}

\newcommand{\impMtdDecMiddleShap}{$-$0.034}

\newcommand{\impMtdQuadT}{1.41}

\newcommand{\impMtdDecBottomShapCI}{[0.01, 0.11]}

\newcommand{\impMtdDecTopShapCI}{[0.12, 0.14]}

\newcommand{\ruSerialMinLines}{50}

\newcommand{\ruBZeroThreeMean}{2.5\%}

\newcommand{\ruBZeroThreePrPos}{22.4\%}

\newcommand{\ruBTenTwentyfiveMean}{7.2\%}

\newcommand{\ruBFiftyHundredN}{100}

\newcommand{\ruBFiftyHundredMean}{$-$5.3\%}
\newcommand{\ruBFiftyHundredCI}{[$-$14.9\%, 4.4\%]}

\newcommand{\ruLowMeanCAR}{4.0\%}
\newcommand{\ruLowPrPos}{27.2\%}

\newcommand{\ruLowMedVol}{0.58}
\newcommand{\ruLowMedFtwh}{$-$41.8\%}

\newcommand{\ruHighMeanCAR}{6.1\%}
\newcommand{\ruHighPrPos}{38.8\%}

\newcommand{\ruHighMedVol}{0.94}
\newcommand{\ruHighMedFtwh}{$-$50.2\%}

\newcommand{\ruAllPrDiff}{11.6}
\newcommand{\ruAllPrTCl}{4.88}

\newcommand{\ruTWelch}{2.62}
\newcommand{\ruPWelch}{0.009}
\newcommand{\ruTCl}{1.60}
\newcommand{\ruPCl}{0.109}
\newcommand{\ruDiffPP}{2.1}
\newcommand{\ruNClusters}{1,072}
\newcommand{\ruTClTwoWay}{1.35}

\newcommand{\ruDiffBootCILo}{$-$0.4}
\newcommand{\ruDiffBootCIHi}{4.7}

\newcommand{\ruOutCarTwentyTCl}{1.55}

\newcommand{\ruOutCarSixtyTCl}{0.71}

\newcommand{\ruOutBharTCl}{0.97}

\newcommand{\ruOutMarDiff}{3.1}

\newcommand{\ruOutMarTCl}{2.54}

\newcommand{\ruOutCarThirtyVOneDiff}{1.7}

\newcommand{\ruOutCarThirtyVOneTCl}{1.28}

\newcommand{\ruSsTestPrDiff}{9.5}

\newcommand{\ruSsAddvLowDiff}{3.0}

\newcommand{\ruSsAddvMidDiff}{0.1}

\newcommand{\ruSsAddvHighDiff}{1.6}

\newcommand{\ruSsBiotechZeroDiff}{0.3}

\newcommand{\ruSsBiotechOneDiff}{6.7}

\newcommand{\ruSsBiotechOneTCl}{2.65}

\newcommand{\ruSsRoleDirectorOnlyDiff}{3.5}

\newcommand{\ruSsRoleDirectorOnlyTCl}{2.16}

\newcommand{\ruSsRoleTenPctDiff}{5.5}

\newcommand{\ruSsRoleTenPctTCl}{2.45}

\newcommand{\ruSsFirstZeroDiff}{3.2}

\newcommand{\ruSsFirstZeroTCl}{2.14}

\newcommand{\ruDiffPPTest}{2.0}
\newcommand{\ruTClTest}{0.74}

\newcommand{\ruPrTClTest}{1.77}

\newcommand{\ruNYearsDiffPos}{4}
\newcommand{\ruNYearsDiffSig}{0}

\newcommand{\ruDsQFiveDiff}{6.2}
\newcommand{\ruDsQFiveTCl}{2.26}

\newcommand{\ruDsNQuintilesNeg}{2}
\newcommand{\ruDsNQuintilesSig}{1}

\newcommand{\ruDsPooledCoefPP}{1.2}
\newcommand{\ruDsPooledTCl}{0.92}
\newcommand{\ruPanelBaseCoef}{1.6}
\newcommand{\ruPanelBaseT}{1.20}

\newcommand{\ruPanelBaseN}{7,354}
\newcommand{\ruPanelDecCoef}{0.6}
\newcommand{\ruPanelDecT}{0.42}

\newcommand{\ruPanelCtrlCoef}{$-$0.2}
\newcommand{\ruPanelCtrlT}{$-$0.12}

\newcommand{\ruPanelNControls}{12}
\newcommand{\ruPanelNDeciles}{10}
\newcommand{\ruArtNAtClip}{1,353}

\newcommand{\ruArtShareGtHundred}{15.1\%}

\newcommand{\ruArtHighShareRatioGtTwo}{72.4\%}
\newcommand{\ruArtHighShareActualHigh}{38.3\%}

\newcommand{\ruArtAdjDiff}{0.2}
\newcommand{\ruArtAdjTWelch}{0.33}

\newcommand{\ruArtAdjTCl}{0.17}

\newcommand{\ruArtAdjPrDiff}{12.1}
\newcommand{\ruArtAdjPrTCl}{6.33}

\newcommand{\ruArtVOneDiff}{4.0}
\newcommand{\ruArtVOneTWelch}{5.13}

\newcommand{\ruArtVOneTCl}{2.59}

\newcommand{\ruDivEighteenFinAdj}{$-$21.8\%}

\newcommand{\ruDivTwentyFourFinAdj}{$-$7.1\%}

\newcommand{\ruDivFinActMaxAbsMedian}{0.2\%}
\newcommand{\ruDivFinShareLowAdj}{92.2\%}
\newcommand{\ruDivFinShareLowAct}{49.2\%}

\newcommand{\mechQoneAlphaDailyBp}{$-$21.6}
\newcommand{\mechQoneAlphaCarryPp}{6.48}
\newcommand{\mechQoneMeanCARVOne}{15.1\%}

\newcommand{\mechQoneMeanDist}{$-$77.0\%}
\newcommand{\mechQonePrPos}{43.0\%}
\newcommand{\mechQonePrNeg}{25.2\%}
\newcommand{\mechQoneMeanCAR}{8.6\%}

\newcommand{\mechQoneSdCAR}{29.8\%}
\newcommand{\mechQoneMeanVol}{1.19}
\newcommand{\mechQonePrZ}{13.7\%}

\newcommand{\mechQfiveAlphaDailyBp}{8.9}
\newcommand{\mechQfiveAlphaCarryPp}{$-$2.68}
\newcommand{\mechQfiveMeanCARVOne}{$-$0.9\%}

\newcommand{\mechQfiveMeanDist}{$-$10.4\%}
\newcommand{\mechQfivePrPos}{16.5\%}
\newcommand{\mechQfivePrNeg}{11.3\%}
\newcommand{\mechQfiveMeanCAR}{1.7\%}

\newcommand{\mechQfiveSdCAR}{14.1\%}
\newcommand{\mechQfiveMeanVol}{0.41}
\newcommand{\mechQfivePrZ}{11.7\%}

\newcommand{\mechQoneQfivePrPosRatio}{2.6}

\newcommand{\mechQoneQfiveMeanCARDiffPp}{6.9}
\newcommand{\mechQoneQfiveMeanCARTCl}{4.92}

\newcommand{\mechQoneQfivePrPosDiffPp}{26.4}

\newcommand{\mechQoneQfivePrZDiffPp}{2.0}
\newcommand{\mechQoneQfivePrZTCl}{1.25}

\newcommand{\mechQoneQfivePrNegDiffPp}{13.8}
\newcommand{\mechQoneQfivePrNegTCl}{7.66}

\newcommand{\mechQoneQfiveMeanCARVOneDiffPp}{16.1}
\newcommand{\mechQoneQfiveMeanCARVOneTCl}{11.41}

\newcommand{\mechQoneQfivePrPosVOneDiffPp}{39.6}

\newcommand{\mechQoneQfiveAlphaCarryDiffPp}{9.2}
\newcommand{\mechQoneQfiveAlphaCarryTCl}{22.41}

\newcommand{\mechQoneQfiveCarryShareOfVOneGap}{57\%}

\newcommand{\mechPriceSubOneMeanCAR}{5.2\%}

\newcommand{\mechPriceOneTwoMeanCAR}{12.1\%}

\newcommand{\mechPriceAboveTenMeanCAR}{0.5\%}

\newcommand{\mechSubTwoShareAll}{12.5\%}
\newcommand{\mechSubTwoShareFlagged}{27.0\%}
\newcommand{\mechSubTwoShareRejected}{1.2\%}

\newcommand{\mechSubTwoPrPos}{43.9\%}
\newcommand{\mechAboveTenPrPos}{16.6\%}

\newcommand{\mechRhoTarget}{$-$0.24}

\newcommand{\mechRhoAbsTarget}{$-$0.29}

\newcommand{\mechRhoTargetZ}{$-$0.03}

\newcommand{\mechRhoVol}{$-$0.64}

\newcommand{\mechRhoDropShare}{89\%}
\newcommand{\mechRhoRetainedShare}{11\%}
\newcommand{\mechLogitClsDistCoef}{$-$2.32}
\newcommand{\mechLogitClsDistTCl}{$-$12.61}

\newcommand{\mechLogitZDistCoef}{$-$0.37}
\newcommand{\mechLogitZDistTCl}{$-$1.73}

\newcommand{\mechLogitZDistVolCoef}{$-$1.56}
\newcommand{\mechLogitZDistVolTCl}{$-$5.06}

\newcommand{\mechTgtPrimaryAUCthree}{0.676}

\newcommand{\mechTgtPrimaryDistGain}{6.6\%}
\newcommand{\mechTgtPrimaryDistShap}{17.6\%}
\newcommand{\mechTgtPrimaryDistRankGain}{4}
\newcommand{\mechTgtPrimaryDistRankShap}{2}

\newcommand{\mechTgtVOneDistGain}{24.8\%}
\newcommand{\mechTgtVOneDistShap}{51.5\%}

\newcommand{\mechTgtZAUCthree}{0.641}
\newcommand{\mechTgtZAUCSE}{0.018}

\newcommand{\mechTgtZDistGain}{5.6\%}
\newcommand{\mechTgtZDistShap}{11.0\%}

\newcommand{\mechDecOneMeanCAR}{0.1\%}

\newcommand{\mechDecOneMeanVol}{0.25}

\newcommand{\mechDecTenMeanCAR}{14.5\%}

\newcommand{\mechDecTenMeanVol}{1.51}
\newcommand{\mechDecTopBottomDiffPp}{14.5}
\newcommand{\mechDecTopBottomTCl}{2.99}

\newcommand{\mechDecTopBottomCIpp}{[5.3, 23.8]}

\newcommand{\mechNPairsHigh}{12}

\newcommand{\mechDistVolPearson}{$-$0.54}

\newcommand{\rxShareDriftSixtyAvail}{98.4\%}

\newcommand{\rxQoneArTzeroN}{803}
\newcommand{\rxQoneArTzeroMean}{2.60\%}

\newcommand{\rxQoneArTzeroTCl}{5.70}

\newcommand{\rxQoneArToneMean}{4.13\%}

\newcommand{\rxQoneArToneTCl}{13.19}

\newcommand{\rxQoneArToneTTwo}{13.84}

\newcommand{\rxQoneDriftThirtyMean}{4.52\%}

\newcommand{\rxQoneDriftThirtyTCl}{3.62}

\newcommand{\rxQoneDriftThirtyTTwo}{2.01}

\newcommand{\rxQoneCarThirtyMean}{8.64\%}

\newcommand{\rxQoneDayOneShare}{0.48}
\newcommand{\rxQoneDriftShare}{0.52}

\newcommand{\rxQfiveArTzeroMean}{0.90\%}

\newcommand{\rxQfiveArToneMean}{0.86\%}

\newcommand{\rxQfiveArToneTCl}{6.75}

\newcommand{\rxQfiveDriftThirtyMean}{0.88\%}

\newcommand{\rxQfiveDriftThirtyTCl}{1.87}

\newcommand{\rxAllArTzeroN}{5,408}

\newcommand{\rxAllDayOneShare}{0.53}

\newcommand{\rxDiffArTzeroPp}{1.70}

\newcommand{\rxDiffArTzeroTTwo}{4.02}

\newcommand{\rxDiffArTonePp}{3.27}

\newcommand{\rxDiffArToneTCl}{9.55}

\newcommand{\rxDiffArToneTTwo}{9.77}

\newcommand{\rxDiffDriftThirtyPp}{3.63}

\newcommand{\rxDiffDriftThirtyTCl}{2.73}

\newcommand{\rxDiffDriftThirtyTTwo}{1.70}

\newcommand{\rxDiffDriftSixtyPp}{5.50}

\newcommand{\rxDiffDriftSixtyTTwo}{1.43}

\newcommand{\rxDiffCarThirtyPp}{6.90}

\newcommand{\rxQoneQfiveArToneRatio}{4.79}

\newcommand{\rxNQuintilesDriftSigTwo}{2}
\newcommand{\rxNQuintilesArToneSigTwo}{5}

\newcommand{\rxVoneQoneArToneMean}{4.34\%}

\newcommand{\rxVoneQoneDriftThirtyMean}{10.78\%}

\newcommand{\rxVoneQoneDriftThirtyTCl}{8.54}

\newcommand{\rxVoneDiffDriftThirtyPp}{12.48}

\newcommand{\rxDiffArToneVoneMinusPrimaryPp}{0.31}
\newcommand{\rxDiffDriftThirtyVoneMinusPrimaryPp}{8.85}
\newcommand{\rxBBelowFiftyN}{4,915}

\newcommand{\rxBBelowFiftyArToneMean}{3.46\%}

\newcommand{\rxBAboveFiveArToneMean}{0.78\%}

\newcommand{\rxBDiffArTonePp}{2.68}

\newcommand{\rxBDiffArToneTTwo}{8.67}

\newcommand{\rxQoneCovidN}{245}

\newcommand{\rxQoneArToneCovidMean}{4.13\%}

\newcommand{\rxQoneDriftThirtyCovidMean}{26.54\%}

\newcommand{\rxQoneDriftThirtyCovidTCl}{3.21}

\newcommand{\rxQoneExCovidN}{2,462}

\newcommand{\rxQoneArToneExCovidMean}{4.13\%}

\newcommand{\rxQoneDriftThirtyExCovidMean}{2.32\%}

\newcommand{\rxQoneDriftThirtyExCovidTCl}{2.44}

\newcommand{\rxQoneDriftThirtyExCovidTTwo}{2.11}

\newcommand{\rxIssuerDayN}{10,858}

\newcommand{\rxIssuerDayQoneArToneMean}{3.48\%}

\newcommand{\rxIssuerDayQfiveArToneMean}{0.80\%}

\newcommand{\rxIssuerDayDiffArTonePp}{2.68}

\newcommand{\rxIssuerDayDiffArToneTCl}{11.00}

\newcommand{\rxPreQoneN}{803}

\newcommand{\rxPreQoneArTzeroMean}{2.60\%}

\newcommand{\rxPreQoneArTzeroTCl}{5.70}

\newcommand{\rxPreQoneArToneMean}{1.74\%}

\newcommand{\rxPreQoneArToneTCl}{3.16}

\newcommand{\rxPostQoneN}{1,904}

\newcommand{\rxPostQoneArTzeroMean}{0.14\%}

\newcommand{\rxPostQoneArTzeroTCl}{0.48}

\newcommand{\rxPostQoneArToneMean}{5.14\%}

\newcommand{\rxPostQoneArToneTCl}{14.89}

\newcommand{\rxPostQoneGapMean}{3.27\%}

\newcommand{\rxPostQoneGapTCl}{15.31}

\newcommand{\rxPostQoneIntradayMean}{1.68\%}

\newcommand{\rxPostQoneIntradayTCl}{5.72}

\newcommand{\rxPostDiffGapPp}{2.35}

\newcommand{\rxPostDiffGapTTwo}{8.14}

\newcommand{\rxPreQoneTwoDayMean}{4.34\%}

\newcommand{\rxPreQoneTwoDayTCl}{6.93}

\newcommand{\rxPreQoneTwoDayMinusPostDayOnePp}{$-$0.80}

\newcommand{\rxSharePreClose}{40.0\%}
\newcommand{\rxSharePostClose}{60.0\%}

\newcommand{\rxQoneRoleOfficerArToneMean}{4.23\%}

\newcommand{\rxQoneRoleOfficerDriftThirtyMean}{4.96\%}

\newcommand{\rxQoneRoleOfficerDriftThirtyTCl}{2.74}

\newcommand{\rxQoneRoleTenPctArToneMean}{3.85\%}

\newcommand{\rxQoneRoleTenPctDriftThirtyMean}{2.45\%}

\newcommand{\rxQoneRoleTenPctDriftThirtyTCl}{1.38}

\newcommand{\rxQoneRoleDirectorArToneMean}{3.97\%}

\newcommand{\rxQoneRoleDirectorDriftThirtyMean}{5.25\%}

\newcommand{\rxQoneRoleDirectorDriftThirtyTCl}{3.31}

\newcommand{\rxQoneRoleOfficerMinusDirectorArTonePp}{0.26}

\newcommand{\rxQoneRoleOfficerMinusDirectorArToneTCl}{0.44}

\newcommand{\rxCtNwLags}{5}
\newcommand{\rxCtMonthDays}{21}

\newcommand{\rxCtQoneEwBaseAlphaMonthly}{4.26\%}
\newcommand{\rxCtQoneEwBaseAlphaT}{3.41}

\newcommand{\rxCtQoneEwSkipAlphaMonthly}{0.95\%}
\newcommand{\rxCtQoneEwSkipAlphaT}{0.79}

\newcommand{\rxCtQoneVwBaseAlphaMonthly}{4.24\%}
\newcommand{\rxCtQoneVwBaseAlphaT}{3.34}

\newcommand{\rxCtQoneVwSkipAlphaMonthly}{1.01\%}
\newcommand{\rxCtQoneVwSkipAlphaT}{0.82}

\newcommand{\rxCtAllEwBaseAlphaMonthly}{2.14\%}
\newcommand{\rxCtAllEwBaseAlphaT}{5.54}

\newcommand{\rxCtAllEwSkipAlphaMonthly}{0.68\%}
\newcommand{\rxCtAllEwSkipAlphaT}{1.78}

\newcommand{\rxCtLsEwBaseAlphaMonthly}{2.30\%}
\newcommand{\rxCtLsEwBaseAlphaT}{1.74}

\newcommand{\rxCtLsEwSkipAlphaMonthly}{$-$0.17\%}
\newcommand{\rxCtLsEwSkipAlphaT}{$-$0.14}

\newcommand{\rxCtLsVwBaseAlphaMonthly}{3.04\%}
\newcommand{\rxCtLsVwBaseAlphaT}{2.22}

\newcommand{\rxCtLsVwSkipAlphaMonthly}{0.48\%}
\newcommand{\rxCtLsVwSkipAlphaT}{0.36}

\newcommand{\rxCtQoneEwSkipShareOfBase}{0.22}

\newcommand{\rxCtQoneVwSkipShareOfBase}{0.24}

\newcommand{\rxFloorFiveN}{9,095}
\newcommand{\rxFloorFiveShare}{67.2\%}
\newcommand{\rxFloorFiveNTickers}{840}
\newcommand{\rxFloorFiveQoneShareOfQone}{26.8\%}
\newcommand{\rxFloorFiveQfiveShareOfQfive}{91.4\%}
\newcommand{\rxFloorFiveQoneN}{725}

\newcommand{\rxFloorFiveQoneArToneMean}{4.17\%}

\newcommand{\rxFloorFiveQoneArToneTCl}{7.16}

\newcommand{\rxFloorFiveQoneArToneTTwo}{7.98}

\newcommand{\rxFloorFiveQoneDriftThirtyMean}{2.09\%}

\newcommand{\rxFloorFiveQoneDriftThirtyTCl}{1.64}

\newcommand{\rxFloorFiveQoneDriftThirtyTTwo}{1.23}

\newcommand{\rxFloorFiveQoneDayOneShare}{0.67}

\newcommand{\rxFloorFiveQfiveArToneMean}{0.57\%}

\newcommand{\rxFloorFiveDiffArTonePp}{3.60}

\newcommand{\rxFloorFiveDiffArToneTCl}{6.08}

\newcommand{\rxFloorFiveDiffArToneTTwo}{6.57}

\newcommand{\rxFloorFiveDiffDriftThirtyPp}{2.09}

\newcommand{\rxFloorFiveDiffDriftThirtyTTwo}{1.22}

\newcommand{\rxFloorFiveQoneQfiveArToneRatio}{7.30}

\newcommand{\rxFloorFiveCtQoneEwBaseAlphaMonthly}{3.30\%}
\newcommand{\rxFloorFiveCtQoneEwBaseAlphaT}{1.71}

\newcommand{\rxFloorFiveCtQoneEwSkipAlphaMonthly}{0.96\%}
\newcommand{\rxFloorFiveCtQoneEwSkipAlphaT}{0.56}

\newcommand{\rxFloorFiveLiqN}{4,131}
\newcommand{\rxFloorFiveLiqShare}{30.5\%}
\newcommand{\rxFloorFiveLiqNTickers}{643}
\newcommand{\rxFloorFiveLiqQoneShareOfQone}{20.2\%}

\newcommand{\rxFloorFiveLiqQoneN}{546}

\newcommand{\rxFloorFiveLiqQoneArToneMean}{4.73\%}

\newcommand{\rxFloorFiveLiqQoneGapMean}{2.91\%}

\newcommand{\rxFloorFiveLiqQoneGapTCl}{6.99}

\newcommand{\rxFloorFiveLiqQoneDriftThirtyMean}{1.70\%}

\newcommand{\rxFloorFiveLiqQoneDriftThirtyTCl}{1.05}

\newcommand{\rxFloorFiveLiqQoneDriftThirtyTTwo}{0.70}

\newcommand{\rxFloorFiveLiqQoneDayOneShare}{0.74}

\newcommand{\rxFloorFiveLiqQfiveArToneMean}{1.09\%}

\newcommand{\rxFloorFiveLiqQfiveGapMean}{0.80\%}

\newcommand{\rxFloorFiveLiqDiffArTonePp}{3.64}

\newcommand{\rxFloorFiveLiqDiffArToneTCl}{4.88}

\newcommand{\rxFloorFiveLiqDiffArToneTTwo}{5.65}

\newcommand{\rxFloorFiveLiqDiffDriftThirtyPp}{1.27}

\newcommand{\rxFloorFiveLiqDiffDriftThirtyTTwo}{0.50}

\newcommand{\rxFloorFiveLiqCtQoneEwBaseAlphaMonthly}{2.47\%}
\newcommand{\rxFloorFiveLiqCtQoneEwBaseAlphaT}{1.26}

\newcommand{\rxFloorFiveLiqCtQoneEwSkipAlphaMonthly}{$-$0.28\%}
\newcommand{\rxFloorFiveLiqCtQoneEwSkipAlphaT}{$-$0.16}

\newcommand{\rxFloorFiveLiqCtQoneVwBaseAlphaMonthly}{2.82\%}
\newcommand{\rxFloorFiveLiqCtQoneVwBaseAlphaT}{1.41}

\newcommand{\rxFloorFiveLiqCtQoneVwSkipAlphaMonthly}{0.13\%}
\newcommand{\rxFloorFiveLiqCtQoneVwSkipAlphaT}{0.07}

\newcommand{\plNPlacebo}{151,622}
\newcommand{\plNTickers}{1,179}

\newcommand{\plDrawsPerYear}{24}

\newcommand{\plYearFirst}{2018}
\newcommand{\plYearLast}{2024}
\newcommand{\plExclDays}{30}
\newcommand{\plNRawLines}{31,520}

\newcommand{\plNRowsExcluded}{562,183}
\newcommand{\plNRowsPanel}{2,144,144}

\newcommand{\plEventsPerPlacebo}{11.2}

\newcommand{\plQoneCeboMeanDist}{$-$78.6\%}
\newcommand{\plQoneEventMeanDist}{$-$77.0\%}
\newcommand{\plQoneCeboAlphaCarryPp}{7.58}

\newcommand{\plQoneEventPrimary}{8.64\%}

\newcommand{\plQoneCeboPrimary}{1.47\%}

\newcommand{\plQoneCeboVOne}{9.05\%}

\newcommand{\plQoneCeboArToneMean}{0.14\%}
\newcommand{\plQoneCeboArToneTTwo}{1.77}

\newcommand{\plQfiveCeboAlphaCarryPp}{$-$3.73}

\newcommand{\plQfiveCeboVOne}{$-$3.54\%}

\newcommand{\plGapCarryPp}{11.31}
\newcommand{\plGapCarryT}{28.27}

\newcommand{\plEventGapPrimaryPp}{6.90}

\newcommand{\plEventGapPrimaryT}{3.15}

\newcommand{\plGapPrimaryPp}{1.27}

\newcommand{\plGapPrimaryT}{0.95}

\newcommand{\plEventGapVOnePp}{16.05}

\newcommand{\plGapVOnePp}{12.58}

\newcommand{\plGapVOneTCl}{17.29}

\newcommand{\plGapVOneT}{9.27}

\newcommand{\plEventGapMarPp}{6.60}

\newcommand{\plEventGapMarT}{2.93}

\newcommand{\plGapMarPp}{1.07}

\newcommand{\plGapMarT}{0.63}

\newcommand{\plGapVOneOverEventVOne}{0.78}
\newcommand{\plGapCarryShareOfVOne}{90\%}

\newcommand{\plQoneDiffPrimaryPp}{7.18}

\newcommand{\plQoneDiffPrimaryT}{3.36}

\newcommand{\plRatioQonePrimary}{0.17}

\newcommand{\plRatioQoneVOne}{0.60}

\newcommand{\plWinQoneCeboPrimary}{$-$0.58\%}

\newcommand{\plMedQoneCeboPrimary}{$-$3.29\%}

\newcommand{\plWinQoneEventPrimary}{8.06\%}
\newcommand{\plMedQoneEventPrimary}{5.54\%}
\newcommand{\plWinGapPrimary}{$-$0.70}
\newcommand{\plWinGapPrimaryT}{$-$0.69}

\newcommand{\plWinGapVOne}{10.11}
\newcommand{\plWinGapVOneT}{9.70}

\newcommand{\plQoneCeboShareSubFive}{76.2\%}

\newcommand{\plQoneEventShareSubFive}{73.2\%}

\newcommand{\robPrimaryAUCthree}{0.676}

\newcommand{\robRollNTrainTwenty}{1,122}

\newcommand{\robRollAUCmeanThree}{0.655}

\newcommand{\robRollAUCminThree}{0.623}

\newcommand{\robRollAUCmaxThree}{0.688}
\newcommand{\robRollAUCsd}{0.030}
\newcommand{\robRollAUCminYear}{2020}
\newcommand{\robRollAUCmaxYear}{2023}

\newcommand{\robQAUCminThree}{0.633}

\newcommand{\robQAUCmaxThree}{0.722}

\newcommand{\robVolLowPosRate}{16.8\%}

\newcommand{\robVolLowAUCthree}{0.606}

\newcommand{\robVolHighPosRate}{35.5\%}

\newcommand{\robVolHighAUCthree}{0.610}

\newcommand{\robMktVolWindow}{21}

\newcommand{\robMktVolLowAUCthree}{0.660}

\newcommand{\robMktVolHighAUCthree}{0.684}

\newcommand{\robShareTestSeenTrainVal}{83.5\%}

\newcommand{\robBlockNDropRows}{5,858}
\newcommand{\robBlockNDropTickers}{376}

\newcommand{\robBlockAllAUCthree}{0.676}

\newcommand{\robBlockDeLongP}{0.983}

\newcommand{\robBlockVsCtrlIssAUC}{$-$0.001}

\newcommand{\robBlockVsCtrlRowAUC}{$-$0.006}

\newcommand{\robFirmDayN}{1,681}

\newcommand{\robFirmDayAUCthree}{0.664}
\newcommand{\robFirmDayAUCci}{[0.615, 0.706]}

\newcommand{\robGridNConfigs}{16}

\newcommand{\robGridAprioriRank}{6th}
\newcommand{\robGridAprioriRankOf}{17}

\newcommand{\robGridBestDepth}{3}
\newcommand{\robGridBestLR}{0.03}
\newcommand{\robGridBestRounds}{100}

\newcommand{\robGridBestTestAUCthree}{0.682}

\newcommand{\robGridBestVsPrimaryDeLongP}{0.139}

\newcommand{\robGridTestAUCmin}{0.662}
\newcommand{\robGridTestAUCmax}{0.683}

\newcommand{\robXgbAUCnoESthree}{0.672}

\newcommand{\robXgbAUCmaxRounds}{0.662}

\newcommand{\robTestAUCbyRoundsMax}{0.678}
\newcommand{\robTestAUCbyRoundsArgmax}{40}

\newcommand{\robTestCurveRangeTenToHundred}{0.007}

\newcommand{\robEsFullTestAUCthree}{0.675}
\newcommand{\robEsFullThr}{0.220}

\newcommand{\robEsFullTestFone}{0.478}

\newcommand{\robEmbOverlapRowsList}{143, 444, 229, 210, 386}

\newcommand{\robEmbValLateN}{253}
\newcommand{\robEmbValLatePct}{10.3\%}
\newcommand{\robEmbValKeptN}{2,204}

\newcommand{\robEmbPrimTestFone}{0.469}

\newcommand{\robEmbPurgedRounds}{15}

\newcommand{\robEmbPurgedTestAUCthree}{0.674}

\newcommand{\robEmbPurgedVsPrimaryDeLongP}{0.483}

\newcommand{\robEmbPurgedTestFone}{0.474}

\newcommand{\robNBigCAR}{69}

\newcommand{\robNBigCARTickers}{21}

\newcommand{\robBharTradeDiffPP}{4.00}
\newcommand{\robBharTradeTCl}{2.53}

\newcommand{\robTimingTradeDiffPP}{1.97}
\newcommand{\robTimingTradeTCl}{1.28}

\newcommand{\robVOneAUCthree}{0.658}

\newcommand{\robVOneTradeDiffPP}{4.20}
\newcommand{\robVOneTradeTCl}{2.51}

\newcommand{\robVariantTradeDiffMinPP}{1.97}
\newcommand{\robVariantTradeDiffMaxPP}{4.45}
\newcommand{\robVariantContrastMinPP}{1.50}
\newcommand{\robVariantContrastMaxPP}{2.41}
\newcommand{\robVariantContrastTClMin}{0.97}
\newcommand{\robVariantContrastTClMax}{2.01}
\newcommand{\robVariantTradeTClMin}{1.28}
\newcommand{\robVariantTradeTClMax}{2.85}

\newcommand{\robBharRefitAUCthree}{0.640}
\newcommand{\robBharRefitAUCci}{[0.585, 0.683]}

\newcommand{\robTimingRefitAUCthree}{0.671}
\newcommand{\robTimingRefitAUCci}{[0.621, 0.713]}

\newcommand{\robVOneRefitAUCthree}{0.673}
\newcommand{\robVOneRefitAUCci}{[0.623, 0.715]}
\newcommand{\robVOneRefitPrimaryAUCthree}{0.658}

\newcommand{\robVOneRefitDistGain}{24.8\%}
\newcommand{\robVOneRefitDistShap}{51.5\%}

\newcommand{\econThreshold}{0.184}

\newcommand{\econNFlag}{1,205}
\newcommand{\econShareFlag}{59.3\%}
\newcommand{\econTickersFlag}{356}
\newcommand{\econMeanFlag}{5.2\%}
\newcommand{\econMedianFlag}{3.0\%}
\newcommand{\econHitFlag}{55.9\%}

\newcommand{\econNRej}{828}

\newcommand{\econMeanRej}{0.7\%}
\newcommand{\econMedianRej}{0.2\%}
\newcommand{\econHitRej}{51.1\%}

\newcommand{\econMeanAll}{3.3\%}

\newcommand{\econMeanFlagVOne}{5.3\%}

\newcommand{\econMeanRejVOne}{1.1\%}

\newcommand{\econDiffFlagRejVOnePp}{4.2}
\newcommand{\econDiffVOneTCl}{2.51}

\newcommand{\econDiffFlagRejPp}{4.4}
\newcommand{\econDiffTWelch}{5.20}

\newcommand{\econDiffTCl}{2.85}

\newcommand{\econDiffNClusters}{479}

\newcommand{\econDiffTTwoWay}{2.76}

\newcommand{\econMonthsFlagGtRej}{10}
\newcommand{\econMonthsTotal}{12}

\newcommand{\econTopTwentyShare}{36.6\%}

\newcommand{\econNFlagGtFifty}{56}
\newcommand{\econMeanFlagExFifty}{1.6\%}

\newcommand{\econMeanRejExFifty}{0.5\%}

\newcommand{\econMeanAllExFifty}{1.2\%}

\newcommand{\econNetBuyAllThree}{0.3\%}
\newcommand{\econNetCILoBuyAllThree}{$-$1.4\%}
\newcommand{\econNetCIHiBuyAllThree}{2.2\%}

\newcommand{\econGrossFlag}{5.2\%}
\newcommand{\econGrossCILoFlag}{2.5\%}
\newcommand{\econGrossCIHiFlag}{8.0\%}

\newcommand{\econNetFlagThree}{2.2\%}
\newcommand{\econNetCILoFlagThree}{$-$0.5\%}
\newcommand{\econNetCIHiFlagThree}{5.0\%}
\newcommand{\econNetFlagFive}{0.2\%}
\newcommand{\econNetCILoFlagFive}{$-$2.5\%}
\newcommand{\econNetCIHiFlagFive}{3.0\%}

\newcommand{\econBreakEvenFlag}{5.2\%}

\newcommand{\econBreakEvenBuyAll}{3.3\%}

\newcommand{\econCsWindowDays}{30}

\newcommand{\econShareFiftyKOverFivePctAll}{53.3\%}
\newcommand{\econShareFiftyKOverTenPctAll}{29.1\%}
\newcommand{\econCsMedianAll}{1.33\%}

\newcommand{\econShareFiftyKOverTenPctFlag}{24.1\%}
\newcommand{\econCsMedianFlag}{1.73\%}

\newcommand{\econCsMedianRej}{0.81\%}

\newcommand{\econOrderSize}{\$50,000}

\newcommand{\econNetFlagCsMedian}{3.4\%}

\newcommand{\econCtAllAlphaMonthly}{2.10\%}
\newcommand{\econCtAllAlphaT}{5.28}

\newcommand{\econCtAllBetaSmb}{1.16}

\newcommand{\econCtFlagAlphaMonthly}{3.48\%}
\newcommand{\econCtFlagAlphaT}{1.89}

\newcommand{\econCtFlagAlphaP}{0.059}

\newcommand{\econCtRejAlphaMonthly}{$-$0.13\%}
\newcommand{\econCtRejAlphaT}{$-$0.22}

\newcommand{\econCtFlagMinusRejAlphaMonthly}{3.62\%}
\newcommand{\econCtFlagMinusRejAlphaT}{1.78}

\newcommand{\econCtFlagMinusRejAlphaP}{0.076}

\newcommand{\econCtFlagMinusRejArMonthlyVOne}{3.37\%}
\newcommand{\econCtFlagMinusRejArTVOne}{1.68}

\newcommand{\econNwLags}{5}

\newcommand{\econFullAnnualised}{30.2\%}
\newcommand{\econFullMedianAnnualised}{11.1\%}

\newcommand{\roleCfoMean}{6.0\%}
\newcommand{\roleCfoMedian}{3.2\%}
\newcommand{\roleCfoPos}{35.8\%}
\newcommand{\roleCfoDiffPp}{2.6}
\newcommand{\roleCfoTCl}{2.88}

\newcommand{\roleOfficerAnyMean}{4.5\%}

\newcommand{\roleOfficerAnyTCl}{2.25}

\newcommand{\roleDirOnlyMean}{3.4\%}

\newcommand{\roleTenPctMean}{2.7\%}

\newcommand{\roleTenPctDiffPp}{$-$1.2}
\newcommand{\roleTenPctTCl}{$-$1.34}

\newcommand{\roleTenPctTWelch}{$-$2.40}

\newcommand{\roleCfoDiffPpVOne}{3.0}
\newcommand{\roleCfoTClVOne}{3.32}

\newcommand{\roleTenPctDiffPpVOne}{$-$1.4}
\newcommand{\roleTenPctTClVOne}{$-$1.52}

\newcommand{\roleRoutineYears}{2021--2024}

\newcommand{\roleOppMean}{2.6\%}

\newcommand{\roleRoutineN}{178}
\newcommand{\roleRoutineShare}{2.0\%}
\newcommand{\roleRoutineMean}{$-$1.8\%}

\newcommand{\roleRoutineDiffPp}{$-$4.3}
\newcommand{\roleRoutineDiffTCl}{$-$5.23}

\newcommand{\roleRoutineVOneDiffPp}{$-$5.6}
\newcommand{\roleRoutineVOneDiffTCl}{$-$6.69}

\newcommand{\roleClusterZeroMean}{2.7\%}

\newcommand{\roleClusterAnyMean}{4.4\%}

\newcommand{\roleClusterAnyVsRefTCl}{2.50}

\newcommand{\roleRepeatMean}{2.7\%}

\newcommand{\roleFirstMean}{5.0\%}

\newcommand{\roleFirstDiffTCl}{3.87}

\newcommand{\roleTenPctVsNonDiffCI}{[$-$2.8, 0.6]}
\newcommand{\roleRoutineDiffCI}{[$-$5.8, $-$2.3]}

\newcommand{\roleImpRoleShap}{1.4\%}

\newcommand{\roleImpHistoryShap}{4.6\%}

\newcommand{\roleImpClusterShap}{4.3\%}

\newcommand{\roleImpWhoShap}{10.4\%}

\newcommand{\roleOlsN}{13,534}

\newcommand{\roleOlsRsq}{0.017}

\newcommand{\roleOlsCfoCoef}{2.35}
\newcommand{\roleOlsCfoT}{2.46}

\newcommand{\roleOlsTenPctCoef}{$-$1.01}
\newcommand{\roleOlsTenPctT}{$-$0.96}

\newcommand{\roleOlsFirstCoef}{1.68}
\newcommand{\roleOlsFirstT}{2.78}

\newcommand{\roleOlsClusterCoef}{1.47}
\newcommand{\roleOlsClusterT}{2.49}

\newcommand{\roleOlsLogValueCoef}{0.80}
\newcommand{\roleOlsLogValueT}{3.71}

\newcommand{\roleOlsTwoRoutineCoef}{$-$3.40}
\newcommand{\roleOlsTwoRoutineT}{$-$3.46}

\newcommand{\figCarRangeLo}{$-$60}
\newcommand{\figCarRangeHi}{80}
\newcommand{\figCarBinWidth}{2}

\newcommand{\figCarN}{13,534}

\newcommand{\figCarClipShare}{1.0\%}

\newcommand{\figCarTestN}{2,033}

\newcommand{\figCarTestClipShare}{1.3\%}

\newcommand{\figRocXgbAUCthree}{0.676}

\newcommand{\figPrXgbAPthree}{0.419}

\newcommand{\figRocRfAUCthree}{0.674}

\newcommand{\figPrRfAPthree}{0.419}

\newcommand{\figRocLrAUCthree}{0.653}

\newcommand{\figPrLrAPthree}{0.390}
\newcommand{\figPrBaseRate}{0.262}
\newcommand{\figRocN}{2,033}
\newcommand{\figRocNPos}{532}

\newcommand{\figBeeN}{2,033}
\newcommand{\figBeeTopFeature}{30-day volatility (annualized)}
\newcommand{\figBeeTopMeanAbs}{0.371}
\newcommand{\figBeeTopShare}{30.3\%}
\newcommand{\figBeeSecondFeature}{Distance from 52-week high}
\newcommand{\figBeeSecondMeanAbs}{0.215}

\newcommand{\figBeeExpectedProb}{0.270}

\newcommand{\figArtAdjGapGtTenLe}{0.2}

\newcommand{\figArtActGapGtTenLe}{2.1}

\newcommand{\figArtAdjGtTenShare}{20.9\%}
\newcommand{\figArtActGtTenShare}{9.5\%}

\newcommand{\figRcCvRounds}{26}
\newcommand{\figRcEsRounds}{10}
\newcommand{\figRcMaxRounds}{300}
\newcommand{\figRcTestAUCatCv}{0.676}
\newcommand{\figRcTestAUCatEs}{0.675}
\newcommand{\figRcTestAUCMax}{0.678}
\newcommand{\figRcTestAUCArgmax}{40}

\newcommand{\figRcTestAUCRangeTenOn}{0.016}

%% file: outputs/extension_metrics.tex
\newcommand{\ExtPersonalPositions}{302}

\newcommand{\ExtPersonalFlowAnn}{\$56,298,527}
\newcommand{\ExtPersonalNetTwentyFive}{\$1,778,434}

\newcommand{\ExtPersonalMaxDD}{\$800,183}
\newcommand{\ExtPersonalExtraCash}{\$125,189}
\newcommand{\ExtPersonalEdgeFiveHundred}{89}
\newcommand{\ExtPersonalAvgInventory}{\$4,351,529}
\newcommand{\ExtClusterPositions}{156}

\newcommand{\ExtClusterFlowAnn}{\$29,075,559}
\newcommand{\ExtClusterNetTwentyFive}{\$739,498}

\newcommand{\ExtClusterMaxDD}{\$698,390}
\newcommand{\ExtClusterExtraCash}{\$55,294}
\newcommand{\ExtClusterEdgeFiveHundred}{172}
\newcommand{\ExtClusterAvgInventory}{\$2,230,833}
\newcommand{\ExtSecondPositions}{125}

\newcommand{\ExtSecondFlowAnn}{\$23,264,761}
\newcommand{\ExtSecondNetTwentyFive}{\$486,974}

\newcommand{\ExtSecondMaxDD}{\$539,077}
\newcommand{\ExtSecondExtraCash}{\$0}
\newcommand{\ExtSecondEdgeFiveHundred}{215}
\newcommand{\ExtSecondAvgInventory}{\$1,794,592}
\newcommand{\ExtPersonalMatched}{197}
\newcommand{\ExtPersonalCoverage}{65.2\%}
\newcommand{\ExtPersonalExcess}{\$411,013}
\newcommand{\ExtPersonalExcessFifty}{\$288,654}
\newcommand{\ExtPersonalGatePass}{fails}
\newcommand{\ExtClusterMatched}{87}
\newcommand{\ExtClusterCoverage}{55.9\%}
\newcommand{\ExtClusterExcess}{\$59,764}
\newcommand{\ExtClusterExcessFifty}{\$5,580}
\newcommand{\ExtClusterGatePass}{fails}
\newcommand{\ExtSecondMatched}{9}
\newcommand{\ExtSecondCoverage}{7.3\%}
\newcommand{\ExtSecondExcess}{$-$\$1,532}
\newcommand{\ExtSecondExcessFifty}{$-$\$7,157}
\newcommand{\ExtSecondGatePass}{fails}
\newcommand{\ExtPersonalCiLow}{$-$\$499,967}
\newcommand{\ExtPersonalCiHigh}{\$1,233,343}
\newcommand{\ExtClusterCiLow}{$-$\$523,680}
\newcommand{\ExtClusterCiHigh}{\$678,126}
\newcommand{\ExtSecondCiLow}{$-$\$164,356}
\newcommand{\ExtSecondCiHigh}{\$170,867}
\newcommand{\ExtPersonalTopQuarter}{2025Q2}
\newcommand{\ExtPersonalTopQuarterShare}{40.8\%}
\newcommand{\ExtPersonalExTopQuarter}{\$1,052,981}
\newcommand{\ExtClusterTopQuarter}{2026Q2}
\newcommand{\ExtClusterTopQuarterShare}{85.3\%}
\newcommand{\ExtClusterExTopQuarter}{\$108,506}
\newcommand{\ExtSecondTopQuarter}{2026Q2}
\newcommand{\ExtSecondTopQuarterShare}{97.3\%}
\newcommand{\ExtSecondExTopQuarter}{\$13,311}
\newcommand{\ExtPersonalXmlVerified}{79}
\newcommand{\ExtPersonalXmlUnverified}{223}
\newcommand{\ExtPersonalXmlUnverifiedFlowShare}{74\%}

\newcommand{\ExtClusterNarrativeConflict}{3}
\newcommand{\ExtClusterNarrativeConflictFlow}{\$747,732}

%% file: outputs/d18_metrics.tex
\newcommand{\dxSigmaRatio}{3.10}
\newcommand{\dxZRatio}{2.51}
\newcommand{\dxZQone}{0.63}
\newcommand{\dxZQfive}{0.25}
\newcommand{\dxZDiff}{0.38}
\newcommand{\dxZDiffTTwo}{7.49}
\newcommand{\dxPosQone}{68.3\%}
\newcommand{\dxPosQfive}{56.7\%}
\newcommand{\dxPosDiffPp}{11.6}
\newcommand{\dxPosDiffT}{6.09}
\newcommand{\dxCtrlQoneCoefPp}{0.94}
\newcommand{\dxCtrlQoneTTwo}{2.11}
\newcommand{\dxCtrlZQoneCoef}{0.26}
\newcommand{\dxCtrlZQoneTTwo}{3.59}
\newcommand{\dxRawGapPp}{3.27}
\newcommand{\dxRawGapTTwo}{9.77}
\newcommand{\dxVolOnlyGapPp}{1.72}
\newcommand{\dxVolOnlyGapTTwo}{3.85}
\newcommand{\dxPriceOnlyGapPp}{2.34}
\newcommand{\dxPosCtrlPp}{5.6}
\newcommand{\dxPosCtrlT}{2.52}
\newcommand{\dxSurvHiGapPp}{3.29}
\newcommand{\dxSurvHiT}{5.24}

\newcommand{\dxSurvLoGapPp}{3.42}
\newcommand{\dxSurvLoT}{5.73}

\newcommand{\dxSurvMinYearGapPp}{2.16}
\newcommand{\dxPurAcN}{8,126}
\newcommand{\dxPurAcQone}{5.14\%}
\newcommand{\dxPurAcQfive}{1.38\%}
\newcommand{\dxPurAcGapPp}{3.76}
\newcommand{\dxPurAcGapTTwo}{8.43}
\newcommand{\dxPurAcRatio}{3.73}
\newcommand{\dxSaleAcN}{18,250}
\newcommand{\dxSaleAcQone}{$-$0.10\%}
\newcommand{\dxSaleAcQfive}{$-$0.31\%}
\newcommand{\dxSaleAcGapPp}{0.22}
\newcommand{\dxSaleAcGapTTwo}{0.71}

\newcommand{\dxEarnAcN}{4,060}
\newcommand{\dxEarnAcQone}{1.50\%}
\newcommand{\dxEarnAcQfive}{1.50\%}
\newcommand{\dxEarnAcGapPp}{0.00}
\newcommand{\dxEarnAcGapTTwo}{0.00}
\newcommand{\dxEarnAcRatio}{1.00}
\newcommand{\dxEarnTopAcN}{2,561}
\newcommand{\dxEarnTopAcQone}{4.74\%}
\newcommand{\dxEarnTopAcQfive}{2.06\%}
\newcommand{\dxEarnTopAcGapPp}{2.68}
\newcommand{\dxEarnTopAcGapTTwo}{2.24}

\newcommand{\dxEarnBadAcN}{3,568}
\newcommand{\dxEarnBadAcQone}{$-$0.60\%}
\newcommand{\dxEarnBadAcQfive}{$-$1.51\%}
\newcommand{\dxEarnBadAcGapPp}{0.91}
\newcommand{\dxEarnBadAcGapTTwo}{1.22}

\newcommand{\dxKSale}{$-$0.06}
\newcommand{\dxKSaleLo}{$-$0.27}
\newcommand{\dxKSaleHi}{0.09}

\newcommand{\dxSaleEvents}{24,893}
\newcommand{\dxEarnRatioLo}{$-$0.39}
\newcommand{\dxEarnRatioHi}{2.51}
\newcommand{\dxEarnNSue}{15,097}
\newcommand{\dxEarnNEightK}{26,528}
\newcommand{\dxEarnTopRatio}{2.30}
\newcommand{\dxEarnTopRatioLo}{1.01}
\newcommand{\dxEarnTopRatioHi}{3.94}
\newcommand{\dxZPurRatio}{2.13}
\newcommand{\dxZPurGapTTwo}{5.77}
\newcommand{\dxZPurGap}{0.43}

\newcommand{\dxZEarnGoodRatio}{0.31}
\newcommand{\dxZEarnGoodGapTTwo}{$-$1.93}

\newcommand{\dxZEarnTopRatio}{0.92}
\newcommand{\dxZEarnTopGapTTwo}{$-$0.24}

\newcommand{\dxZSaleRatio}{0.33}
\newcommand{\dxZSaleGapTTwo}{1.26}
\newcommand{\dxZSaleGap}{0.06}

\newcommand{\dxZEarnTopLo}{0.33}
\newcommand{\dxZEarnTopHi}{1.58}
\newcommand{\dnCikTickers}{1,124}
\newcommand{\dnNoCikLines}{632}
\newcommand{\dnNoCikTickers}{68}
\newcommand{\dnDeadCikLines}{50}
\newcommand{\dnDeadCikTickers}{8}
\newcommand{\dnFlagShare}{31.1\%}
\newcommand{\dnBroadShare}{12.9\%}

\newcommand{\dnErnShare}{21.8\%}
\newcommand{\dnFlagShareQone}{32.5\%}
\newcommand{\dnFlagShareQfive}{29.1\%}
\newcommand{\dnUnflaggedN}{8,856}

\newcommand{\dnUnflaggedQone}{3.55\%}
\newcommand{\dnUnflaggedQfive}{0.94\%}

\newcommand{\dnUnflaggedGapPp}{2.61}
\newcommand{\dnUnflaggedGapTTwo}{7.13}
\newcommand{\dnUnflaggedZQone}{0.54}
\newcommand{\dnUnflaggedZQfive}{0.26}
\newcommand{\dnUnflaggedZRatio}{2.10}
\newcommand{\dnUnflaggedZGap}{0.28}
\newcommand{\dnUnflaggedZGapTTwo}{4.71}

\newcommand{\dnFlaggedQone}{5.49\%}

\newcommand{\dnUnflaggedAcQone}{4.34\%}
\newcommand{\dnUnflaggedAcQfive}{1.42\%}

\newcommand{\dnUnflaggedAcGapTTwo}{6.97}

\newcommand{\dnCheckedN}{12,852}

\newcommand{\dnQoneFlagDiffPp}{1.94}
\newcommand{\dnQoneFlagDiffT}{2.77}
\newcommand{\dnCtrlUnflaggedQoneCoefPp}{0.75}
\newcommand{\dnCtrlUnflaggedQoneTTwo}{1.19}
\newcommand{\dnCtrlUnflaggedZQoneCoef}{0.22}
\newcommand{\dnCtrlUnflaggedZQoneTTwo}{2.52}

\newcommand{\dpTtwoQone}{0.77\%}
\newcommand{\dpTtwoQoneTCl}{3.83}
\newcommand{\dpTtwoQoneTTwo}{2.86}
\newcommand{\dpTtwoQfive}{0.05\%}

\newcommand{\dpTtwoGapPp}{0.72}
\newcommand{\dpTtwoGapTTwo}{2.58}

\newcommand{\dpTthreeQone}{0.34\%}

\newcommand{\dpTthreeQoneTTwo}{2.60}
\newcommand{\dpTthreeQfive}{0.07\%}

\newcommand{\dpTthreeGapPp}{0.28}
\newcommand{\dpTthreeGapTTwo}{1.93}

\newcommand{\dpTtwoTenQone}{2.33\%}

\newcommand{\dpTtwoTenQoneTTwo}{2.18}
\newcommand{\dpTtwoTenQfive}{0.38\%}

\newcommand{\dpTtwoTenGapPp}{1.95}
\newcommand{\dpTtwoTenGapTTwo}{1.73}

\newcommand{\dpTtwoTthreeQoneSumPp}{1.11}

\newcommand{\dpQoneShareTthree}{61\%}

\newcommand{\drMeanZRatio}{2.51}
\newcommand{\drMeanZLo}{1.94}
\newcommand{\drMeanZHi}{3.35}

\newcommand{\drMedianZRatio}{3.50}
\newcommand{\drMedianZLo}{2.21}
\newcommand{\drMedianZHi}{7.19}
\newcommand{\drWinsOneRatio}{2.44}
\newcommand{\drWinsFiveRatio}{2.16}

\newcommand{\drMeanArMeanSigRatio}{1.63}

\newcommand{\drOriginSlopeQone}{0.43}
\newcommand{\drOriginSlopeQfive}{0.44}
\newcommand{\drOriginSlopeRatio}{0.98}
\newcommand{\drMatchedZRatio}{1.28}
\newcommand{\drMatchedRawRatio}{1.11}

\newcommand{\drFeZDiff}{0.36}
\newcommand{\drFeZTTwo}{4.68}
\newcommand{\drFeRawDiffPp}{1.78}
\newcommand{\drFeRawTTwo}{4.09}

\newcommand{\drAcZRatio}{2.02}

\newcommand{\drExTzeroZRatio}{2.79}

\newcommand{\drIssuerDayZRatio}{2.30}

\newcommand{\drTercLoQoneN}{44}

\newcommand{\drTercMidZRatio}{1.50}

\newcommand{\drTercMidZTTwo}{2.14}

\newcommand{\drTercHiZRatio}{1.70}

\newcommand{\drTercHiZTTwo}{2.40}

\newcommand{\drPosTTwo}{6.00}

\newcommand{\dcPurQoneN}{1,904}
\newcommand{\dcPurQoneIssuers}{478}

\newcommand{\dcSaleQoneN}{2,171}

\newcommand{\dcSaleQfiveN}{7,629}

\newcommand{\dcDigGoodRawPp}{3.75}
\newcommand{\dcDigGoodRawTTwo}{3.42}

\newcommand{\dcDigGoodZ}{0.93}
\newcommand{\dcDigGoodZTTwo}{3.60}

\newcommand{\dcDigTopRawTTwo}{0.90}

\newcommand{\dcDigTopZ}{0.51}
\newcommand{\dcDigTopZTTwo}{1.65}

\newcommand{\dcDigSaleRawPp}{3.98}
\newcommand{\dcDigSaleRawTTwo}{7.28}

\newcommand{\dcDigSaleZ}{0.50}
\newcommand{\dcDigSaleZTTwo}{5.73}

\newcommand{\dcBootTopRatioDiffLo}{$-$0.55}
\newcommand{\dcBootTopRatioDiffHi}{3.54}

\newcommand{\dcBootTopZRatioDiffLo}{0.32}
\newcommand{\dcBootTopZRatioDiffHi}{2.15}

\newcommand{\dcSurvDiffPp}{0.13}

\newcommand{\dcSurvDiffTTwo}{0.15}
\newcommand{\dcSurvDiffLoPp}{$-$1.55}
\newcommand{\dcSurvDiffHiPp}{1.81}
\newcommand{\dcEarnEvents}{15,097}

\newcommand{\dcEarnNonFirst}{1,128}

\newcommand{\dcEpsAfterShare}{57\%}
\newcommand{\dcEpsSameDayShare}{39\%}
\newcommand{\dcEpsDilutedShare}{99\%}

\newcommand{\dcEarnGoodAcQoneN}{289}

\newcommand{\dcEarnGoodFirstRatio}{1.03}

\newcommand{\dcEarnTopAcQoneN}{156}

\newcommand{\dcEarnTopFirstGapPp}{3.03}
\newcommand{\dcEarnTopFirstGapTTwo}{2.41}

\newcommand{\dcEarnBadAcQoneN}{458}

\newcommand{\dcEarnSpreadQonePp}{2.11}
\newcommand{\dcEarnSpreadQfivePp}{3.01}
\newcommand{\dcEarnSpreadDiffPp}{$-$0.90}
\newcommand{\dcEarnSpreadDiffTTwo}{$-$0.80}
\newcommand{\dcEarnGoodQoneIssuers}{181}
\newcommand{\dcEarnTopQoneIssuers}{113}

\newcommand{\dcSaleAboveFiveHundredShare}{51\%}

\newcommand{\dcSaleAboveFiveHundredShareQone}{21\%}
\newcommand{\dcSaleAboveFiveHundredShareQfive}{66\%}

\newcommand{\dcSaleCapN}{8,654}
\newcommand{\dcSaleCapQoneN}{1,567}

\newcommand{\dcSaleCapQone}{$-$0.13\%}
\newcommand{\dcSaleCapQfive}{$-$0.29\%}
\newcommand{\dcSaleCapGapPp}{0.16}
\newcommand{\dcSaleCapGapTTwo}{0.36}
\newcommand{\dcSaleCapK}{$-$0.04}
\newcommand{\dcSaleCapKLo}{$-$0.30}
\newcommand{\dcSaleCapKHi}{0.16}

\newcommand{\mfAllFlowAnnM}{\$32.5 million}
\newcommand{\mfAllStockDaysPerYear}{1,551}

\newcommand{\mfAllMedianTicket}{\$8,798}

\newcommand{\mfClusterFlowAnnM}{\$4.4 million}

\newcommand{\cfFloorDropped}{9}
\newcommand{\cfAlsoJump}{4}
\newcommand{\cfRestored}{5}
\newcommand{\cfMinCarPct}{$-$170\%}
\newcommand{\cfQone}{4.12\%}
\newcommand{\cfQfive}{0.86\%}

\newcommand{\cfGapTTwo}{9.76}
\newcommand{\cfZRatio}{2.50}

\newcommand{\saN}{50}
\newcommand{\saTickers}{48}
\newcommand{\saSingleRow}{34}
\newcommand{\saAggRows}{16}
\newcommand{\saExplicitOpen}{2}
\newcommand{\saMultiTrade}{17}
\newcommand{\saUnspecified}{27}
\newcommand{\saFinancing}{2}
\newcommand{\saDrip}{1}
\newcommand{\saPlanHeld}{1}
\newcommand{\saIndirect}{14}
\newcommand{\saMultiOwner}{7}
\newcommand{\saPlanAffirmative}{2}
\newcommand{\saCtxPrior}{6}
\newcommand{\saCtxRoutine}{2}
\newcommand{\saCtxUnresolved}{42}
\newcommand{\smNullIntBias}{$+$7}
\newcommand{\smNullIntRmse}{386}
\newcommand{\smNullLoadBias}{$+$1}
\newcommand{\smNullLoadRmse}{84}
\newcommand{\smTransientIntBias}{$+$46}
\newcommand{\smTransientIntRmse}{392}
\newcommand{\smTransientLoadBias}{$+$1}
\newcommand{\smTransientLoadRmse}{85}
\newcommand{\smSelectedIntBias}{$+$1,043}
\newcommand{\smSelectedIntRmse}{1,080}
\newcommand{\smSelectedLoadBias}{$+$2}
\newcommand{\smSelectedLoadRmse}{136}
\newcommand{\smPersPosIntBias}{$+$13}
\newcommand{\smPersPosIntRmse}{390}
\newcommand{\smPersPosLoadBias}{$+$203}
\newcommand{\smPersPosLoadRmse}{221}
\newcommand{\smPersNegIntBias}{$-$6}
\newcommand{\smPersNegIntRmse}{391}
\newcommand{\smPersNegLoadBias}{$-$201}
\newcommand{\smPersNegLoadRmse}{219}
\newcommand{\smPersEventIntBias}{$-$7}
\newcommand{\smPersEventIntRmse}{387}
\newcommand{\smPersEventLoadBias}{$+$200}
\newcommand{\smPersEventLoadRmse}{217}
\newcommand{\smPaths}{10,000}
\newcommand{\smSelectedDraws}{100,000}

%% file: insider_v2.bbl
\begin{thebibliography}{65}
\providecommand{\natexlab}[1]{#1}
\providecommand{\url}[1]{\texttt{#1}}
\expandafter\ifx\csname urlstyle\endcsname\relax
  \providecommand{\doi}[1]{doi: #1}\else
  \providecommand{\doi}{doi: \begingroup \urlstyle{rm}\Url}\fi

\bibitem[Ahern(2009)]{ahern2009}
Kenneth~R. Ahern.
\newblock Sample selection and event study estimation.
\newblock \emph{Journal of Empirical Finance}, 16\penalty0 (3):\penalty0 466--482, 2009.
\newblock \doi{10.1016/j.jempfin.2009.01.003}.

\bibitem[Ahn(2026)]{ahn2026}
Hyun Ahn.
\newblock Reading the insider cluster: Concurrent disclosure-tone classification predicts the cross-section of post-filing returns.
\newblock SSRN Working Paper No.\ 7178138, 2026.

\bibitem[Aktas et~al.(2007)Aktas, de~Bodt, and Cousin]{aktas2007}
Nihat Aktas, Eric de~Bodt, and Jean-Gabriel Cousin.
\newblock Event studies with a contaminated estimation period.
\newblock \emph{Journal of Corporate Finance}, 13\penalty0 (1):\penalty0 129--145, 2007.
\newblock \doi{10.1016/j.jcorpfin.2006.09.001}.

\bibitem[Alldredge and Blank(2019)]{alldredge2019}
Dallin~M. Alldredge and Brian Blank.
\newblock Do insiders cluster trades with colleagues? evidence from daily insider trading.
\newblock \emph{Journal of Financial Research}, 42\penalty0 (2):\penalty0 331--360, 2019.
\newblock \doi{10.1111/jfir.12172}.

\bibitem[Aroussi(2025)]{yfinance}
Ran Aroussi.
\newblock yfinance: Download market data from {Y}ahoo! {F}inance's {API} ({P}ython package), version 0.2.58.
\newblock \url{https://pypi.org/project/yfinance/0.2.58/}, 2025.
\newblock Released 2025-05-02. Source: \url{https://github.com/ranaroussi/yfinance}. Accessed 2026-09-17.

\bibitem[Baber and Kang(2002)]{baber2002}
William~R. Baber and Sok-Hyon Kang.
\newblock The impact of split adjusting and rounding on analysts' forecast error calculations.
\newblock \emph{Accounting Horizons}, 16\penalty0 (4):\penalty0 277--289, 2002.
\newblock \doi{10.2308/acch.2002.16.4.277}.

\bibitem[Barber and Lyon(1997)]{barber1997}
Brad~M. Barber and John~D. Lyon.
\newblock Detecting long-run abnormal stock returns: The empirical power and specification of test statistics.
\newblock \emph{Journal of Financial Economics}, 43\penalty0 (3):\penalty0 341--372, 1997.
\newblock \doi{10.1016/S0304-405X(96)00890-2}.

\bibitem[Batebi and Elnahas(2026)]{batebi2026}
Solmaz Batebi and Ahmed Elnahas.
\newblock Are machines better predictors of insider trading?
\newblock \emph{Global Finance Journal}, 69:\penalty0 101237, 2026.
\newblock \doi{10.1016/j.gfj.2026.101237}.

\bibitem[Blume and Stambaugh(1983)]{blume1983}
Marshall~E. Blume and Robert~F. Stambaugh.
\newblock Biases in computed returns.
\newblock \emph{Journal of Financial Economics}, 12\penalty0 (3):\penalty0 387--404, 1983.
\newblock \doi{10.1016/0304-405X(83)90056-9}.

\bibitem[Brochet(2010)]{brochet2010}
Francois Brochet.
\newblock Information content of insider trades before and after the {S}arbanes-{O}xley {A}ct.
\newblock \emph{The Accounting Review}, 85\penalty0 (2):\penalty0 419--446, 2010.
\newblock \doi{10.2308/accr.2010.85.2.419}.

\bibitem[Cameron et~al.(2011)Cameron, Gelbach, and Miller]{cameron2011}
A.~Colin Cameron, Jonah~B. Gelbach, and Douglas~L. Miller.
\newblock Robust inference with multiway clustering.
\newblock \emph{Journal of Business \& Economic Statistics}, 29\penalty0 (2):\penalty0 238--249, 2011.
\newblock \doi{10.1198/jbes.2010.07136}.

\bibitem[Chakravorty and Elsayed(2025)]{chakravorty2025}
Amitabh Chakravorty and Nelly Elsayed.
\newblock A comparative study of machine learning algorithms for stock price prediction using insider trading data.
\newblock arXiv preprint arXiv:2502.08728, 2025.
\newblock URL \url{https://arxiv.org/abs/2502.08728}.

\bibitem[Chen et~al.(2024)Chen, Pelger, and Zhu]{chen2024}
Luyang Chen, Markus Pelger, and Jason Zhu.
\newblock Deep learning in asset pricing.
\newblock \emph{Management Science}, 70\penalty0 (2):\penalty0 714--750, 2024.
\newblock \doi{10.1287/mnsc.2023.4695}.

\bibitem[Chen and Guestrin(2016)]{chen2016}
Tianqi Chen and Carlos Guestrin.
\newblock {XGBoost}: A scalable tree boosting system.
\newblock In \emph{Proceedings of the 22nd {ACM} {SIGKDD} International Conference on Knowledge Discovery and Data Mining ({KDD} '16)}, pages 785--794, New York, NY, 2016. ACM.
\newblock \doi{10.1145/2939672.2939785}.

\bibitem[Cohen et~al.(2012)Cohen, Malloy, and Pomorski]{cohen2012}
Lauren Cohen, Christopher Malloy, and Lukasz Pomorski.
\newblock Decoding inside information.
\newblock \emph{The Journal of Finance}, 67\penalty0 (3):\penalty0 1009--1043, 2012.
\newblock \doi{10.1111/j.1540-6261.2012.01740.x}.

\bibitem[Corwin and Schultz(2012)]{corwin2012}
Shane~A. Corwin and Paul Schultz.
\newblock A simple way to estimate bid-ask spreads from daily high and low prices.
\newblock \emph{The Journal of Finance}, 67\penalty0 (2):\penalty0 719--760, 2012.
\newblock \doi{10.1111/j.1540-6261.2012.01729.x}.

\bibitem[DeLong et~al.(1988)DeLong, DeLong, and Clarke-Pearson]{delong1988}
Elizabeth~R. DeLong, David~M. DeLong, and Daniel~L. Clarke-Pearson.
\newblock Comparing the areas under two or more correlated receiver operating characteristic curves: A nonparametric approach.
\newblock \emph{Biometrics}, 44\penalty0 (3):\penalty0 837--845, 1988.
\newblock \doi{10.2307/2531595}.

\bibitem[Fama(1998)]{fama1998}
Eugene~F. Fama.
\newblock Market efficiency, long-term returns, and behavioral finance.
\newblock \emph{Journal of Financial Economics}, 49\penalty0 (3):\penalty0 283--306, 1998.
\newblock \doi{10.1016/S0304-405X(98)00026-9}.

\bibitem[Fama and French(1993)]{fama1993}
Eugene~F. Fama and Kenneth~R. French.
\newblock Common risk factors in the returns on stocks and bonds.
\newblock \emph{Journal of Financial Economics}, 33\penalty0 (1):\penalty0 3--56, 1993.
\newblock \doi{10.1016/0304-405X(93)90023-5}.

\bibitem[Fidrmuc et~al.(2006)Fidrmuc, Goergen, and Renneboog]{fidrmuc2006}
Jana~P. Fidrmuc, Marc Goergen, and Luc Renneboog.
\newblock Insider trading, news releases, and ownership concentration.
\newblock \emph{The Journal of Finance}, 61\penalty0 (6):\penalty0 2931--2973, 2006.
\newblock \doi{10.1111/j.1540-6261.2006.01008.x}.

\bibitem[Francis et~al.(2018)Francis, Mubako, and Olsen]{francis2018}
Rick~N. Francis, Grace Mubako, and Lori Olsen.
\newblock Archival research considerations for {CRSP} data.
\newblock \emph{Accounting Research Journal}, 31\penalty0 (3):\penalty0 360--370, 2018.
\newblock \doi{10.1108/ARJ-06-2016-0065}.

\bibitem[French(2026)]{french_library}
Kenneth~R. French.
\newblock Kenneth {R}. {F}rench data library.
\newblock \url{https://mba.tuck.dartmouth.edu/pages/faculty/ken.french/data_library.html}, 2026.
\newblock Tuck School of Business, Dartmouth College. Accessed 2026-09-17.

\bibitem[George and Hwang(2004)]{george2004}
Thomas~J. George and Chuan-Yang Hwang.
\newblock The 52-week high and momentum investing.
\newblock \emph{The Journal of Finance}, 59\penalty0 (5):\penalty0 2145--2176, 2004.
\newblock \doi{10.1111/j.1540-6261.2004.00695.x}.

\bibitem[Goldsmith-Pinkham and Lyu(2025)]{goldsmithpinkham2025}
Paul Goldsmith-Pinkham and Tianshu Lyu.
\newblock Causal inference in financial event studies.
\newblock arXiv preprint arXiv:2511.15123, 2025.
\newblock URL \url{https://arxiv.org/abs/2511.15123}.

\bibitem[Gu et~al.(2020)Gu, Kelly, and Xiu]{gu2020}
Shihao Gu, Bryan Kelly, and Dacheng Xiu.
\newblock Empirical asset pricing via machine learning.
\newblock \emph{The Review of Financial Studies}, 33\penalty0 (5):\penalty0 2223--2273, 2020.
\newblock \doi{10.1093/rfs/hhaa009}.

\bibitem[Hong and Stein(1999)]{hong1999}
Harrison Hong and Jeremy~C. Stein.
\newblock A unified theory of underreaction, momentum trading, and overreaction in asset markets.
\newblock \emph{The Journal of Finance}, 54\penalty0 (6):\penalty0 2143--2184, 1999.
\newblock \doi{10.1111/0022-1082.00184}.

\bibitem[Hong et~al.(2000)Hong, Lim, and Stein]{hong2000}
Harrison Hong, Terence Lim, and Jeremy~C. Stein.
\newblock Bad news travels slowly: Size, analyst coverage, and the profitability of momentum strategies.
\newblock \emph{The Journal of Finance}, 55\penalty0 (1):\penalty0 265--295, 2000.
\newblock \doi{10.1111/0022-1082.00206}.

\bibitem[Hou(2007)]{hou2007}
Kewei Hou.
\newblock Industry information diffusion and the lead-lag effect in stock returns.
\newblock \emph{The Review of Financial Studies}, 20\penalty0 (4):\penalty0 1113--1138, 2007.
\newblock \doi{10.1093/revfin/hhm003}.

\bibitem[Huang et~al.(2025)Huang, Ma, Gao, Liu, Liu, Ma, Moe, Zhang, Xie, Han, Wang, Wang, and Yu]{huang2025}
Cheng Huang, Yao Ma, Fan Gao, Yutong Liu, Yadi Liu, Xiaoli Ma, Ye~Aung Moe, Yuhan Zhang, Weizheng Xie, Zeyu Han, Xiangxiang Wang, Hao Wang, and Yongbin Yu.
\newblock Modeling insider filing delays in financial markets with an interpretable {XGBoost} framework.
\newblock arXiv preprint arXiv:2507.20162, 2025.
\newblock URL \url{https://arxiv.org/abs/2507.20162}.

\bibitem[Jaffe(1974)]{jaffe1974}
Jeffrey~F. Jaffe.
\newblock Special information and insider trading.
\newblock \emph{The Journal of Business}, 47\penalty0 (3):\penalty0 410--428, 1974.
\newblock \doi{10.1086/295655}.

\bibitem[Jegadeesh and Titman(1993)]{jegadeesh1993}
Narasimhan Jegadeesh and Sheridan Titman.
\newblock Returns to buying winners and selling losers: Implications for stock market efficiency.
\newblock \emph{The Journal of Finance}, 48\penalty0 (1):\penalty0 65--91, 1993.
\newblock \doi{10.1111/j.1540-6261.1993.tb04702.x}.

\bibitem[Jeng et~al.(2003)Jeng, Metrick, and Zeckhauser]{jeng2003}
Leslie~A. Jeng, Andrew Metrick, and Richard Zeckhauser.
\newblock Estimating the returns to insider trading: A performance-evaluation perspective.
\newblock \emph{The Review of Economics and Statistics}, 85\penalty0 (2):\penalty0 453--471, 2003.
\newblock \doi{10.1162/003465303765299936}.

\bibitem[Jenter(2005)]{jenter2005}
Dirk Jenter.
\newblock Market timing and managerial portfolio decisions.
\newblock \emph{The Journal of Finance}, 60\penalty0 (4):\penalty0 1903--1949, 2005.
\newblock \doi{10.1111/j.1540-6261.2005.00783.x}.

\bibitem[Kang et~al.(2018)Kang, Kim, and Wang]{kang2018}
Chang-Mo Kang, Donghyun Kim, and Qinghai Wang.
\newblock Cluster trading of corporate insiders.
\newblock Working paper, 2018.

\bibitem[Ke et~al.(2003)Ke, Huddart, and Petroni]{ke2003}
Bin Ke, Steven Huddart, and Kathy Petroni.
\newblock What insiders know about future earnings and how they use it: Evidence from insider trades.
\newblock \emph{Journal of Accounting and Economics}, 35\penalty0 (3):\penalty0 315--346, 2003.
\newblock \doi{10.1016/S0165-4101(03)00036-3}.

\bibitem[Kolari and Pynn{\"o}nen(2010)]{kolari2010}
James~W. Kolari and Seppo Pynn{\"o}nen.
\newblock Event study testing with cross-sectional correlation of abnormal returns.
\newblock \emph{The Review of Financial Studies}, 23\penalty0 (11):\penalty0 3996--4025, 2010.
\newblock \doi{10.1093/rfs/hhq072}.

\bibitem[Kong et~al.(2025)Kong, Wei, and Zhang]{kong2025}
Tongyang Kong, Chishen Wei, and Lei Zhang.
\newblock Can machines discover new signals from trading data?
\newblock SSRN Working Paper No.\ 5967134, 2025.
\newblock URL \url{https://ssrn.com/abstract=5967134}.

\bibitem[Kothari and Warner(2007)]{kothari2007}
S.~P. Kothari and Jerold~B. Warner.
\newblock Econometrics of event studies.
\newblock In B.~Espen Eckbo, editor, \emph{Handbook of Empirical Corporate Finance}, volume~1, chapter~1, pages 3--36. Elsevier/North-Holland, Amsterdam, 2007.
\newblock \doi{10.1016/B978-0-444-53265-7.50015-9}.

\bibitem[Kravet and Lee(2026)]{kravet2026}
Todd~D. Kravet and Minah Lee.
\newblock When are insider purchases credible signals of private information? {E}vidence from acquisitions.
\newblock SSRN Working Paper No.\ 7320700, 2026.

\bibitem[Lakonishok and Lee(2001)]{lakonishok2001}
Josef Lakonishok and Inmoo Lee.
\newblock Are insider trades informative?
\newblock \emph{The Review of Financial Studies}, 14\penalty0 (1):\penalty0 79--111, 2001.
\newblock \doi{10.1093/rfs/14.1.79}.

\bibitem[Lasfer and Ye(2024)]{lasfer2024}
Meziane Lasfer and Xiaoke Ye.
\newblock Corporate insiders' exploitation of investors' anchoring bias at the 52-week high and low.
\newblock \emph{The Financial Review}, 59\penalty0 (2):\penalty0 391--432, 2024.
\newblock \doi{10.1111/fire.12371}.

\bibitem[Lee and Piqueira(2019)]{lee2019}
Eunju Lee and Natalia Piqueira.
\newblock Behavioral biases of informed traders: Evidence from insider trading on the 52-week high.
\newblock \emph{Journal of Empirical Finance}, 52:\penalty0 56--75, 2019.
\newblock \doi{10.1016/j.jempfin.2019.02.007}.

\bibitem[Lesmond et~al.(2004)Lesmond, Schill, and Zhou]{lesmond2004}
David~A. Lesmond, Michael~J. Schill, and Chunsheng Zhou.
\newblock The illusory nature of momentum profits.
\newblock \emph{Journal of Financial Economics}, 71\penalty0 (2):\penalty0 349--380, 2004.
\newblock \doi{10.1016/S0304-405X(03)00206-X}.

\bibitem[Lorie and Niederhoffer(1968)]{lorie1968}
James~H. Lorie and Victor Niederhoffer.
\newblock Predictive and statistical properties of insider trading.
\newblock \emph{The Journal of Law and Economics}, 11\penalty0 (1):\penalty0 35--53, 1968.
\newblock \doi{10.1086/466642}.

\bibitem[Lundberg and Lee(2017)]{lundberg2017}
Scott~M. Lundberg and Su-In Lee.
\newblock A unified approach to interpreting model predictions.
\newblock In \emph{Advances in Neural Information Processing Systems 30 ({NIPS} 2017)}, volume~30, 2017.
\newblock URL \url{https://papers.nips.cc/paper_files/paper/2017/hash/8a20a8621978632d76c43dfd28b67767-Abstract.html}.

\bibitem[Lundberg et~al.(2020)Lundberg, Erion, Chen, DeGrave, Prutkin, Nair, Katz, Himmelfarb, Bansal, and Lee]{lundberg2020}
Scott~M. Lundberg, Gabriel Erion, Hugh Chen, Alex DeGrave, Jordan~M. Prutkin, Bala Nair, Ronit Katz, Jonathan Himmelfarb, Nisha Bansal, and Su-In Lee.
\newblock From local explanations to global understanding with explainable {AI} for trees.
\newblock \emph{Nature Machine Intelligence}, 2\penalty0 (1):\penalty0 56--67, 2020.
\newblock \doi{10.1038/s42256-019-0138-9}.

\bibitem[Lyon et~al.(1999)Lyon, Barber, and Tsai]{lyon1999}
John~D. Lyon, Brad~M. Barber, and Chih-Ling Tsai.
\newblock Improved methods for tests of long-run abnormal stock returns.
\newblock \emph{The Journal of Finance}, 54\penalty0 (1):\penalty0 165--201, 1999.
\newblock \doi{10.1111/0022-1082.00101}.

\bibitem[Mandelker(1974)]{mandelker1974}
Gershon Mandelker.
\newblock Risk and return: The case of merging firms.
\newblock \emph{Journal of Financial Economics}, 1\penalty0 (4):\penalty0 303--335, 1974.
\newblock \doi{10.1016/0304-405X(74)90012-9}.

\bibitem[Mitchell and Stafford(2000)]{mitchell2000}
Mark~L. Mitchell and Erik Stafford.
\newblock Managerial decisions and long-term stock price performance.
\newblock \emph{The Journal of Business}, 73\penalty0 (3):\penalty0 287--329, 2000.
\newblock \doi{10.1086/209645}.

\bibitem[Neupane and Griva(2024)]{neupane2024}
Krishna Neupane and Igor Griva.
\newblock A random forest approach to detect and identify unlawful insider trading.
\newblock arXiv preprint arXiv:2411.13564, 2024.
\newblock URL \url{https://arxiv.org/abs/2411.13564}.

\bibitem[Newey and West(1987)]{newey1987}
Whitney~K. Newey and Kenneth~D. West.
\newblock A simple, positive semi-definite, heteroskedasticity and autocorrelation consistent covariance matrix.
\newblock \emph{Econometrica}, 55\penalty0 (3):\penalty0 703--708, 1987.
\newblock \doi{10.2307/1913610}.

\bibitem[Novy-Marx and Velikov(2016)]{novymarx2016}
Robert Novy-Marx and Mihail Velikov.
\newblock A taxonomy of anomalies and their trading costs.
\newblock \emph{The Review of Financial Studies}, 29\penalty0 (1):\penalty0 104--147, 2016.
\newblock \doi{10.1093/rfs/hhv063}.

\bibitem[Oenschl{\"a}ger and M{\"o}llenhoff(2025)]{oenschlager2025}
Eike Oenschl{\"a}ger and Steffen M{\"o}llenhoff.
\newblock Insider filings as trading signals --- {D}oes it pay to be fast?
\newblock \emph{Finance Research Letters}, 72:\penalty0 106514, 2025.
\newblock \doi{10.1016/j.frl.2024.106514}.

\bibitem[{OpenInsider}(2026)]{openinsider}
{OpenInsider}.
\newblock {OpenInsider}: {SEC} {F}orm 4 insider trading screener.
\newblock \url{http://openinsider.com/}, 2026.
\newblock Accessed 2026-09-17.

\bibitem[Payne and Thomas(2003)]{payne2003}
Jeff~L. Payne and Wayne~B. Thomas.
\newblock The implications of using stock-split adjusted {I/B/E/S} data in empirical research.
\newblock \emph{The Accounting Review}, 78\penalty0 (4):\penalty0 1049--1067, 2003.
\newblock \doi{10.2308/accr.2003.78.4.1049}.

\bibitem[Petersen(2009)]{petersen2009}
Mitchell~A. Petersen.
\newblock Estimating standard errors in finance panel data sets: Comparing approaches.
\newblock \emph{The Review of Financial Studies}, 22\penalty0 (1):\penalty0 435--480, 2009.
\newblock \doi{10.1093/rfs/hhn053}.

\bibitem[Piotroski and Roulstone(2005)]{piotroski2005}
Joseph~D. Piotroski and Darren~T. Roulstone.
\newblock Do insider trades reflect both contrarian beliefs and superior knowledge about future cash flow realizations?
\newblock \emph{Journal of Accounting and Economics}, 39\penalty0 (1):\penalty0 55--81, 2005.
\newblock \doi{10.1016/j.jacceco.2004.01.003}.

\bibitem[Rozeff and Zaman(1988)]{rozeff1988}
Michael~S. Rozeff and Mir~A. Zaman.
\newblock Market efficiency and insider trading: New evidence.
\newblock \emph{The Journal of Business}, 61\penalty0 (1):\penalty0 25--44, 1988.
\newblock \doi{10.1086/296418}.

\bibitem[Rozeff and Zaman(1998)]{rozeff1998}
Michael~S. Rozeff and Mir~A. Zaman.
\newblock Overreaction and insider trading: Evidence from growth and value portfolios.
\newblock \emph{The Journal of Finance}, 53\penalty0 (2):\penalty0 701--716, 1998.
\newblock \doi{10.1111/0022-1082.275500}.

\bibitem[Seyhun(1986)]{seyhun1986}
H.~Nejat Seyhun.
\newblock Insiders' profits, costs of trading, and market efficiency.
\newblock \emph{Journal of Financial Economics}, 16\penalty0 (2):\penalty0 189--212, 1986.
\newblock \doi{10.1016/0304-405X(86)90060-7}.

\bibitem[Tang et~al.(2025)Tang, Zhang, and Wang]{tang2025}
Chun Tang, Ditian Zhang, and Yacang Wang.
\newblock Do insider characteristics matter in opportunistic selling? {E}vidence from interpretable machine learning.
\newblock \emph{Finance Research Letters}, 85:\penalty0 107817, 2025.
\newblock \doi{10.1016/j.frl.2025.107817}.

\bibitem[{U.S. Securities and Exchange Commission}(2002)]{sec2002}
{U.S. Securities and Exchange Commission}.
\newblock Ownership reports and trading by officers, directors and principal security holders.
\newblock Final Rule, Release No.\ 34-46421, August 2002.
\newblock Adopted August 27, 2002; effective August 29, 2002. \url{https://www.sec.gov/rule-release/34-46421}, accessed 2026-09-17.

\bibitem[{U.S. Securities and Exchange Commission}(2024)]{sec_form4}
{U.S. Securities and Exchange Commission}.
\newblock Officers, directors and 10\% shareholders ({S}ection 16 of the {E}xchange {A}ct; {F}orms 3, 4 and 5).
\newblock \url{https://www.sec.gov/resources-small-businesses/going-public/officers-directors-10-shareholders}, 2024.
\newblock Last reviewed or updated June 6, 2024. Accessed 2026-09-17.

\bibitem[{U.S. Securities and Exchange Commission}(2025)]{sec_xbrl_companyfacts}
{U.S. Securities and Exchange Commission}.
\newblock {EDGAR} application programming interfaces: {XBRL} company facts.
\newblock \url{https://www.sec.gov/search-filings/edgar-application-programming-interfaces}, 2025.
\newblock Endpoint \url{https://data.sec.gov/api/xbrl/companyfacts/CIK\#\#\#\#\#\#\#\#\#\#.json}. Page last modified April 8, 2025. Accessed 2026-09-17.

\bibitem[Veenman(2012)]{veenman2012}
David Veenman.
\newblock Disclosures of insider purchases and the valuation implications of past earnings signals.
\newblock \emph{The Accounting Review}, 87\penalty0 (1):\penalty0 313--342, 2012.
\newblock \doi{10.2308/accr-10162}.

\end{thebibliography}
